\documentclass[%
nofootinbib,
 amsmath,amssymb,
 aps,
floatfix,
10pt,
prd,twocolumn,
superscriptaddress]{revtex4-2}

\usepackage[normalem]{ulem}
\usepackage{graphicx}% Include figure files
\usepackage{dcolumn}% Align table columns on decimal point
\usepackage{bm}% bold math
\usepackage[dvipsnames]{xcolor}
\usepackage{soul}
\usepackage{graphicx}  
\usepackage[colorlinks]{hyperref}
\hypersetup{
  colorlinks=true,
  linkcolor={MidnightBlue},
  citecolor={OliveGreen},
  urlcolor={gray}
}

\newcommand{\SwitchableClearpage}{} % put \clearpage in the second bracket to turn on page clearing

\begin{document}

%\preprint{APS/123-QED}

\title{Neutrino transport and flavor instabilities in a post-merger disk}

\author{Erick Urquilla}
\affiliation{Department of Physics and Astronomy, University of Tennessee, Knoxville, TN 37996, USA.}
\email{eurquill@vols.utk.edu} 

\author{Swapnil Shankar}
\affiliation{Faculty of Mathematics, Informatics and Natural Sciences, University of Hamburg, Gojenbergsweg 112, 21029 Hamburg, Germany}

\author{Debraj Kundu}
\affiliation{Department of Physics and Astronomy, University of Tennessee, Knoxville, TN 37996, USA.}

\author{Julien Froustey}
\affiliation{Institut de Física Corpuscular (IFIC), CSIC-Universitat de València, Parc Científic UV, C/ Catedrático José Beltrán 2, E-46980 Paterna (València), Spain}

\author{Sherwood Richers}
\affiliation{Department of Physics and Astronomy, University of Tennessee, Knoxville, TN 37996, USA.}

\author{Jonah M. Miller}
\affiliation{Michigan SPARC, Los Alamos National Laboratory, Ann Arbor, MI 48109, USA}
\affiliation{Computational Physics and Methods, Los Alamos National Laboratory, Los Alamos, NM 87545, USA}
\affiliation{Center for Theoretical Astrophysics, Los Alamos National Laboratory, Los Alamos, NM 87545, USA}

\author{Gail C. McLaughlin}
\affiliation{Department of Physics and Astronomy, North Carolina State University, Raleigh, NC 27695, USA}

\author{James P. Kneller}
\affiliation{Department of Physics and Astronomy, North Carolina State University, Raleigh, NC 27695, USA}

\author{Francois Foucart}
\affiliation{Department of Physics \& Astronomy, University of New Hampshire, 9 Library Way, Durham, NH 03824, USA}

\date{\today}% It is always \today, today, 

\begin{abstract}

    Neutron star mergers are multimessenger sources whose dynamics and signals depend critically on neutrinos and their flavor transformations. We investigate whether fast and collisional neutrino flavor instabilities (FFIs and CFIs) arise in a GW170817-like post-merger accretion disk using the six-dimensional phase-space particle-in-cell transport code \texttt{EMU}. In the accretion disk, the neutrino radiation field naturally develops electron-lepton-number crossings through the interplay between the more isotropic electron-neutrino field and the more anisotropic electron-antineutrino field. The disk's neutrino field is also unstable to CFI, though on longer timescales than FFI. We study how instabilities relax by performing quantum-kinetic simulations in centimeter-wide domains with periodic boundary conditions at selected points. We find that the isolated effects of FFI and CFI increase the heavy-lepton neutrino fluxes. CFI breaks the initial symmetry of the heavy-flavor sector by raising the average energy of heavy-flavor antineutrinos above that of heavy-flavor neutrinos. We perform global quantum-kinetic simulations using a Hamiltonian attenuated by factors between $10^{-12}$ and $10^{-5}$. Although we resolve unstable flavor waves, attenuation increases the instability growth timescale by orders of magnitude relative to the advection timescale; therefore, neutrinos leave the domain before FFI and CFI can saturate. The in-medium effective mixing angles and mass splittings, driven by the combined vacuum, matter, and neutrino-neutrino potentials, then govern the quantum-kinetic evolution. These results highlight the resolution and scaling requirements for future global simulations that capture instability growth, saturation, and advection simultaneously.

\end{abstract}

\maketitle

\SwitchableClearpage
\section{Introduction}
    
    Neutron star mergers (NSMs) are a prime example of multimessenger sources, emitting gravitational waves, short gamma-ray bursts, kilonovae, electromagnetic afterglows, and neutrinos \cite{burns2020neutron,fernandez2016electro,Margutii2021firstmul,radice_dynamics_2020,Radice_2018,Margalit_2019}. Substantial progress has been made in connecting NSM observables to fundamental physics via general-relativistic magnetohydrodynamic simulations. Computational efforts focus on predicting the ejecta mass and composition, the kilonova signal, neutrino luminosities, the gravitational-wave signature, and the nature of the remnant \cite{Foucart2021estimating,Mosta_2020_amagnetar,Miller2019,Nedora_2022,Zhu2021fullygeneral,Metzger_2021,just2022neutrino}. The gravitational-wave event GW170817 \cite{Abbott2017gw170817} was identified as the merger of two neutron stars and was observed in both gravitational waves and electromagnetic emission \cite{Abbott2017multimessengerobs,Cowperthwaite_2017}. Among its electromagnetic counterparts, a bright kilonova was detected, whose light curve was powered by the radioactive decay of freshly synthesized nuclei in the merger ejecta. The properties of this kilonova, such as its luminosity and color evolution, are consistent with the production of heavy elements through the rapid neutron-capture ($r$-) process \cite{metzger2017kilonovae,barnes2020physics}. This provided the first direct observational evidence that NSMs are sites of $r$-process nucleosynthesis and contribute significantly to the origin of the heaviest elements in the Universe.
    
    Neutrinos are central to the physics of NSMs (see Ref.~\cite{Foucart:2024cjr} for a recent comprehensive review). In NSMs with neutron-star mass ratios much larger than unity, the lower-mass star is tidally disrupted. When the mass ratio is close to unity, however, both stars disrupt each other, forming a hot, dense accretion disk, partly through the fallback of bound ejecta, around a compact remnant that may be either a hypermassive neutron star or a black hole. Photons in the disk and hypermassive remnant are strongly coupled to the fluid. Neutrinos are also trapped deep inside the hypermassive remnant, but they decouple near the surface and in the disk, making neutrino emission the primary cooling mechanism. Charged–current reactions interconvert neutrons and protons and therefore regulate the electron fraction $Y_{e}$ of the outflows. The resulting changes in neutron richness shape $r$-process nucleosynthesis and imprint themselves on kilonova observables \cite{Radice2020thedynamics,curtis2023nucleosynthesis,wu2015effects,Lippuner_2017}. Because heavy-lepton neutrinos do not directly alter the neutron and proton balance, flavor conversion of electron (anti)neutrinos into heavy flavors can modify $Y_{e}$ and, in turn, the final heavy-element yields, the kilonova signal, the disk dynamics and gravitational wave emission \cite{lund2025angle, just2022,qiu2025neutrino}.

    The origin of the interest for neutrino flavor transformation in NSMs can be traced back to the resolution of the solar neutrino problem through flavor oscillations~\cite{Kajita_1999,SNO2001}. Neutrinos propagate as quantum superpositions of weak-interaction eigenstates (electron $\nu_e$, muon $\nu_\mu$, and tau $\nu_\tau$), and their evolution is modified by forward scattering on electrons, known as the MSW mechanism \cite{Wolfenstein1978,MikheevSmirnov_1985,Mikheyev_and_Smirnov}. In dense astrophysical neutrino gases, coherent neutrino–neutrino forward scattering also induces nonlinear, collective flavor dynamics \cite{Pantaleone:1992eq, sigl1993general, qianfuller}. Astrophysical flavor transformations can exhibit a wide range of collective phenomena, including fast flavor instabilities (FFIs)~\cite{richers2021neutrino, richers2022code, sawyer2016neutrino, wu2017fast, morinaga2020fast, george2020fast, nagakura2021where, tamborra2021new, ehring2023fast, Grohs:2023pgq, mukhopadhyay2024time}, collisional flavor instabilities (CFIs)~\cite{johns2021collisional, johns2022collisional, xiong2022evolution, xiong2022collisional, liu2023systematic, liu2023universality, akaho2023collisional, shalgar2023neutrinos, kato2023collisional, zaizen2025spectral, Froustey:2025nbi, wang2025effectcfi}, matter–neutrino resonances~\cite{vaananen2016uncovering, zhu2016matter, vlasenko2018matter, 2012PhRvD..86h5015M, malkus2016symmetric, wu2016physics, padilla2024symmetry, faiz2025can}, and slow flavor instabilities~\cite{fiorillo2025first, fiorillo2025theoryslow, fiorillo2025theoryslow2, shalgar2024neutrino, duan2006collective, duan2010collective, dasgupta2015temporal, chakraborty2016collective,Padilla-Gay:2025tko}. Among these, FFIs have received particular attention because their rapid dynamics can rearrange the neutrino angular distributions on timescales of order $\sim 10^{-9}\,\mathrm{s}$.
    FFIs naturally emerge in the classical neutrino fields of post-merger disk simulations. It was found in Ref.~\cite{nagakura2025neutrino} that they arise transiently and locally through multiple mechanisms as the disk evolves in the simulations, and that they are highly sensitive to the nature of the central compact object, whether a hypermassive neutron start or a black hole, although their study focused only the hypermassive neutron star case. On the other hand, they also found that CFI occurs persistently and over a wide region of the disk, and does not depend strongly on the central compact object. In contrast, Ref.~\cite{mukhopadhyay2024time} found persistent FFI conditions during the evolution of a post-merger accretion disk designed to reproduce the GW170817 event. References~\cite{froustey2024neutrino, froustey25cfi}  also found persistent appearance of FFI and CFI with an angular moment-based linear stability framework in the simulations of NSM mergers with a hypermassive neutron star \cite{PhysRevD.94.123016, PhysRevD.110.083028}, and they also quantified the impact of additional physical effects, including nuclear many-body corrections and scattering opacities.
    Beyond identifying and quantifying the FFI, multiple studies have already attempted to incorporate its effects into global astrophysical radiation-hydrodynamics simulations using an approximate description of neutrino flavor dynamics.
    These efforts can be broadly classified into two schemes: effective classical transport~\cite{lund2025angle, mori2025, xiong2024robust, ehring2023fast, just2022, fernandez2022, Li2021neutrinofast} and Bhatnagar–Gross–Krook (BGK) \cite{nagakura2024bhatnagar} subgrid models~\cite{qiu2025impact, qiu2025neutrino, wang2025, wang2025effectcfi, Wang_2026, akaho2026bifurcatedimpactneutrinofast}.

    Within the effective classical transport framework, Ref.~\cite{lund2025angle} presents the first three-dimensional radiation-hydrodynamics simulations of post-merger disks that include in situ angle-dependent modeling of FFC and go beyond the standard assumption of flavor equipartition. They couple general relativistic magnetohydrodynamics and Monte Carlo neutrino radiation transport to a flavor conversion routine in a two-step process. First, the classical neutrino radiation field is evolved until angular crossings in the electron-lepton-number minus heavy-lepton-number distribution (ELN-XLN) emerge. The classical field is then converted using a common prescription for FFC~\cite{PhysRevD.107.103022,Zaizen:2023ihz,Richers:2024zit}, in which angular regions on the ``shallow'' side of ELN–XLN crossings undergo flavor equipartition, while other angular regions adjust their flavor content to satisfy conservation laws. Flavor conversion is assumed to occur instantaneously because local simulations of the FFI, which solve for neutrino quantum kinetics starting from highly unstable states, modify the neutrino flavor content on subnanosecond timescales and subcentimeter length scales, which are much shorter than the hydrodynamical and interaction scales of classical processes. Ref.~\cite{lund2025angle} shows that FFC tends to reduce the abundances of electron neutrinos and antineutrinos by converting them into heavy-lepton flavors, thereby cooling the disk more efficiently and weakening the early thermally driven wind. Reduced re-leptonization makes this cooler wind more neutron rich, leading to a more robust $r$-process in the outflow. Ref.~\cite{Li2021neutrinofast} employed a simplified description of FFC in which any locally unstable region was assumed to undergo complete flavor mixing when the unstable growth-rate was estimated to be quicker than $10^{-7}~\mathrm{s}$. Within this framework, they reported widespread flavor conversion, which also drove the outflows to more neutron-rich conditions and thereby favored lanthanide synthesis as well as third-peak $r$-process production. Ref.~\cite{fernandez2022} instead modeled FFI parametrically by altering the absorbed neutrino fluxes and temperatures to represent different degrees of flavor equilibration. They found that, when a black hole formed promptly, FFI reduced the mean electron fraction of the disk outflow. In contrast, in the presence of a long-lived hypermassive neutron star, the outflow developed a broader electron-fraction distribution with a more proton-rich peak. Cases with intermediate hypermassive neutron-star lifetimes exhibited behavior between these two extremes. The changes in ejecta mass, mean velocity, and average electron fraction were at the level of about $\sim 10\%$, while the lanthanide and actinide abundances could vary by as much as a factor of $\sim 2$. Ref.~\cite{just2022} also investigated idealized parametric scenarios designed to mimic dynamically self-consistent fast mixing and found that such conversions increased disk cooling and lowered the electron fraction in both the disk and the ejecta. In their models, the kilonova signal was correspondingly extended, owing to the combined effects of higher lanthanide opacities and stronger radioactive heating.

    Within the approaches using the BGK subgrid model, Refs.~\cite{qiu2025neutrino,qiu2025impact} introduced the first numerical-relativity radiation-hydrodynamics simulations of NSMs that include effects of flavor mixing. Instead of instantaneously converting the flavor field to a prescribed asymptotic state, this model relaxes the neutrino moments towards an asymptotic configuration on a defined timescale. The asymptotic states used are motivated either by quantum many-body neutrino effects or by complete flavor equipartition, both of which conserve the electron-lepton number and the total lepton number. Refs.~\cite{qiu2025neutrino,qiu2025impact} found that neutrino flavor mixing drives the ejecta toward more neutron-rich conditions, significantly enhance the $r$-process yields, and can also lead to stronger gravitational-wave emission. 
    
    In spite of these significant efforts, important uncertainties remain in current subgrid models. In particular, these approaches require assumptions about the neutrino radiation field after flavor instabilities have relaxed. Such asymptotic states have been studied in many localized simulations motivated by dense astrophysical environment setups, often with periodic boundary conditions, and several prescriptions have been proposed based on the patterns and conservation laws found in those calculations \cite{PhysRevD.107.103022, PhysRevD.107.123021, PhysRevD.108.063003, PhysRevD.110.123018, Richers:2024zit}. However, these prescriptions do not fully capture the global setting, where ELN or XLN crossings are generated continuously by advection and incoherent neutrino interactions with the fluid while the resulting instabilities relax through quantum-kinetic evolution. Because of this, there is a need for self-consistent global neutrino quantum-kinetic simulations in order to test and calibrate the subgrid prescriptions, even if they are computationally challenging. This need is reinforced by evidence that the way ELN crossings are generated can affect the final flavor content of the neutrino field \cite{fiorillo2024fast}. Therefore, global neutrino quantum-kinetic simulations, or carefully designed local calculations that reproduce both the instability-driving and instability-relaxation mechanisms, are needed to robustly predict the post-relaxation neutrino radiation field and to build more reliable subgrid models for radiation-hydrodynamic simulations.    
    
    To help fill these gaps through numerical experiments, we have developed \texttt{Emu} \cite{Particle-in-cell}, an open source neutrino quantum kinetic particle-in-cell transport code that is fully parallel and portable across GPUs and CPUs within the AMReX framework. \texttt{Emu} is designed to simulate neutrino flavor transformations in explosive astrophysical environments such as core-collapse supernovae and NSMs. We implement for the first time multi-energy neutrino-nucleon absorption and emission, as well as pair annihilation. We couple the code with neutrino opacities using tables from \texttt{NuLib}~\cite{OConnor:2014sgn} and use a finite-temperature tabulated equation of state, which is \texttt{SFHo}~\cite{Steiner:2012rk} in the current work. This allows us to compute steady-state, multi-energy global neutrino solutions including flavor transformation.

    Using global and local neutrino quantum-kinetic simulations, together with linear stability analysis (LSA), we investigate the implications of FFI and CFI in a post-merger accretion disk built upon data from a simulation designed to resemble the GW170817 event. In Sec.~\ref{sec:quantum_kinetics}, we present the theoretical framework for neutrino quantum kinetics. In Sec.~\ref{sec:emu}, we describe the particle-in-cell framework for neutrino quantum kinetics implemented in the \texttt{Emu} code. In Sec.~\ref{sec:flui_prof}, we detail the fluid properties of the accretion disk under investigation, and in Sec.~\ref{sec:simpar}, we specify the simulation parameters used. In Sec.~\ref{sec:global_description_class}, we first present the classical transport solution of the post-merger disk and describe the global properties of the steady-state neutrino radiation field. Then, in Sec.~\ref{sec:global_ffi}, we address the emergence of ELN angular crossings and show how FFI naturally arises in this environment. We map the regions of the disk that exhibit FFI and estimate the growth rates of unstable modes. In addition, using local simulations at a representative location in the disk, we study the nonlinear evolution of fast flavor conversion in Sec.~\ref{sec:local_study_of_the_ffi}. In Sec.~\ref{sec:global_cfi}, we perform an analogous analysis for CFI and quantify the instability growth rates using monochromatic and multi-energy approaches. We then study the nonlinear evolution through CFI saturation and describe the properties of the asymptotic energy spectra in Sec.~\ref{sec:local_study_of_the_cfi}. In Sec.~\ref{sec:global_qke_simulations}, we discuss full neutrino quantum-kinetic simulations of the post-merger disk with attenuated Hamiltonians. Finally, in Sec.~\ref{sec:sumcon}, we summarize our main findings and conclusions.

\SwitchableClearpage
\section{Methods}
    
    \subsection{Neutrino Quantum Kinetics \label{sec:quantum_kinetics}}
        
        The state of a neutrino field is described by the seven-dimensional $3\times3$ matrix distribution function $f_{ab}(t,\,\textbf{x},\,\textbf{p})$. The diagonal terms represent occupation numbers, while the off-diagonal terms capture flavor correlations. The matrix distribution function and the number
        and flux density matrices are related by\footnote{We emphasize the use of the arrow for the flux $\vec{f}_{ab}$, while the distribution function matrix is $f_{ab}$.}
        \begin{align}
            n_{ab}(t,\,\mathbf{x}) & = \frac{1}{(hc)^3} \int \, d^3\mathbf{p} \, f_{ab}(t,\,\mathbf{x},\,\mathbf{p}).\\
            \vec{f}_{ab}(t,\,\mathbf{x}) & = \frac{1}{(hc)^3} \int \, d^3\mathbf{p} \, f_{ab}(t,\,\mathbf{x},\,\mathbf{p})\, \hat{\mathbf{p}}.
        \end{align}
        The time evolution of $f_{ab}$ is governed by the quantum kinetic equation (QKE)~\cite{sigl1993general, Vlasenko:2013fja, volpe2015neutrino} 
        \begin{equation}
            (\partial_t+\textbf{v}\cdot \nabla) f_{ab} = C_{ab} - \eta \frac{i}{\hbar}\left[H, f\right]_{ab} \, ,
            \label{eq:QKE}
        \end{equation}
        with $\mathbf{v}$ the neutrino velocity. The Hamiltonian $H_{ab}$ is written
        \begin{equation}
        \begin{aligned}
            H_{ab}(t, \mathbf{x}, \mathbf{p}) = \  &\eta_{\mathrm{vacuum}}H^{\mathrm{vacuum}}_{ab} \ + \\  &\eta_{\mathrm{matter}}H^{\mathrm{matter}}_{ab} \ + \\
            &\eta_{\mathrm{self}}H^{\nu-\nu}_{ab} \, .
        \end{aligned}
        \end{equation}

        Following an approach used for instance in~\cite{xiong2022evolution,Nagakura:2022kic,nagakura2023global,shalgar2024neutrino}, we introduce a global parameter $\eta$ to attenuate the strength of the Hamiltonian responsible for flavor conversion. This parameter helps to bring the advection and flavor-conversion scales closer together, making global simulations feasible. Furthermore, we introduce component-specific attenuation parameters ($\eta_\mathrm{vacuum}$, $\eta_\mathrm{matter}$, $\eta_\mathrm{self}$) for specific tests requiring a term-dependent attenuation. %The Hamiltonian $H_{ab}$ combines contributions from the neutrino vacuum mixing energy 
        These different contributions to the Hamiltonian are the neutrino vacuum mixing energy
        \begin{eqnarray}
        H^{\mathrm{vacuum}}_{ab} = U_{ac}\left[\sqrt{\textbf{p}^2 c^2 + m_c^2 c^4}\,\delta_{cd}\right]U_{db}^{\dagger},
        \end{eqnarray}
        where $U$ is the Pontecorvo–Maki–Nakagawa–Sakata matrix; the neutrino-matter forward scattering potential
        \begin{equation}
        \begin{aligned}
        H^{\mathrm{matter}}_{ab}(t, \mathbf{x}, \mathbf{p}) 
        &= \frac{\sqrt{2}\, G_F}{(h c)^3}
           \int d^3\mathbf{q}\,
           (1 - \hat{\mathbf{p}} \cdot \hat{\mathbf{q}} ) \\[4pt]
        &\quad\times 
           \left[l_{ab}(t, \mathbf{x}, \mathbf{q})
           - \bar{l}^{*}_{ab}(t, \mathbf{x}, \mathbf{q})\right],
        \end{aligned}
        \label{nuelectronham}
        \end{equation}
        where $l_{ab}$ and $\bar{l}_{ab}$ are the distribution function matrices for the heavy leptons and anti-leptons (which do not oscillate in flavor space and so are diagonal); and the neutrino-neutrino forward scattering potential \cite{qianfuller}
        \begin{equation}
        \begin{aligned}
        H^{\nu-\nu}_{ab}(t, \mathbf{x}, \mathbf{p}) 
        &= \frac{\sqrt{2}\, G_F}{(h c)^3}
           \int d^3\mathbf{q}\,
           (1 - \hat{\mathbf{p}} \cdot \hat{\mathbf{q}} ) \\[4pt]
        &\quad\times 
           \left[f_{ab}(t, \mathbf{x}, \mathbf{q})
           - \bar{f}^{*}_{ab}(t, \mathbf{x}, \mathbf{q})\right].
        \end{aligned}
        \label{nunuham}
        \end{equation}
        The antineutrino kinematics are analogous, with  
        \begin{align}
        \bar{H}^{\mathrm{vacuum}}_{ab} &= H^{\mathrm{vacuum}}_{ab}, \\
        \bar{H}^{\mathrm{matter}}_{ab} &= -{H^{\mathrm{matter}}_{ab}}^*, \\
        \bar{H}^{\nu\text{--}\nu}_{ab} &= -{H^{\nu\text{--}\nu}_{ab}}^*.
        \end{align}

        Incoherent neutrino-matter interactions are captured by the collision term $C_{ab}$. In this work, we only consider absorption and emission by neutrino-nucleon and pair-annihilation (via an effective absorption opacity) interactions, which allows us to write the collision term as \cite{sigl1993general,Vlasenko:2013fja,blaschke2016neutrino,Richers:2019grc}
        \begin{equation}
        \label{eq:collision_term}
            C_{ab} = \left\{\Gamma, f^{\text{eq}} - f\right\}_{ab},
        \end{equation}
        where $\Gamma = \text{diag}(\Gamma_e, \Gamma_\mu, \Gamma_\tau)/2$, and $\Gamma_a$ is the total absorption rate for neutrino flavor $a$, related to the mean free path $\lambda_a$ of the interactions by $\Gamma_a = c/\lambda_a$. The curly brackets denote the anticommutator, and $f^{\text{eq}}_{ab} = \text{diag}(f^{\text{eq}}_e, f^{\text{eq}}_\mu, f^{\text{eq}}_\tau)$, where $f^{\text{eq}}_a$ is the Fermi-Dirac equilibrium distribution function for neutrino flavor $a$.
        
        When the vacuum Hamiltonian is included, we adopt the mixing angles $\theta_{12}=33.82^\circ$, $\theta_{23}=8.61^\circ$, and $\theta_{13}=48.3^\circ$, and the normal hierarchy mass-squared differences $\Delta m^2_{21}=7.39\times10^{-5}\,\mathrm{eV}^2$ and $\Delta m^2_{32}=2.449\times10^{-3}\,\mathrm{eV}^2$.         
        The values assigned to $\theta_{13}$ and $\theta_{23}$ in our calculations were inadvertently interchanged relative to the PDG convention \cite{particle2020review}. We assess the effect of this interchange on our results in Sec.~\ref{sec:global_qke_simulations}.

    \SwitchableClearpage
    \subsection{EMU \label{sec:emu}}
        
        \texttt{Emu}~\cite{Particle-in-cell} solves the QKEs by discretizing the neutrino field into packets (particles) moving through a static Cartesian spatial grid. Each particle carries two quantum states defined by the Hermitian number matrices $N_{ab}$ ($\bar{N}_{ab}$), representing the physical neutrinos (antineutrinos). \texttt{Emu} employs a deposition and interpolation algorithm to estimate the neutrino number and flux densities of the full neutrino field needed to compute the Hamiltonian in Eq.~\eqref{nunuham} at every particle position. Each particle is considered to have an extended shape centered on its position, of a size comparable to a grid cell. This shape function is used to compute the neutrino number density matrices at the cell centers. The cell centers also have an extended shape and a hydrodynamic state (matter density $\rho$, temperature $T$, electron fraction $Y_e$). The number densities, fluxes and matter properties are interpolated from the mesh to the location of each particle. Neutrino and antineutrino emission rates and chemical potentials are interpolated from a \texttt{NuLib} table~\cite{OConnor:2014sgn} and the \texttt{SFHo} equation of state~\cite{Steiner:2012rk}. The QKE is integrated using a fourth-order Runge-Kutta method.  
    
    \SwitchableClearpage
    \subsection{GW170817 Post-Merger Snapshot\label{sec:flui_prof}}      

        We perform non-general relativistic 6+1 dimensional classical and quantum neutrino transport simulations in a three-dimensional snapshot of a post-merger accretion disk designed to reproduce observables from GW170817 and associated electromagnetic counterparts \cite{Miller:2019gig}. The post-merger matter profile was generated with the general relativistic magnetohydrodynamics code $\mathtt{\nu bhlight}$ \cite{Miller:2019gig} using the \texttt{SFHo} equation of state and a Monte-Carlo neutrino transport scheme with neutrino emission, absorption, and scattering.
        The setup involves a Kerr black hole spacetime with a mass of $M_{\text{BH}} = 2.58M_\odot$ and spin $a = 0.69$, along with a torus initially in hydrostatic equilibrium, entropy $s = 4k_\mathrm{B}/$baryon, electron fraction $Y_e = 0.1$, and disk mass $M_d = 0.12M_\odot$ \cite{Miller2019,mukhopadhyay2024time}.
        
\begin{figure*}[!htbp]
            \centering
            \includegraphics[width=0.8\textwidth]{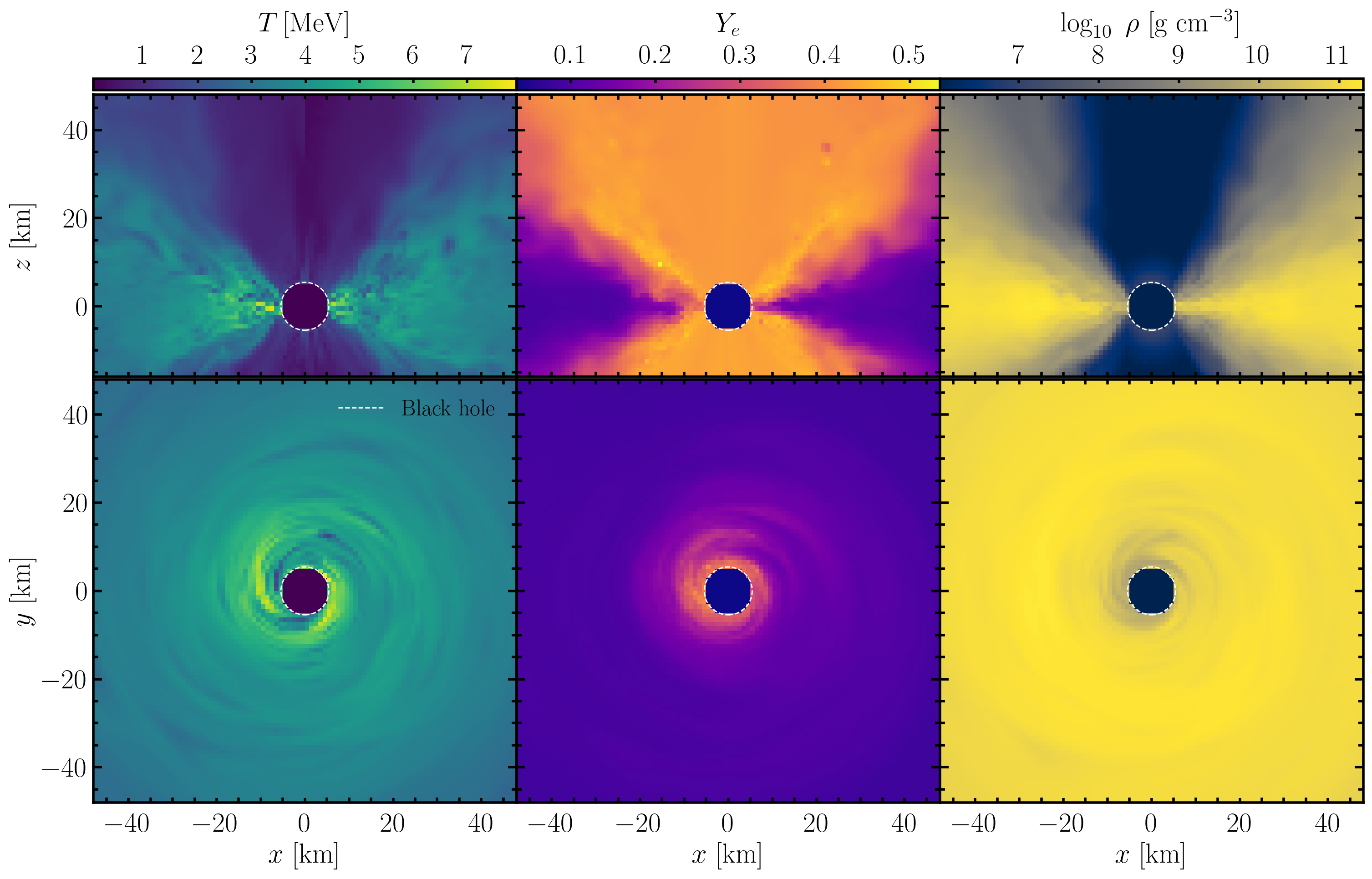}
            \caption{\label{fig:gw170817_1.00ye_profile} Temperature ($T$), electron fraction ($Y_e$), and density ($\rho$) of a post-merger accretion disk designed to reproduce observables from GW170817 at $t=24\,\mathrm{ms}$ in \cite{Miller:2019gig}. The upper (lower) panels show a polar (equatorial) slice. This is the fluid profile used in the {\tt HAR-*} and {\tt LAR-LSR-*} simulations in Table~\ref{tab:resolution}. The {\tt HSR-*} simulations use the same fluid properties but are interpolated onto a more-refined Cartesian mesh with the $z$ domain extending up to $16$ km.}
        \end{figure*}

        For this system, the classical neutrino radiation field produced by the Monte Carlo transport~\cite{Miller:2019gig} was post-processed, averaged over azimuthal angles, and analyzed to identify the time evolution of ELN crossings and the appearance of FFI, as reported in Ref.~\cite{mukhopadhyay2024time}. We take here a different approach and reconstruct the neutrino radiation field with higher angular resolution to characterize the angular distribution of each neutrino flavor at every location in the 3D post-merger disk to assess the emergence of ELN crossings and FFIs. We focus on the fluid snapshot at $t=24\,\mathrm{ms}$ after the start of the simulation. Density, temperature, and electron fractions are shown in Fig.~\ref{fig:gw170817_1.00ye_profile}. This is an early stage of the disk evolution, when the disk remains sufficiently hot and dense to provide favorable conditions for both FFI and CFI. Nevertheless, the properties of the neutrino radiation field and the associated flavor-instability conditions evolve significantly with time, and the conclusions drawn from this snapshot should therefore be extrapolated to other stages of the disk evolution with caution. The GW170817-like post-merger disk considered in this work is relatively massive compared with other post-merger disk models. Its large mass increases the optical depths of neutrinos and antineutrinos, thereby favoring the development of both FFI and CFI. As shown in Sec.~\ref{sec:global_ffi}, larger opacities and differences between neutrino and antineutrino transport generate angular structure in the ELN distribution and help the formation of ELN crossings.
        
        A $t=24\,\mathrm{ms}$ the accretion disk is strongly inhomogeneous (see Fig.~\ref{fig:gw170817_1.00ye_profile}). In the inner few tens of kilometers in the disk around the black hole, the gas is hot ($T \gtrsim 6$--$7~\mathrm{MeV}$), dense ($\rho \gtrsim 10^{10}$--$10^{11}~\mathrm{g \, cm^{-3}}$) with high electron fraction ($Y_e \simeq 0.3$--$0.5$). A cooler ($T \lesssim 2$--$3~\mathrm{MeV}$), more neutron rich ($Y_e \simeq 0.1$--$0.3$), yet still high-density disk extends outward. The polar funnel shows high $Y_e$ ($\simeq 0.3$--$0.5$), and exhibits lower temperatures and densities. Both polar and equatorial slices reveal pronounced non-axisymmetric structures. High temperature spiral arms with $Y_e \simeq 0.4$ are imprinted in the inner disk, as well as temperature fluctuations. These inhomogeneities will modulate the local neutrino emission/absorption rates and the angular distributions of $\nu_e$ and $\bar{\nu}_e$.
        
        The black hole is located at the origin of our coordinate system and has a radius of $5.43\,\mathrm{km}$. When a particle enters the black hole, we set the neutrino number of all flavors to zero, while retaining the particle's record of the position, direction, and phase-space volume it represents. When the particle exits the black-hole, neutrino emissivities interpolated from the background fluid grid naturally repopulate its phase-space volume. We impose outflow boundary conditions in the boundary cells of the domain. For efficiency, particles wrap around the domain in both position and momentum, as in periodic boundary conditions, but we enforce zero neutrino number and flux densities for all flavors whenever a particle crosses the domain boundary. This procedure ensures that the full phase-space volume of the domain remains continuously represented by the particles up to the 103 MeV maximum neutrino energy of the NuLib table.
    
    \subsection{Simulation Parameters\label{sec:simpar}}
        
        We interpolate the density $\rho$, temperature $T$, and electron fraction $Y_e$ from the snapshot onto our Cartesian simulation mesh and evolve 13 neutrino energy bins, logarithmically spaced up to $103\,\mathrm{MeV}$. In the high–angular-resolution (\texttt{HAR-*}) runs, the domain is $(96,96,64)\,\mathrm{km}$, discretized with $1\times 1 \times 1\,\mathrm{km}$ cubic cells, with 1506 particles per energy bin per cell. In the high–spatial–resolution (\texttt{HSR-*}) runs, the domain is $(96,96,32)\,\mathrm{km}$, discretized with $0.5\times 0.5\times 0.5\,\mathrm{km}$ cubic cells, with 92 particles per energy bin per cell. The low angular resolution and low spatial resolution ({\tt LAR-LSR-*}) runs have a domain of $(96,96,64)\,\mathrm{km}$, discretized with $1\times 1 \times 1\,\mathrm{km}$ cubic cells, with 92 particles per energy bin per cell. The global simulation parameters (for classical kinetics, ``\texttt{class}'' and quantum kinetics, ``\texttt{QKE}'') are summarized in Table~\ref{tab:resolution}. Because we neglect general relativistic effects and scattering, the resulting radiation field is expected to differ from that in Ref.~\cite{Miller:2019gig}, but it provides a global, energy-dependent, inhomogeneous, and anisotropic backdrop to analyze whether the neutrino radiation field is unstable. The unstable distributions are used as input for local centimeter-wide domain QKE simulations, providing insightful demonstrations of the asymptotic state of FFI and CFI within the limitations of periodic boundary conditions. Our global QKE simulations, even though they do not experience flavor conversion due to instabilities, highlight effective in-medium flavor evolution and the resolution and scaling requirements for future global simulations that capture instability growth, saturation, and advection simultaneously.
        \begin{table*}
            \centering
            \begin{tabular}{lccccclccc}
                Name & $\Delta x$ & $z_\mathrm{max}$ & $N_\mathrm{ppepc}$ & $c\Delta t$ & $ct$ & $\eta$ & $\eta_{\mathrm{vacuum}}$ & $\eta_{\mathrm{matter}}$ & $\eta_{\mathrm{self}}$ \\
                & (km) & (km) & & (km) & (km) & & & & \\\hline
                {\tt HSR-class} & 0.5 & 16 & 92 & 0.250 & 96 & 0 & $1$ & $1$ & $1$ \\
                {\tt HSR-QKE-reducedmatter} & 0.5 & 16 & 92 & 0.100 & 96 & $10^{-5}$ & $1$ & $10^{-2}$ & $1$ \\
                {\tt HAR-class} & 1 & 48 & 1506 & 0.250 & 96 & 0 & $1$ & $1$ & $1$ \\
                {\tt HAR-QKE} & 1 & 48 & 1506 & 0.001 & 96 & $10^{-5}$ & $1$ & $1$ & $1$ \\
                {\tt LAR-LSR-QKE-1e-6} & $1$ & 48 & 92 & 0.100 & 96 & $10^{-6}$ & $1$ & $1$ & $1$ \\
                {\tt LAR-LSR-QKE-1e-7} & $1$ & 48 & 92 & 0.250 & 96 & $10^{-7}$ & $1$ & $1$ & $1$ \\
                {\tt LAR-LSR-QKE-1e-8} & $1$ & 48 & 92 & 0.250 & 96 & $10^{-8}$ & $1$ & $1$ & $1$ \\
                {\tt LAR-LSR-QKE-1e-9} & $1$ & 48 & 92 & 0.250 & 96 & $10^{-9}$ & $1$ & $1$ & $1$ \\
                {\tt LAR-LSR-QKE-1e-10} & $1$ & 48 & 92 & 0.250 & 96 & $10^{-10}$ & $1$ & $1$ & $1$ \\
                {\tt LAR-LSR-QKE-1e-11} & $1$ & 48 & 92 & 0.250 & 96 & $10^{-11}$ & $1$ & $1$ & $1$ \\
                {\tt LAR-LSR-QKE-1e-12} & $1$ & 48 & 92 & 0.250 & 96 & $10^{-12}$ & $1$ & $1$ & $1$ \\
            \end{tabular}
            \caption{
                Parameters of our global neutrino-kinetics simulations: cartesian grid spacing ($\Delta x$); the domain’s maximum vertical coordinate $z_\mathrm{max}$ (with $z_\mathrm{min} = -16\,\mathrm{km}$ for all runs); the number of particles per energy bin per cell ($N_\mathrm{ppepc}$); the time step ($c\Delta t$); the total simulated time ($ct$); and the attenuation factors ($\eta$, $\eta_{\mathrm{vacuum}}$, $\eta_{\mathrm{matter}}$, $\eta_{\mathrm{self}}$). In \texttt{HSR-QKE-reducedmatter}, the matter term in the Hamiltonian is scaled by $\eta\cdot \eta_\mathrm{matter} = 10^{-7}$, whereas all other Hamiltonian terms are scaled by $\eta=10^{-5}$.
            }
            \label{tab:resolution}
        \end{table*}

\SwitchableClearpage
\section{Classical Radiation Field\label{sec:clas_solu}}   

    In this section, we present simulations of classical neutrino transport (i.e., without including flavor mixing) in the GW170817 matter snapshot described in Sec.~\ref{sec:flui_prof}. We used three neutrino flavors and three antineutrino flavors Although our calculations are non-general relativistic and thus differ from the physics used to generate the snapshot, they resolve anisotropies and inhomogeneities in the full six-dimensional phase space, which is necessary to assess the emergence of flavor instabilities. We post-process these simulations to predict where FFIs (Sec.~\ref{sec:global_ffi}) and CFIs (Sec.~\ref{sec:global_cfi}) are likely to occur. We also follow up with local simulations to clarify the asymptotic states of the FFI and multi-energy CFI in the inner regions of the disk.
    
    \subsection{Global Description\label{sec:global_description_class}}      
    
\begin{figure*}[!htbp]
            \centering
            \includegraphics[width=0.8\textwidth]{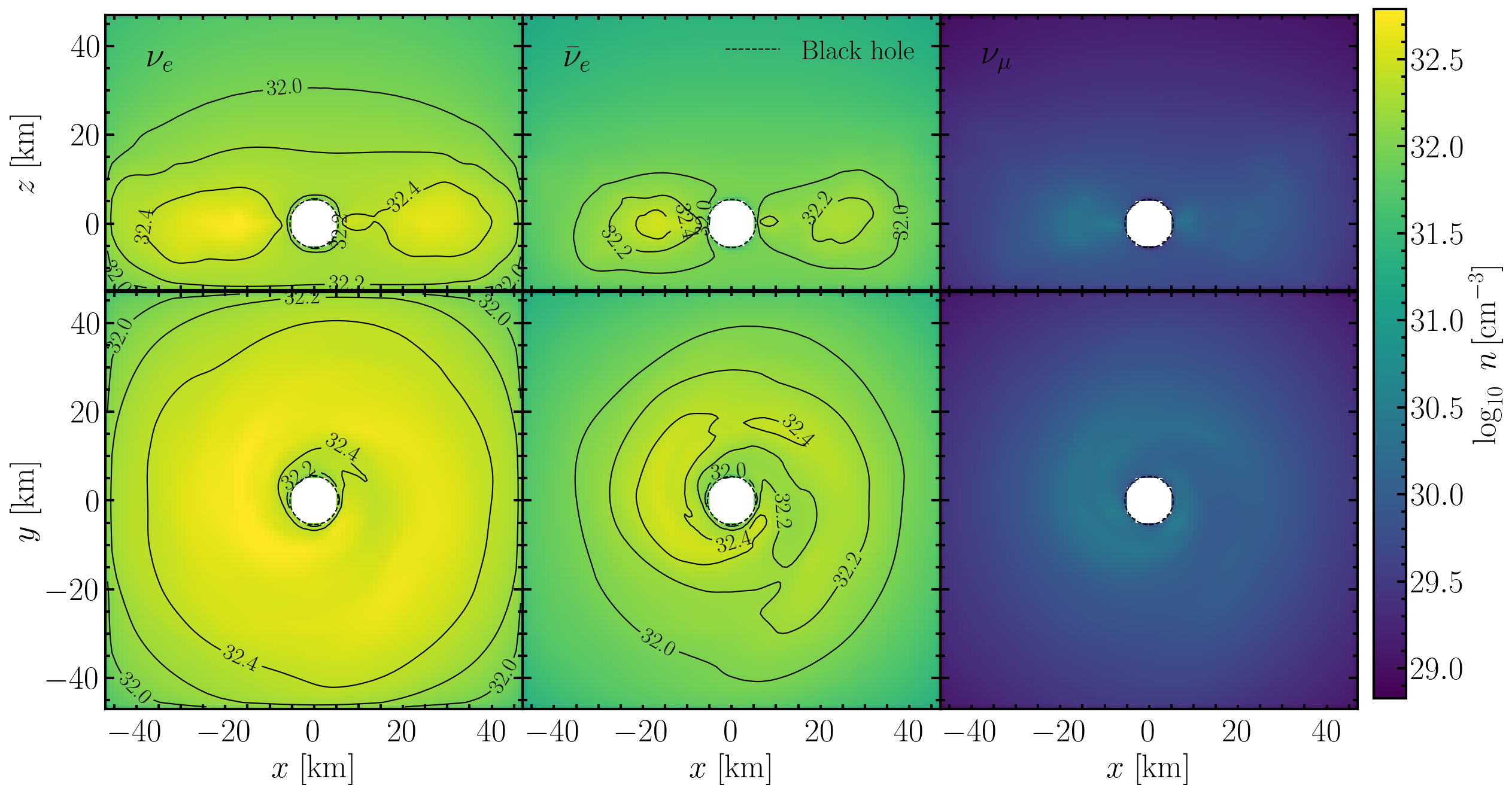}
            \caption{\label{fig:n_diagonal_densities} 
                Number densities of electron neutrinos (left), electron antineutrinos (center), and muon neutrinos (right) for the {\tt HAR-class} simulation. Other heavy-flavor neutrinos and antineutrinos follow the same trend as the muon-neutrino panel. The upper (lower) panels show polar (equatorial) slices.}
        \end{figure*}

        Fig.~\ref{fig:n_diagonal_densities} shows the number densities of each neutrino species in the high-angular resolution, classical ({\tt HAR-class}) simulation. All heavy flavors look identical because we do not differentiate between heavy lepton neutrino or antineutrino opacities in \texttt{NuLib} and the equilibrium chemical potentials given by the \texttt{SFHo} equation of state. The neutrino and antineutrino number densities peak in the accretion disk and decline toward the poles, with distinct internal three dimensional patterns that reflect the emission structure combined with the effect of advection and absorption. The $10^{32.0}$, $10^{32.2}$, and $10^{32.4}\,\mathrm{cm^{-3}}$ number-density contour lines span nearly the entire disk for $\nu_e$. In contrast, the corresponding contours for $\bar{\nu}_e$ lie in the inner disk, closer to the black hole. The heavy-flavor neutrino densities also peak in the accretion disk and decline toward the poles but remain roughly two orders of magnitude below the electron-flavor densities.
        
\begin{figure*}[!htbp]
            \centering
            \includegraphics[width=0.8\textwidth]{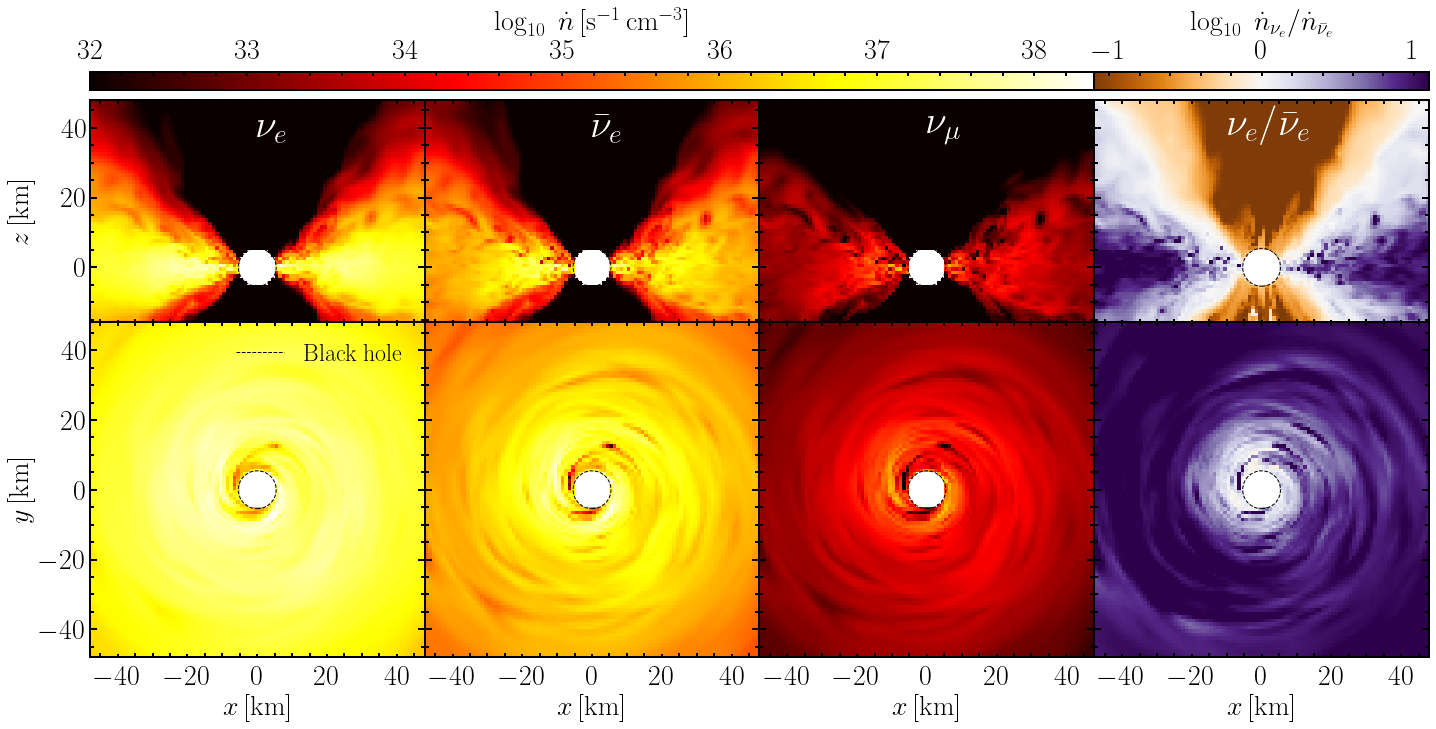}
            \caption{\label{fig:nu_nubar_emission} 
                From left to right: energy-integrated emission rates for electron neutrinos, electron antineutrinos, muon neutrinos and ratio between electron neutrinos and antineutrinos for the fluid profile (see $\rho$, $Y_e$ and $T$ in Fig.\ref{fig:gw170817_1.00ye_profile}) of the {\tt HAR-*} and {\tt LAR-LSR-*} simulations in Table~\ref{tab:resolution}. Other heavy-flavor neutrinos and antineutrinos follow the same trend as the muon-neutrino panel.
                Emission rates are obtained by interpolating the \texttt{SFHo} equation of state and the \texttt{NuLib} opacity tables.
                The upper (lower) panels show polar (equatorial) slices.
            }
        \end{figure*}

        At this stage of the accretion-disk evolution, the disk contains mildly degenerate electrons that, at these densities, enhances the rate of deleptonization \cite{just2022neutrino, de2021igniting}. In the accretion disk, we find degeneracy parameters up to $\eta_e = \mu_e/T \approx 3.6$, which suppress the positron population. The resulting scarcity of positrons reduces the rate of $e^{+}+n \rightarrow p+\bar{\nu}_e$ and therefore lowers the $\bar{\nu}_e$ emissivity relative to $\nu_e$, which is produced predominantly through $p + e^{-} \leftrightarrow n + \nu_{e}$. This trend is consistent with the chemical potentials interpolated from the \texttt{SFHo} equation of state and with \texttt{NuLib} opacities, as shown in Fig.~\ref{fig:nu_nubar_emission}: the $\nu_e$ emission rate exceeds that of $\bar{\nu}_e$ throughout the accretion disk.  Accordingly, the neutrino number densities follow the hierarchy $n_{\nu_e} > n_{\bar{\nu}_e} > n_{\nu_x}$ (see Fig.~\ref{fig:n_diagonal_densities}). Along the poles, by contrast, electrons are not degenerate, positrons are more abundant, and $\bar{\nu}_e$ emission dominates (see the ratio $\dot{n}_{\nu_e}/\dot{n}_{\bar{\nu}_e}$ in the rightmost panel). The overall excess of electron neutrino luminosity is consistent with the original simulation with Monte Carlo transport \cite{Miller2019}.
        
\begin{figure*}[!htbp]
            \centering
            \includegraphics[width=0.8\textwidth]{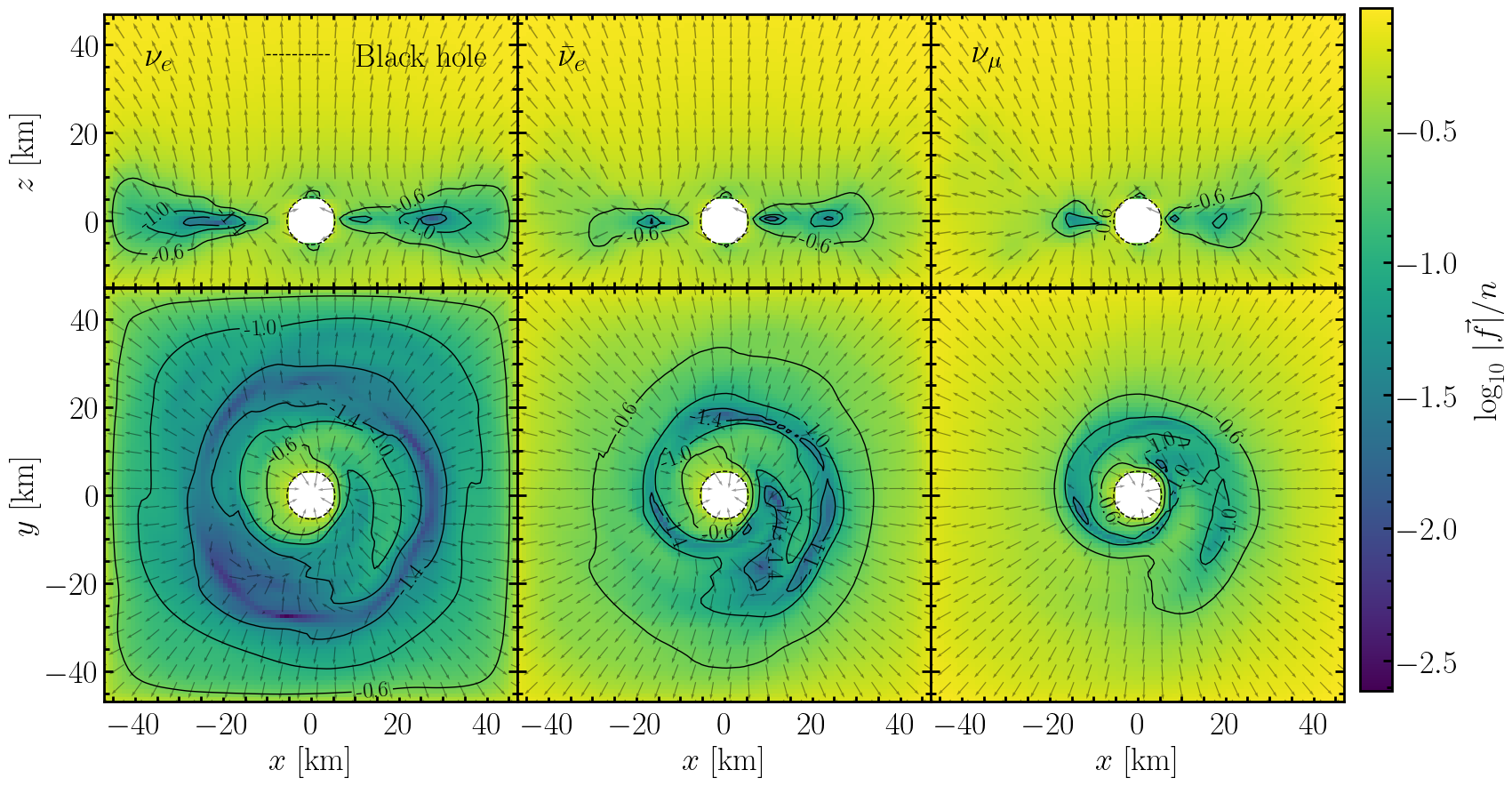}
            \caption{\label{fig:fhat_diagonal_densities} 
                Flux factors of electron neutrinos (left), electron antineutrinos (center), and muon neutrinos (right) for the {\tt HAR-class} simulation. Other heavy-flavor neutrinos and antineutrinos follow the same trend as the muon-neutrino panel. The arrows indicate the local direction of the flux $\vec{f}$. The upper (lower) panels show polar (equatorial) slices.
            }
        \end{figure*}

        Fig.~\ref{fig:fhat_diagonal_densities} shows large flux factors in the polar regions above the disk for all neutrino and antineutrino flavors, indicating a strongly forward–peaked outflow. Within the disk, heavy–lepton neutrinos exhibit relatively small to moderate flux factors in the central regions ($0<r<30~\mathrm{km}$), as suggested by the $|\vec f| / n = 10^{-0.6} \approx 0.25$ contour, and larger flux factors for $r>30~\mathrm{km}$. This implies strong inner–disk emission and absorption that keeps the angular distribution more isotropic there, transitioning to a more forward–peaked field at larger radii. Electron antineutrinos show a similar pattern, with the low flux factor region extending out to $r\lesssim35~\mathrm{km}$, as indicated by the same $-0.6$ contour. For electron neutrinos the corresponding contour appears even farther out, near $r\approx40~\mathrm{km}$. The resulting radiation field enhances the conditions that are conducive to neutrino flavor instabilities. In the disk, the flux factors also show strong three-dimensional effects that break azimuthal symmetry.

        By interpolating the absorption opacities from \texttt{NuLib} onto our fluid profile, we find that the hottest regions of the disk have an absorption optical depth integrated vertically from the mid plane of $\tau_{\mathrm{abs}} < 2/3$ for electron neutrinos with energies up to $8~\mathrm{MeV}$, whereas higher-energy neutrinos experience strong absorption and strongly couple to the fluid. The disk has an optical depth of $\tau_{\mathrm{abs}} < 2/3$ for electron neutrinos with energies up to $27~\mathrm{MeV}$. This implies that a larger number of low-energy $\bar{\nu}_e$ propagate efficiently through the disk compared with $\nu_e$. The disk does  not reach $\tau_{\mathrm{abs}} = 2/3$ for heavy-lepton neutrinos in any energy bin, which therefore interact only weakly with the disk and escape efficiently. 

        We also estimated the impact of scattering in the disk by computing the optical depth of a neutrino emitted at representative points in the accretion disk and propagating upward, perpendicular to the disk. The scattering opacities we used come from direct interpolation from \texttt{NuLib} onto the fluid profile shown in Fig.~\ref{fig:gw170817_1.00ye_profile}. We found a scattering optical depth $\tau_{\mathrm{sca}} < 2/3$ for all neutrino and antineutrino flavors up to $16~\mathrm{MeV}$. Electron neutrinos above $16~\mathrm{MeV}$ are already trapped by absorption processes, and adding scattering would keep them trapped even more effectively. By contrast, electron antineutrinos in the range $16$--$27~\mathrm{MeV}$ are only weakly affected by absorption, so our absorption-only treatment likely overestimates how easily these neutrinos propagate through the disk compared with a model that includes scattering. For heavy-lepton neutrinos, which lack charged-current absorption on nucleons and therefore stream nearly freely in our absorption-only setup, scattering provides their dominant coupling to the disk matter and helps determine their decoupling radius. More generally, scattering increases the effective optical depth, shifting the decoupling radius of each neutrino species outward. Inelastic scattering on electrons and positrons can also redistribute neutrino energies and soften the emerging spectra. We estimate that scattering quantitatively modifies the angular distributions but is unlikely to qualitatively change the formation mechanism of the ELN crossings identified in our \textsc{Emu} calculations that will be presented later. Nevertheless, the absence of scattering is an important limitation of the present study, particularly for a precise quantitative determination of the angular distributions and crossing locations, and a fully self-consistent treatment is deferred to future work.
        
    \subsection{Fast Flavor Instability\label{sec:global_ffi}}
        
        \subsubsection{Global analysis on the classical radiation field}
        
            Following Ref.~\cite{PhysRevResearch.2.012046}, we define the electron lepton number (ELN) -- heavy lepton number (XLN) angular distribution as
            \begin{eqnarray}
            \label{eq:ELN}
            \Delta G = (G_{\nu_e} - G_{\bar{\nu}_e}) - (G_{\nu_x} - G_{\bar{\nu}_x}) \, ,
            \end{eqnarray}
            where the angular distribution of neutrinos of flavor $i$ is
            \begin{eqnarray}
                G_{\nu_i}(\Omega) = \sqrt{2}G_F \, 4\pi\, \int \frac{E^2\, dE}{(hc)^3}\, f_{\nu_i}(E,\Omega).
                \label{eq:angdis}
            \end{eqnarray}            
            If $\Delta G$ changes sign across solid angle space, i.e., there exists a direction where the distribution switches from neutrino to antineutrino dominance (an ELN--XLN angular crossing), then the FFI develops under periodic boundary conditions~\cite{Morinaga:2021vmc,Dasgupta:2021gfs,Fiorillo:2024bzm}. An approximate estimate of the growth rate of the FFI is then given by~\cite{PhysRevResearch.2.012046}
            \begin{equation}
            \sigma_{\text{FFI}}
              = \sqrt{ - \left( \int_{ \Delta G > 0 } 
                    \frac{d\Omega}{4\pi}\, \Delta G \right)
                    \left( \int_{\Delta G<0} 
                    \frac{d\Omega}{4\pi}\, \Delta G \right) } \, .   
                \label{eq:sigma}
            \end{equation}
            If instead $\Delta G$ maintains a constant sign over all directions, the growth rate vanishes, and the neutrino radiation field remains stable against FFC, though other types of instabilities may still occur. The estimate~\eqref{eq:sigma} has clear limitations: it does not identify the most unstable mode and its associated wavevector, which would be obtained by a standard (but more costly) multiangle linear stability analysis~\cite{Banerjee:2011fj,Izaguirre:2016gsx,froustey2024neutrino,Fiorillo:2025zio}. In particular, this estimate could perform more poorly in situations with multiple~\cite{Bhattacharyya:2021klg} and/or shallow~\cite{fiorillo2024fast,Fiorillo:2024uki} crossings.
            
            The estimated FFI growth rate [Eq.~\eqref{eq:sigma}] is shown in the leftmost panels of Fig.~\ref{fig:instabilities} for the {\tt HAR-class} simulation (results for {\tt HSR-class} simulations are shown in Appendix~\ref{con_num_par_ffi}). Within the disk, the FFI growth rate reaches $\sim 10^{9}\,\mathrm{s^{-1}}$, but decreases rapidly toward the poles due to the shallow crossings dominating those regions. These crossing regions agree with the late-time location of crossings found in Ref.~\cite{mukhopadhyay2024time}.
            \begin{figure*}[!htbp]
                \centering
                \includegraphics[width=0.84\textwidth]{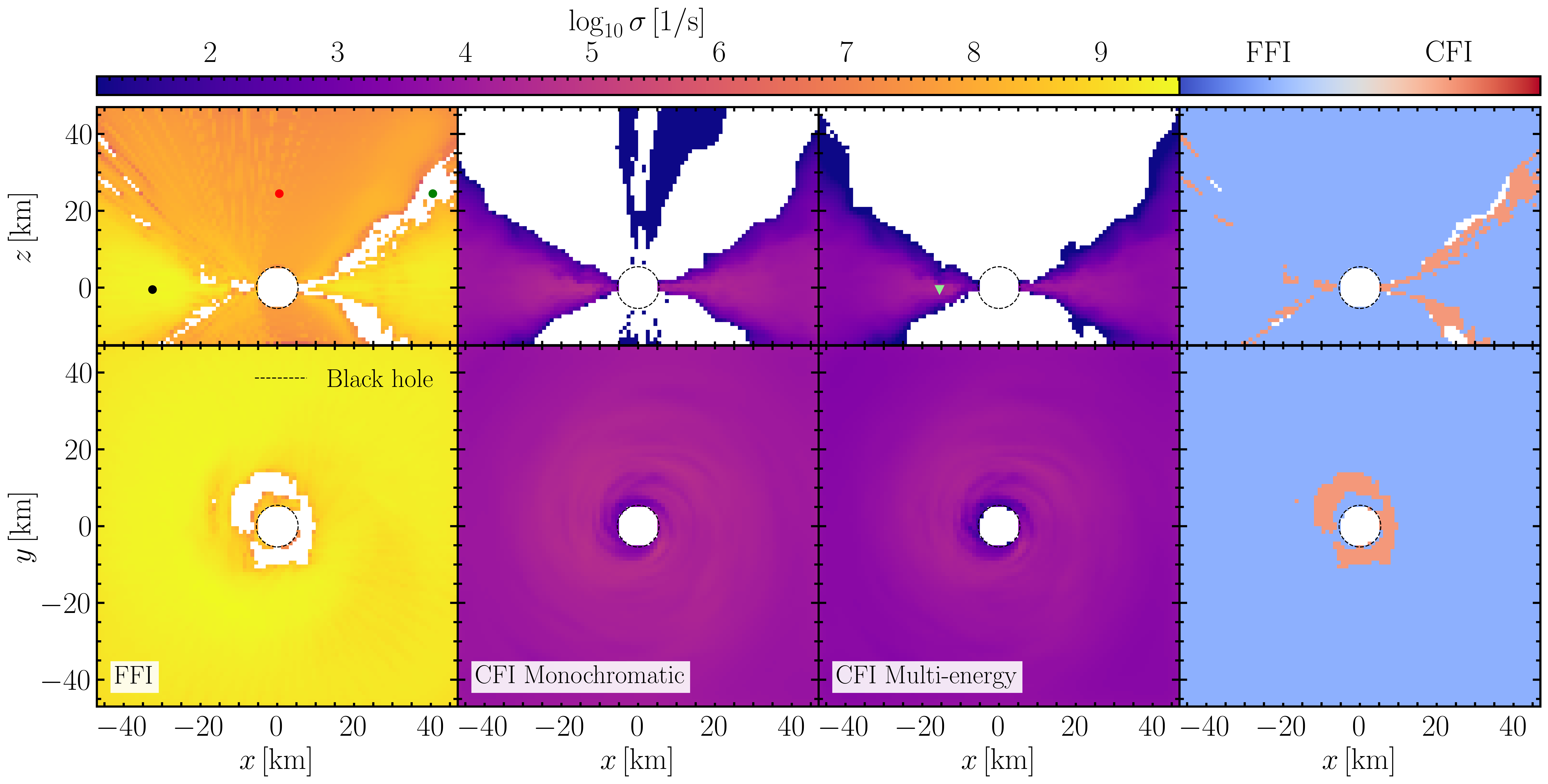}
                \caption{\label{fig:instabilities} 
                    Growth rates of FFI (left), monochromatic CFI (center left) and multi-energy CFI (center right), and the instability that dominates in each region (right) in our GW170817-like post-merger disk snapshot ({\tt HAR-class} simulation in Table~\ref{tab:resolution}). FFIs appear robustly within the accretion disk and decrease toward the polar regions. CFIs permeate the disk but are generally subdominant to FFIs. Monochromatic CFI growth rates tend to overestimate the multi-energy CFI growth rate. Some innermost regions near the black hole and the $45^\circ$ regions above the accretion disk exhibit a CFI without FFI (see right panels). The angular distributions at the locations with black, red and green circular markers are shown in Fig.~\ref{fig:ffi_angular_distributions}. The upper (lower) panels show polar (equatorial) slices.
                }
            \end{figure*}
            
\begin{figure}[!htbp]
                \centering
                \includegraphics[width=\linewidth]{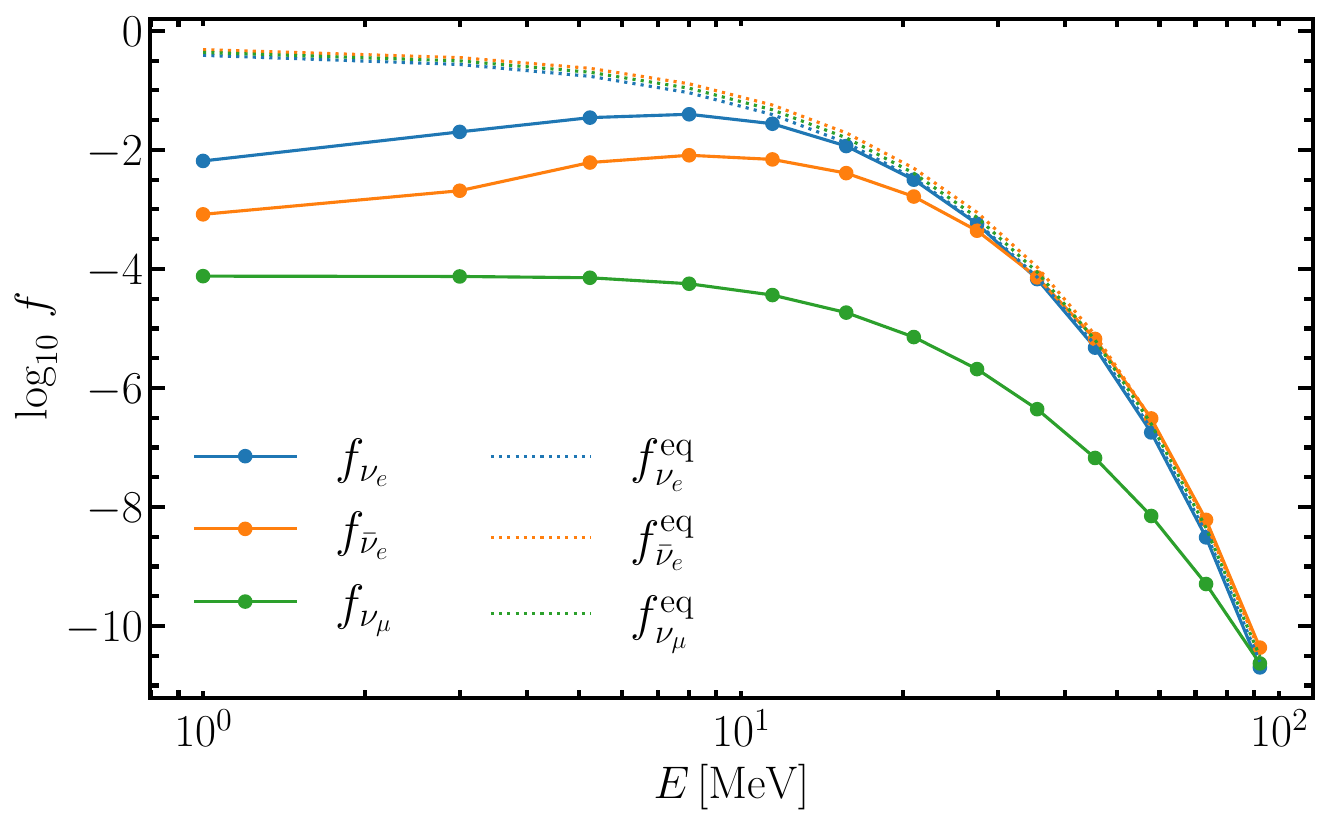}
                \caption{\label{fig:ffi_point_energies} 
                    Distribution functions averaged over solid angle at the location marked by the black circular marker in Fig.~\ref{fig:instabilities}. This corresponds to the {\tt HAR-class} simulation, whose parameters are shown in Table~\ref{tab:resolution}. Equilibrium is shown as Fermi--Dirac distributions, with chemical potentials interpolated from the \texttt{SFHo} equation of state.  Other heavy-flavor neutrinos and antineutrinos follow the same trend as the muon-neutrino panel. Heavy-lepton neutrinos are out of equilibrium with the background fluid, consistent with their weak coupling and efficient escape from the disk. The high-energy portions of the $\nu_e$ and $\bar{\nu}_e$ distributions lie close to their equilibrium values. 
                    $\bar{\nu}_e$ decouple thermally from the fluid at higher energies than $\nu_e$ and their lower-energy population propagates more efficiently through the disk.
                }
            \end{figure}

\begin{figure*}[!htbp]
                \centering
                \includegraphics[width=0.8\textwidth]{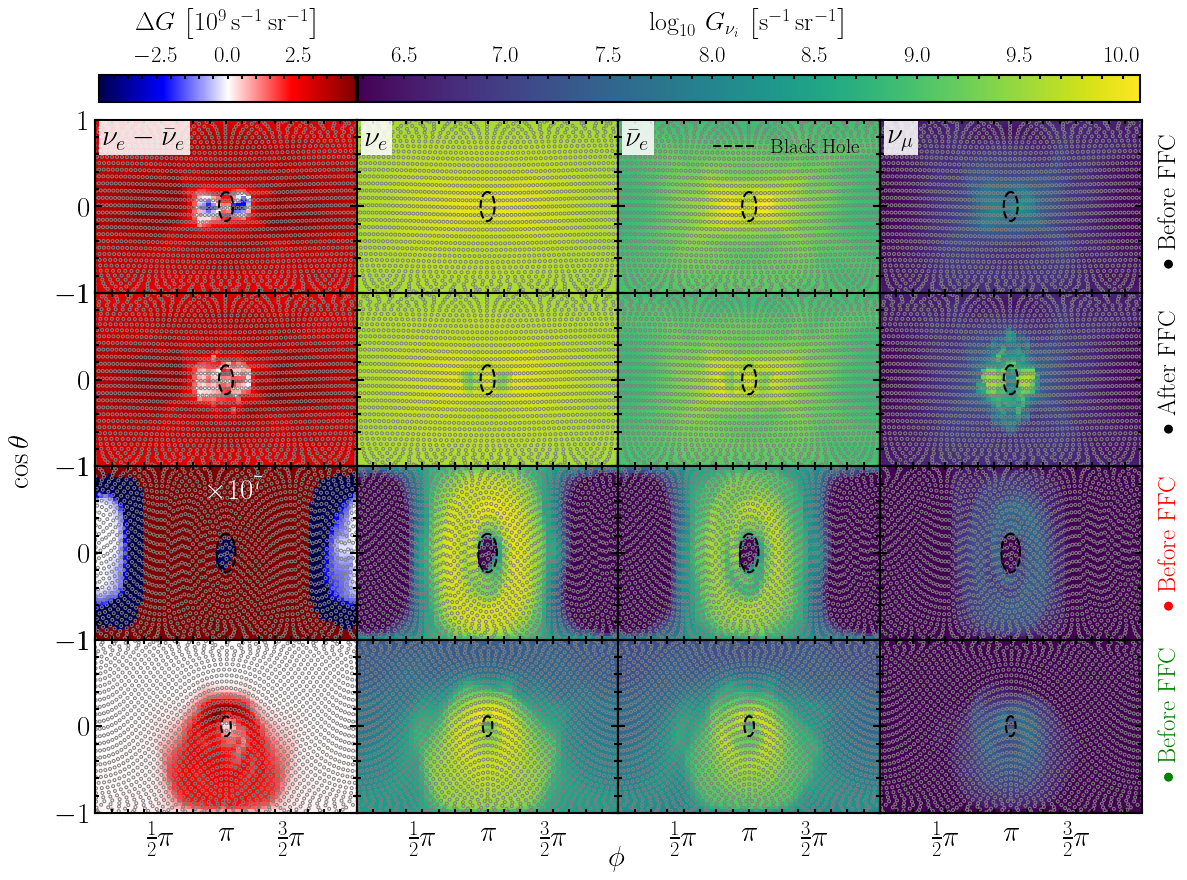}
                \caption{
                    Neutrino and antineutrino angular distributions at the locations corresponding, from top to bottom, to the black, red, and green circular markers in the top-left panel of Fig.~\ref{fig:instabilities}.
                    Data correspond to the {\tt HAR-class} simulation, whose parameters are shown in Table~\ref{tab:resolution}. 
                    From left to right we show: ELN--XLN [Eq.~\eqref{eq:ELN}], $\nu_e$, $\bar{\nu}_e$ and $\nu_\mu$ angular distributions.
                     Other heavy-flavor neutrinos and antineutrinos follow the same trend as the muon-neutrino panel.
                    $\theta$ and $\phi$ represent momentum-space coordinates. We adopt the convention that $\phi$ is the azimuthal angle measured from $\hat{x}_{\mathbf{p}}$, and that the polar angle $\theta$ is the opening angle measured from the $\hat{z}_{\mathbf{p}}$ direction.
                    The momentum-space basis vectors $(\hat{x}_{\mathbf{p}},\,\hat{y}_{\mathbf{p}},\,\hat{z}_{\mathbf{p}})$ are defined to coincide with the position-space basis vectors $(-\hat{r}_{\mathbf{r}},\,\hat{\phi}_{\mathbf{r}},\,\hat{\theta}_{\mathbf{r}})$ at the location $\mathbf{r}$.
                    With this convention, neutrinos moving radially outward (inward) to the black hole correspond to $\phi=\pi$ ($\phi=0$ or $2\pi$) at $\cos\theta=0$.
                    The black hole shadow in angular space is indicated by black dashed lines.
                    Scattered points show the angular resolution in our simulation. The color map was generated by linear interpolation on a finer mesh for visualization.
                }
                \label{fig:ffi_angular_distributions}
            \end{figure*}

            In the disk, the $\bar{\nu}_e$ field is more forward-peaked than the $\nu_e$ field. The $\bar{\nu}_e$ emerging from the emission hotspots undergo relatively few interactions as they propagate through the disk. By contrast, the $\nu_e$ field is more isotropic because $\nu_e$ couples more strongly to the fluid through absorption, keeping them closer to local equilibrium over a broader range of energies. Fig.~\ref{fig:ffi_point_energies} shows the distribution functions at a location in the disk, marked by the black circular marker in Fig.~\ref{fig:instabilities} for the \texttt{HAR-class} simulation. In the trapped, thermalized regime, the high-energy tail of the $\nu_e$ distribution is larger than that of $\bar{\nu}_e$, so the ELN is positive for most directions. At low energies where the neutrinos and antineutrinos are closer to free streaming, the $\bar{\nu}_e$ abundance exceeds the $\nu_e$ abundance. This low-energy $\bar{\nu}_e$ component can propagate more freely, enhancing the forward-peaked character of the $\bar{\nu}_e$ angular distribution relative to $\nu_e$. Along sightlines that intersect the antineutrino hotspots, the beamed $\bar{\nu}_e$ intensity exceeds the $\nu_e$ intensity, making the ELN negative and producing an ELN crossing. This can be seen in Fig.~\ref{fig:ffi_angular_distributions} in the ``Before FFC'' row for the black circular marker. The negative-ELN sectors align with directions toward the emission hotspots, while the more uniformly distributed $\nu_e$ field keeps the ELN positive elsewhere. 

            In our treatment, heavy-lepton neutrinos do not contribute to the ELN--XLN distribution because they are produced predominantly through pair processes. These interactions are flavor independent and generate equal emission and absorption rates for heavy-lepton neutrinos and antineutrinos, resulting in a vanishing net heavy-lepton number. We note that flavor-dependent differences in heavy-lepton neutrino emission may arise in the presence of muons in the accretion disk (see, e.g., Refs.~\cite{gieg2025role,ng2025accurate}). However, incorporating muons consistently would require corresponding extensions of the equation of state and neutrino opacity tables and lies beyond the scope of the present work. Moreover, the temperature in our post-merger accretion disk does not exceed $8~\mathrm{MeV}$ (see Fig.~\ref{fig:gw170817_1.00ye_profile}), which is well below the muon rest-mass energy, $m_\mu c^2 \simeq 106~\mathrm{MeV}$. The thermal production of muons is therefore strongly suppressed, and we expect their impact on the neutrino radiation field in the present model to be small. We do not include such effects here because the \texttt{SFHo} equation of state and the \texttt{NuLib} opacity tables employed in this work do not account for distinct heavy-lepton abundances or distinguish among the individual heavy-lepton flavors. Whether crossings can be generated by an asymmetry between $\nu_\mu$ and $\bar{\nu}_\mu$ produced through charged-current interactions involving muons remains an open question.

            Moving away from the disk into the polar funnel, the $\nu_e$ angular distribution evolves from relatively broad (weakly forward-peaked) to strongly forward-peaked. The $\bar{\nu}_e$ angular distribution exhibits a similar trend. At $\sim 45^\circ$ above the disk plane, the $\nu_e$ and $\bar{\nu}_e$ angular distributions peak at comparable amplitudes and directions, so their ELN nearly cancels and no ELN crossing occurs. This can be seen in Fig.~\ref{fig:ffi_angular_distributions} (``Before FFC'' row for the green circular marker), consistent with the vanishing FFI growth-rate cones in the upper-left panel of Fig.~\ref{fig:instabilities}.
            
            At angles greater than $45^\circ$ above the disk plane, the $\dot{n}_{\nu_e}/\dot{n}_{\bar{\nu}_e}$ panels in Fig.~\ref{fig:nu_nubar_emission} indicate that electron-antineutrino emission dominates. As a result, the downward directions are relatively sparse but are dominated by $\bar{\nu}_e$. In contrast, the upward directions are largely dominated by the broader $\nu_e$ emission from the disk. Along the line of sight intersecting the black-hole shadow, however, $\bar{\nu}_e$ dominate due to emission from the polar atmosphere. This produces a double ELN crossing driven by contributions from both the polar atmosphere and the disk. This behavior is visible in Fig.~\ref{fig:ffi_angular_distributions} in the ``Before FFC'' row for the red circular marker. The angular sectors associated with the black-hole shadow, both inward- and outward-pointing, are dominated by $\bar{\nu}_e$ emission from the polar atmosphere. This resembles the opposite ELN-crossing mechanism, driven by $\nu_e$ contamination, found for accretion disks with a long-lived hypermassive neutron star in Ref.~\cite{nagakura2025neutrino}. In our case, the ELN crossing is generated by $\bar{\nu}_e$ contamination from the polar atmosphere. Interpretation of the neutrino and antineutrino emission from the polar atmosphere requires some caution, since this is the least well-resolved region of the general-relativistic magnetohydrodynamics simulation of Ref.~\cite{Miller:2019gig}, from which our matter snapshot was extracted. Moreover, the original simulation requires an artificial atmosphere in this region for numerical stability, whose properties may not be fully hydrodynamic and may therefore introduce artificial baryon loading. In addition, wherever the fluid quantities fall outside the \texttt{NuLib} and \texttt{SFHo} tables, they are set to the corresponding lower bounds of the tabulated ranges. The number and depth of ELN crossings may also be affected in the polar regions, particularly between radially inward and outward directions, because the matter profile is held fixed in our calculation, which yields isotropic emission there. In a fully dynamical treatment, the polar emission would instead receive a relativistic boost and become forward-peaked along the fluid velocity in the lab frame.
            
            The angular distributions in Fig.~\ref{fig:ffi_angular_distributions} reflect three-dimensional effects arising from the interaction of the neutrino radiation field with a three-dimensional matter profile. In the ``Before FFC'' row for the black circular marker, the ELN angular distribution shows different crossing depths on the two sides of the disk. In the ``Before FFC'' row for the red circular marker, the $\nu_e$, $\bar{\nu}_e$, and heavy-flavor angular distributions reflect the spiral structure of the accretion-disk emission region. Although these three-dimensional patterns are clearly identifiable, they play only a minor role in the appearance of ELN crossings. The main findings of this work are expected to apply equally well to two-dimensional accretion-disk models.
        
        \subsubsection{Local study of the FFI\label{sec:local_study_of_the_ffi}}
        
            To investigate the small-scale behavior of the FFI in our post-merger disk, we perform a local neutrino quantum kinetics simulation at the location identified with a black circular marker in the left panels of Fig.~\ref{fig:instabilities}. Specifically, the full angular data at this location, shown in the top row of Fig.~\ref{fig:ffi_angular_distributions}, is used as initial condition. We solve the QKE in a periodic box of size $48\times1\times1\,\mathrm{cm}$, neglecting vacuum and matter potentials and retaining only the self-interaction Hamiltonian. We ignore the collision term to isolate the effects of the FFI. Spatial flavor structures are resolved by discretizing the domain into cells of size $1$, $0.5$, $0.25$ and $0.125$ cm in the $x$ direction (which is the main direction of the neutrino fluxes) and $1$ cm in the $y$ and $z$ directions. We seed random perturbations in the neutrino flavor states with amplitudes six orders of magnitude smaller than the diagonal number densities.
            
            Studies of FFC asymptotic states suggest that the angular region in the ``shallow" side\footnote{For instance, if the net ELX--XLN is positive, the shallow side corresponds to the regions of negative ELN--XLN.} of the ELN--XLN crossings experience flavor equipartition, while the other regions adjust their flavor conversion to satisfy conservation laws~\cite{PhysRevD.107.103022,Richers:2024zit}. We indeed find this to be the case. The top two rows of Fig.~\ref{fig:ffi_angular_distributions} show the ELN--XLN, electron, electron antineutrino, and heavy-flavor angular distributions before (top row) and after (second row) FFC. In the beams with initially negative ELN--XLN (close to the $-x$ direction, pointing outward from the disk), electron neutrinos and antineutrinos are converted into heavy flavors. This would lead to a stronger flux of heavy neutrinos emerging from the accretion disk, thereby enhancing its cooling as was shown in Ref.~\cite{mukhopadhyay2024time}.
            
\begin{figure}[!htbp]
                \centering
                \includegraphics[width=\linewidth]{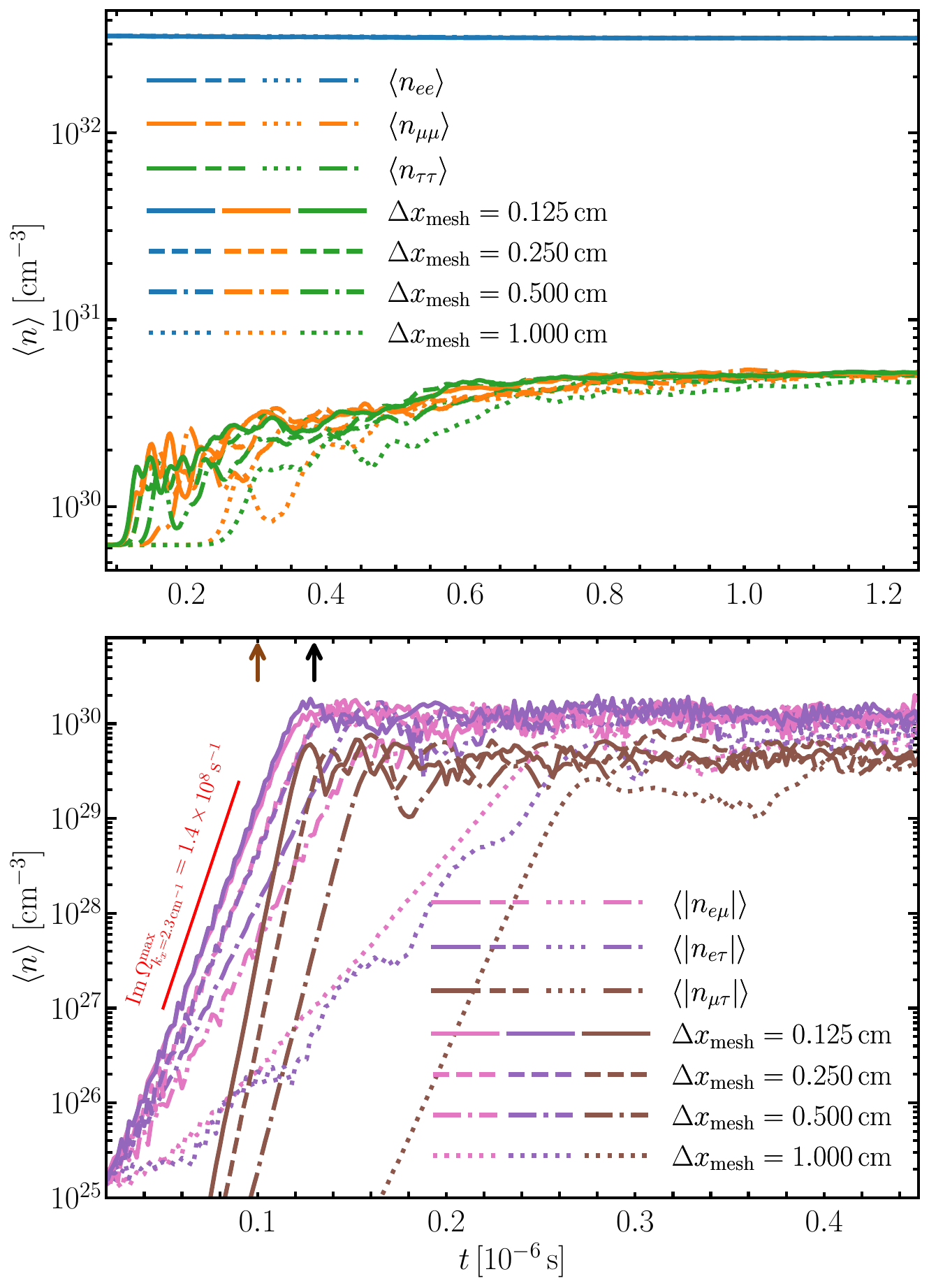}
                \caption{
                    Evolution of the domain averaged neutrino number densities under the FFI at the location with a black circular marker in the left panels of Fig.~\ref{fig:instabilities}. Initial conditions were extracted from the {\tt HAR-class} simulation, with parameters listed in Table~\ref{tab:resolution}. On timescales of $\sim 10^{-7}\,\mathrm{s}$, electron neutrinos and antineutrinos are converted into heavy flavors. This will result in a stronger flux of heavy neutrinos emerging from the accretion disk. 
                    The instability growth is not equally well resolved across spatial resolutions and is underestimated at lower resolution. The red line shows the predicted instability growth rate from a three-dimensional, multiangle linear stability analysis. However, the domain-averaged number densities approach the same asymptotic state for all spatial resolutions considered.
                    The complex Fourier spectra of the spatial flavor inhomogeneities in the number densities and the quantum coherence, evaluated at the times indicated by the black (brown) arrows along the time axis (linear and early-nonlinear regimes, respectively), are shown as black (brown) curves in Fig.~\ref{fig:ffi_fft} for the spatial resolution corresponding to the dashed curves.
                }
                \label{fig:ffi_number_densities}
            \end{figure}

            Figure~\ref{fig:ffi_number_densities} shows the evolution of neutrino number densities under the FFI for this location. Consistent with the smallness and shallowness of the angular region experiencing flavor equipartition (see Fig.~\ref{fig:ffi_angular_distributions}), only a small fraction of electron neutrinos and antineutrinos are converted to heavy flavors. The instability growth is not equally well resolved at all spatial resolutions. Similar to Ref.~\cite{Nagakura:2025brr}, decreasing spatial resolution underestimates the growth of the flavor instability (dotted). The growth nearly converges at the higher resolutions (dashed and solid). Interestingly, at this location, even when low resolution fails to capture all inhomogeneous modes, the asymptotic number densities are still well reproduced across all resolutions.
            
            The FFI exponentially enhances in time the quantum coherence of the neutrino field, with a growth rate of $\sim 10^{8}\,\mathrm{s}^{-1}$ (see the highest spatial resolution run (solid) in lower panel of Fig.~\ref{fig:ffi_number_densities}). Flavor number densities are exchanged on timescales of $\sim 10^{-7}\,\mathrm{s}$ (see upper panel of Fig.~\ref{fig:ffi_number_densities}), meaning that neutrinos undergo flavor conversion if they travel within the disk along paths of $\sim 10\,\mathrm{m}$ or longer. This path length is well below the hydrodynamic scales usually considered in simulations of core-collapse supernovae and binary mergers \cite{wang2025,qiu2025neutrino,lund2025angle}.  
            
\begin{figure*}[!htbp]
                \centering
                \includegraphics[width=0.9\textwidth]{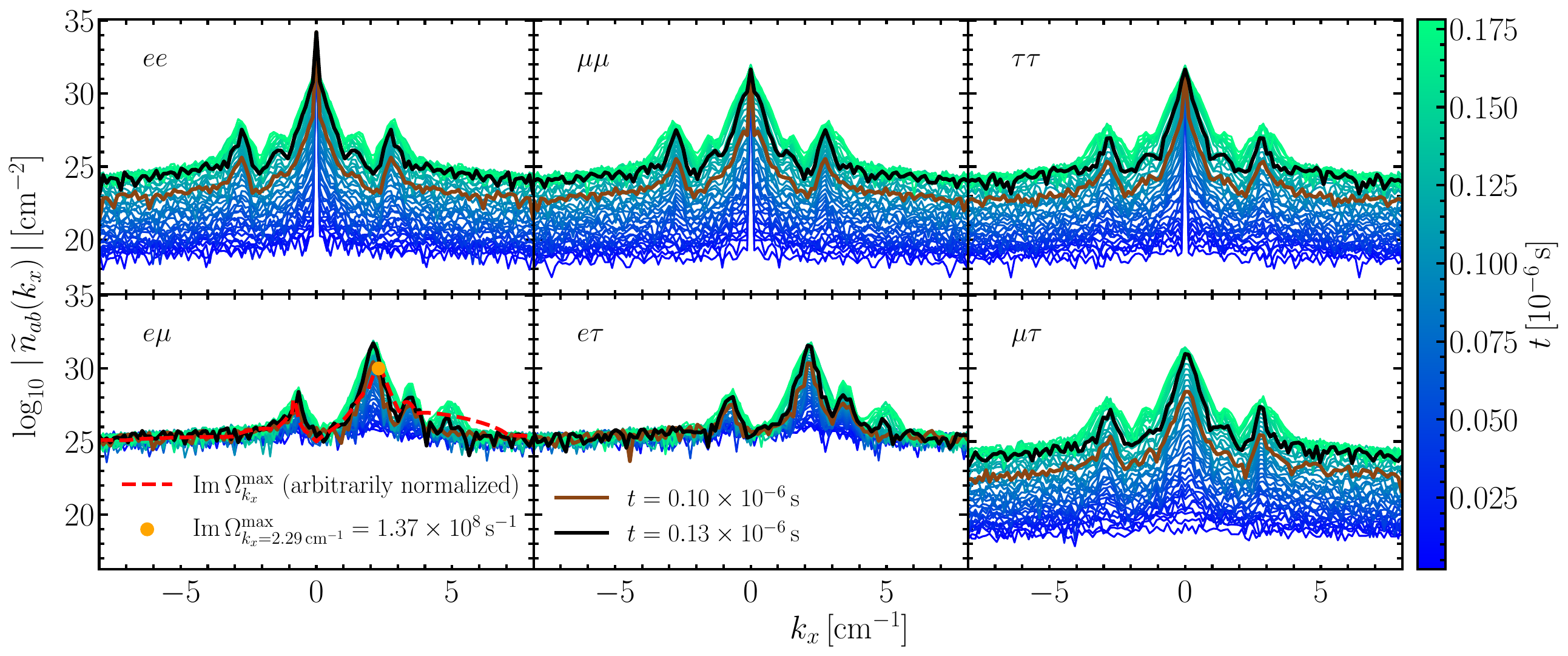}
                \caption{
                    Time evolution of the complex Fourier spectra (denoted $\tilde{n}_{ab}$) of the spatial flavor inhomogeneities in the number densities (top panels) and the quantum coherence (bottom panels) generated by the FFI at the location with a black circular marker in Fig.~\ref{fig:instabilities}, for the spatial resolution corresponding to the dashed curves in Fig.~\ref{fig:ffi_number_densities}. Initial conditions were extracted from the {\tt HAR-class} simulation, with parameters listed in Table~\ref{tab:resolution}.
                    The dashed red line shows the growth rate dispersion from a three-dimensional, multiangle linear stability analysis, arbitrarily normalized for comparison. The most unstable mode is located at the orange circular marker, in good agreement with our simulation.
                    Number densities are dominated by homogeneous modes, while inhomogeneous unstable modes appear in the quantum coherence of the electron–muon and electron–tau sectors. The black (brown) curves show the mode distributions in the linear (early-nonlinear) regime. These times are also indicated in Fig.~\ref{fig:ffi_number_densities} on the time evolution of $\langle n\rangle$, by black (brown) arrows along the time axis.
                }
                \label{fig:ffi_fft}
            \end{figure*}

            Figure~\ref{fig:ffi_fft} shows the complex Fourier spectra of the flavor–number densities and the inter-flavor quantum coherences. The flavor on-diagonal number densities are mostly dominated by the approximately homogeneous ($k_x=0$) initial conditions (upper panels), although $|k_x|>0$ waves are also noticeable. The unstable coherence modes for $n_{e\mu}$ and $n_{e\tau}$, for the simulation in dashed lines in Fig.~\ref{fig:ffi_number_densities}, are dominated by $k_x \approx 2.1\,\mathrm{cm}^{-1}$ with growth rates of $\mathrm{Im} \, \Omega^\mathrm{max}_{k_x\,\approx\,2.1\,\mathrm{cm}^{-1}} \approx 1.1\times 10^{8} \, \mathrm{s}^{-1}$. We confirm that this behavior is expected by performing a multiangle linear stability analysis (LSA) using the full initial angular distribution, generalizing the axially symmetric approach presented in Appendix B of~\cite{froustey2024neutrino}. Because our local QKE simulation is performed in a domain that is highly elongated along the $x$-direction, the unstable spatial modes are effectively restricted to the $k_x$ branch, even though the momentum-space angular distribution is fully three-dimensional. Accordingly, the LSA is used here to predict the unstable modes along $k_x$, rather than to identify the global maximum over all three-dimensional wave vectors. The predicted FFI growth rate as a function of $k_x$, for the two-flavor $e\text{--}\mu$ system, is shown as a dashed red line on the bottom panel of Fig.~\ref{fig:ffi_fft}. The agreement with the Fourier peaks is excellent, validating that the local QKE simulation reproduces the unstable mode expected along the $k_x$ branch. Within this restricted geometry, we predict in particular that the fastest growing mode occurs for $k_x \approx 2.3 \, \mathrm{cm}^{-1}$, with a growth rate $\mathrm{Im} \, \Omega^\mathrm{max} \approx 1.4 \times 10^8 \, \mathrm{s}^{-1}$. Because of the initial symmetry $\mu \leftrightarrow \tau$, we can draw the exact same conclusions in the $e\text{--}\tau$ sector. We emphasize that this corresponds to the fastest-growing mode permitted by the adopted simulation geometry, and not necessarily the global maximum of the full three-dimensional instability. Nevertheless, we corroborated that simulations along the $\hat{k} = (1,0,0)$, $(1/\sqrt{2},0,1/\sqrt{2})$, and $(1/\sqrt{2},1/\sqrt{2},0)$ directions robustly eliminate the ELN--XLN crossing, similarly to the top two rows of Fig.~\ref{fig:ffi_angular_distributions}, and agree in their asymptotic number densities in Fig~\ref{fig:ffi_number_densities}. Furthermore, the latter two directions show instabilities with smaller growth rates than the one along $x$. This also agrees with three-dimensional neutrino quantum kinetic simulations, which have demonstrated that local simulations do not deviate significantly from the one-dimensional counterpart as long as they are aligned with the ELN-XLN flux \cite{PhysRevD.110.123018,richers2021neutrino}.

            In the bottom right panel of Fig.~\ref{fig:ffi_fft}, we see that the Fourier spectrum of $\mu\text{--}\tau$ coherence coincides with the ones of $n_{\mu \mu}$ and $n_{\tau \tau}$. Furthermore, the growth rate of $n_{\mu \tau}$ in Fig.~\ref{fig:ffi_number_densities} is twice as large as the one of $n_{e \mu}$ and $n_{e \tau}$. This can be understood in the polarization vector picture, focusing on the $e\text{--}\mu$ two-flavor system. The instability developing in the $e\text{--}\mu$ sector corresponds to an exponential growth of the transverse part of the polarization vector, that is of $\sin \theta \sim \theta$ the angle with respect to the vertical direction. By conservation of the length of this polarization vector, the flavor-diagonal densities evolve following $1 - \cos \theta \propto \theta^2$ \cite{Particle-in-cell}. As a consequence, the growth of $n_{\mu \mu}(t) - n_{\mu \mu}(0)$ is twice as fast as the growth of $n_{e \mu}$. The same thing happens in the $e\text{--}\tau$ sector. Finally, the growth of $n_{\mu \tau}$ does not correspond to an instability in the $\mu\text{--}\tau$ sector, but simply reflects the growth of $n_{\mu \mu}$ and $n_{\tau \tau}$, which amplifies an initial coherence $n_{\mu \tau} \neq 0$ at the rate $2 \, \mathrm{Im} \, \Omega^\mathrm{max}_{e \mu}$.

            We have pointed out that the on-diagonal number density Fourier spectra are peaked on $k=0$, showing that the occupation numbers remain mostly homogeneous. One can also note some secondary peaks, which appear already in the linear phase (see red line on the top panels of Fig.~\ref{fig:ffi_fft}) but are not aligned with the peaks of the $e-\mu,\tau$ flavor coherences (bottom panel). This is not unexpected: these peaks appear through the self-interaction commutator term in the QKE, which corresponds to a convolution in Fourier space. We thus expect secondary peaks on the top panels to be located around wavevectors given by the differences of the bottom panel peaks. We do not further explore this question here.
            However, we note that flavor inhomogeneities can play a crucial role in the disk evolution as shown in Refs.~\cite{Nagakura:2025brr, urquilla2025testingcommonapproximationsneutrino}. If flavor waves are depleted, the sequence of instabilities that the neutrino field experiences may fail to saturate to the correct asymptotic states \cite{urquilla2025testingcommonapproximationsneutrino}. The extent to which advection and collisions deplete flavor inhomogeneities remains an open question.

    \SwitchableClearpage
    \subsection{Collisional Flavor Instability}
            
        The difference in absorption opacities between neutrinos and antineutrinos can lead to a collisional flavor instability (CFI), even for isotropic distributions~\cite{johns2021collisional}. We study the occurrence of such CFIs in the radiation field of the \texttt{HAR-class} simulation (see Figs.~\ref{fig:n_diagonal_densities} and \ref{fig:fhat_diagonal_densities}) using multienergy neutrino distributions, as well as in a reduced monochromatic setup to explore the differences. We present multienergy local quantum-kinetic simulations of CFI that, to our knowledge, have not been reported elsewhere in the literature. We focus on the post-instability saturation phase in a particular point in the accretion disk, applying different attenuations of the self-interaction Hamiltonian.
    
        \subsubsection{Global analysis on the classical radiation field\label{sec:global_cfi}}
        
            To get some intuition regarding the origin of the CFI, we perform both a monochromatic and a multi-energy LSA in the steady state solution of the neutrino radiation field for the {\tt HAR-class} simulation in Figs.~\ref{fig:n_diagonal_densities} and \ref{fig:fhat_diagonal_densities}. The monochromatic case enables an analytic description of the CFI and helps with building intuition, while the multi-energy calculations are needed for accurate results (although we note that the discrepancy between these approaches was suggested to be much larger in CCSN environments~\cite{wang2025effectcfi} than in NSM ones~\cite{froustey25cfi}). We consider charged-current interactions for $\nu_e$ and $\bar{\nu}_e$, assuming isotropic and homogeneous emission. We also include pair processes as an effective absorption opacity for all neutrino species, interpolated from \texttt{NuLib}.
            
            For the monochromatic analysis we define the energy-averaged neutrino absorption rates similarly to Refs.~\cite{akaho2024collisional,nagakura2025neutrino}, that is,
            \begin{equation}
              \Gamma_{\nu_i}
              \equiv
              \frac{\dot n_{\nu_i}^{\mathrm{abs}}}{n_{\nu_i}},
              \qquad
              \dot n_{\nu_i}^{\mathrm{abs}}
              =  \sum_{ \epsilon}\Gamma_{i,\epsilon}\,n_{\nu_i,\epsilon},
              \label{eq:Gamma_def}
            \end{equation}
            where $\Gamma_{i,\epsilon}$ and $n_{\nu_i,\epsilon}$ are the total absorption rate and neutrino number density, respectively, for neutrino species $i\left(= e,\mu, \tau \right)$ and energy bin $\epsilon$. This energy-averaging method is dubbed ``Method B" in~\cite{wang2025effectcfi}. The monochromatic CFI growth rate is given by (see, e.g.,~\cite{Liu:2023pjw})
            \begin{equation}
            \label{eq:growth_cfi}
              \sigma_{\text{CFI}}
              = \max \left\{
              \operatorname{Im}\left(\omega_{\pm}^{\text{pres}}\right), \operatorname{Im}\left(\omega_{\pm}^{\text{break}}\right)
                 \right\},
            \end{equation}
            where the growth rate of the isotropy-preserving modes is given by the imaginary part of
            \begin{align}
              \omega_{\pm}^{\text{pres}} &= -A -i\gamma \pm \sqrt{A^2-\alpha^2 + i2G\alpha} \, ,
              \label{mode_iso_pres} \\
            \intertext{whereas for the isotropy-breaking modes}
              \omega_{\pm}^{\text{break}} &= \frac{A}{3} -i\gamma \mp \sqrt{\left(\frac{A}{3}\right)^2-\alpha^2 - \frac{i2G\alpha}{3}} \, .
              \label{mode_iso_break}
            \end{align}
            In these equations, the auxiliary quantities $G$, $A$, $\gamma$, and $\alpha$ are
            \begin{equation}
            \begin{aligned}
              G      &\equiv \frac{g + \bar g}{2}, \qquad  &
              A      &\equiv \frac{g - \bar g}{2},  \\
              \gamma &\equiv \frac{\Gamma + \bar\Gamma}{2},  &
              \alpha &\equiv \frac{\Gamma - \bar\Gamma}{2},
              \end{aligned}
            \end{equation}
            with
            \begin{equation}
              g      \equiv \sqrt{2}\,G_F\left(n_{\nu_e} - n_{\nu_x}\right),\qquad
              \bar g \equiv \sqrt{2}\,G_F\left(n_{\bar{\nu}_e} - n_{\bar{\nu}_x}\right),
            \end{equation}
            and
            \begin{equation}
              \Gamma      \equiv \frac{\Gamma_{\nu_e} + \Gamma_{\nu_x}}{2},\qquad
              \bar\Gamma  \equiv \frac{\Gamma_{\bar\nu_e} + \Gamma_{\bar\nu_x}}{2}.
            \end{equation}
            Note that the $\pm$ signs in front of the square root are reversed between Eqs.~\eqref{mode_iso_pres} and \eqref{mode_iso_break}, such that in the relevant limit for NSM environments ($|G \alpha| \ll A^2$),  $\omega_{\pm}^{\text{pres}} \simeq  \omega_{\pm}^{\text{break}}$. With this choice, for $A > 0$ the plus sign (resp. minus sign) corresponds to a gapless (resp. gapped) mode, following the nomenclature of~\cite{Fiorillo:2025zio} (see also the discussion in~\cite{froustey25cfi}).
            
            The monochromatic CFI growth rates, given by Eq.~\eqref{eq:growth_cfi}, are shown in the second-column panels of Fig.~\ref{fig:instabilities}. To obtain the multi-energy LSA predictions, we use the steady-state energy spectrum of the neutrino fields from the {\tt HAR-class} simulation. We adopt the methodology of Refs.~\cite{wang2025effectcfi,froustey25cfi}. The resulting multi-energy CFI growth rates are shown in the third-column panels of Fig.~\ref{fig:instabilities}. The collisional flavor-unstable region extends throughout the accretion disk, reaching values up to $\sim 10^{5}~\mathrm{s^{-1}}$. The growth rate decreases toward the poles, vanishing around $45^{\circ}$ above and below the disk plane. For the snapshot considered here, the number densities satisfy \(n_{\nu_e} > n_{\bar{\nu}_e} > n_{\nu_x}\), and \(\Gamma_{\nu_e} > \Gamma_{\bar{\nu}_e}\) throughout the simulation domain. The potentially unstable branch is therefore of the gapless type, namely the \(+\) branch in the monochromatic expressions~\eqref{mode_iso_pres}--\eqref{mode_iso_break}. This statement, however, need not remain valid over the entire lifetime of the disk. Since we are in the regime $|G \alpha| \ll A^2$ and we only include absorption processes, the isotropy-preserving and breaking modes have almost identical growth rates, the former being ever so slightly larger (we come back to this point in the subsequent local CFI study). The energy-averaged estimate provides overall the correct regions of instability (except very close to the edge of unstable region and the instabilities along the poles, which do not exist for the multi-energy case). We verified that it typically overestimates the growth rates by a factor of a few, which is consistent with the findings of \cite{froustey25cfi} in a completely independent neutron star merger environment, and which is very different from what happens in CCSNe~\cite{wang2025effectcfi}.
            
            Resonant CFI does not occur in the disk (as said above, we are in the regime $|G \alpha| \ll A^2$), a consequence of the clear separation of number densities $n_{\nu_e}>n_{\bar{\nu}_e}>n_{\nu_x}$ visible in Fig.~\ref{fig:n_diagonal_densities}. We emphasize that we solve the complete dispersion relation in the multi-energy case, or use the complete formulas~\eqref{mode_iso_pres}--\eqref{mode_iso_break}, such that the results shown in the second and third columns of Fig.~\ref{fig:instabilities} demonstrate the absence of resonant CFI. As a consequence, CFIs are subdominant compared to FFIs, with growth rates up to four orders of magnitude smaller, although in the innermost disk near the black hole some regions exhibit CFI without FFI as can be observed in the right panels of Fig.~\ref{fig:instabilities}. 
            
        \subsubsection{Local study of the CFI\label{sec:local_study_of_the_cfi}}
        
            \paragraph{Simulation setup}
                
                To assess the potential effect of the CFI in our post-merger disk, we perform a multi-energy collisional neutrino flavor transformation simulation in a small domain inside the post-merger disk. We extracted the neutrino configuration from the point marked with a green upsidedown triangle in the third top panel of Fig.~\ref{fig:instabilities}, which is collisionally unstable. Our focus is on the asymptotic flavor number densities, distribution functions, and the impact of the attenuation factor on them. We simulate a homogeneous periodic box neglecting vacuum and matter potentials and attenuating the self-interaction Hamiltonian setting $\eta$ to $10^{-5}$, $10^{-4}$ and $10^{-3}$ [see Eq.~\eqref{eq:QKE}]. Isotropy is imposed in the initial angular distributions to avoid overlap of the FFI and CFI, since this point also has an ELN crossing. We also allow isotropy-breaking modes to develop by resolving the angular domain into 92 beams per energy bin with 13 energy bins, logarithmically spaced up to $103\,\mathrm{MeV}$. In the off-diagonal number densities we seed small random perturbations six orders of magnitude smaller than the diagonal number densities. Neutrino emission and absorption are included with interaction rates from \texttt{NuLib} for the same fluid properties ($\rho$, $T$ and $Y_e$) from where this neutrino configuration was extracted in the accretion disk.
                
\begin{figure*}[!htbp]
                    \centering
                    \includegraphics[width=0.8\textwidth]{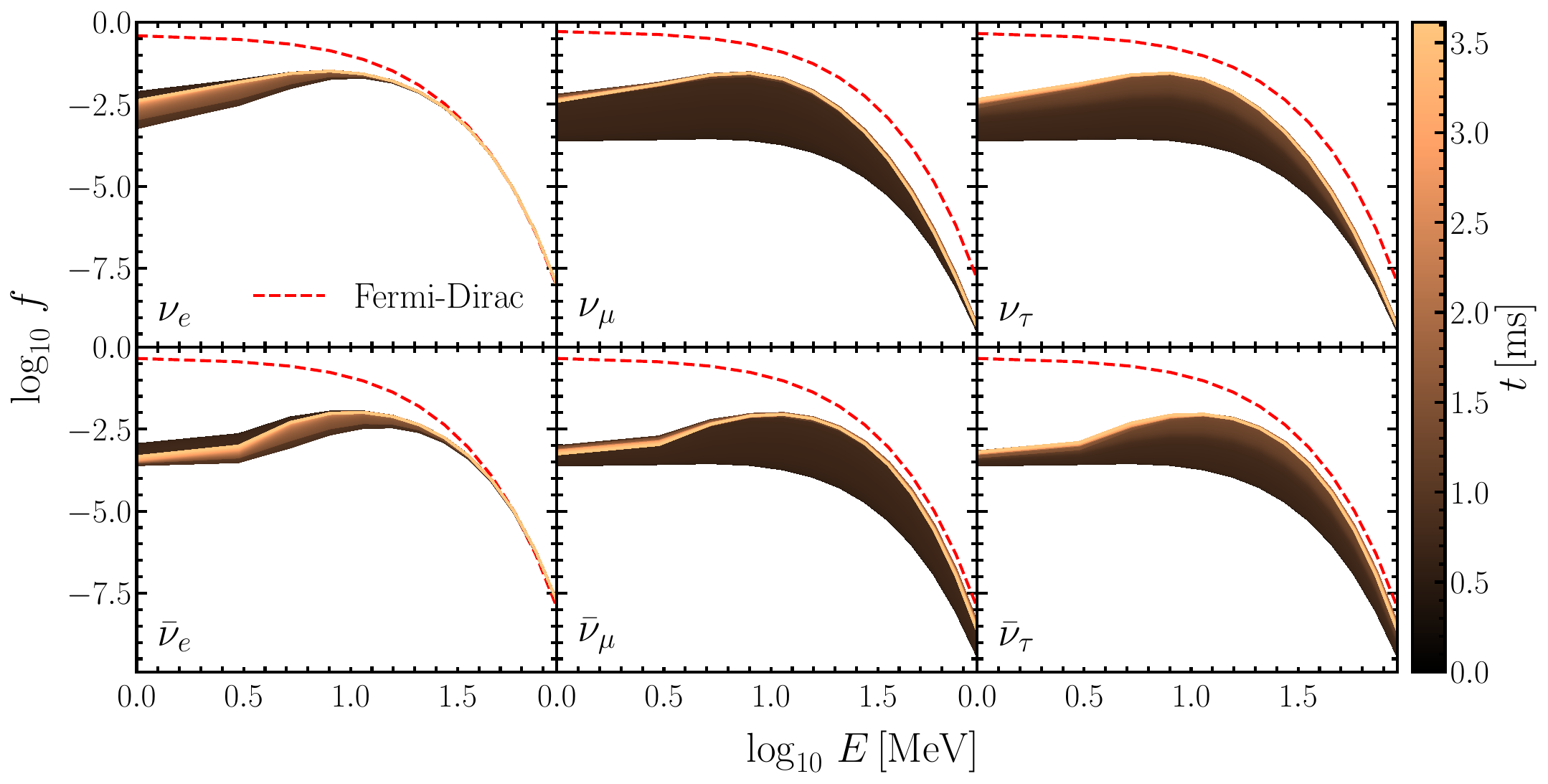}
                    \caption{
                        \label{fig:f_spectrum}
                        Evolution of the distribution functions driven by the CFI in the point marked with a upside down green triangle in the third top panel of Fig.~\ref{fig:instabilities} for an attenuation factor of $\eta=10^{-5}$ (same as the solid lines in Fig.~\ref{fig:n_local_cfi}).
                        Initial conditions were extracted from the {\tt HAR-class} simulation, with parameters listed in Table~\ref{tab:resolution}.
                        By $3.2$ ms, the occupation numbers of electron neutrinos and antineutrinos decrease at low energies, while those at higher energies remain unchanged. The heavy-flavor distribution grows, increasing number densities by two orders of magnitude. The peak of the heavy-neutrino distribution occurs at lower energies than that of heavy antineutrinos.
                    }
                \end{figure*}

                Local collisional neutrino simulations alone cannot capture global advection effects and would naturally relax neutrinos toward thermodynamic and chemical equilibrium with the fluid on collisional timescales. To avoid this and maintain the drive toward a classical steady state given by global advection in the {\tt HAR-class} simulation, we follow the approach of~\cite{Froustey:2025nbi}, generalized to a multi-energy setup, and define an effective ``equilibrium'' spectrum from the classical steady state at the location marked by the green inverted triangle in the third panel of Fig.~\ref{fig:instabilities}. These initial distribution functions for each neutrino species are shown as the darkest brown curves in Fig.~\ref{fig:f_spectrum}. Our results must be considered in light of the limitations of this effective homogeneous setup; we come back to this point at the end of this section.

                \begin{figure}[!htbp]
                    \centering
                    \includegraphics[width=0.428\textwidth]{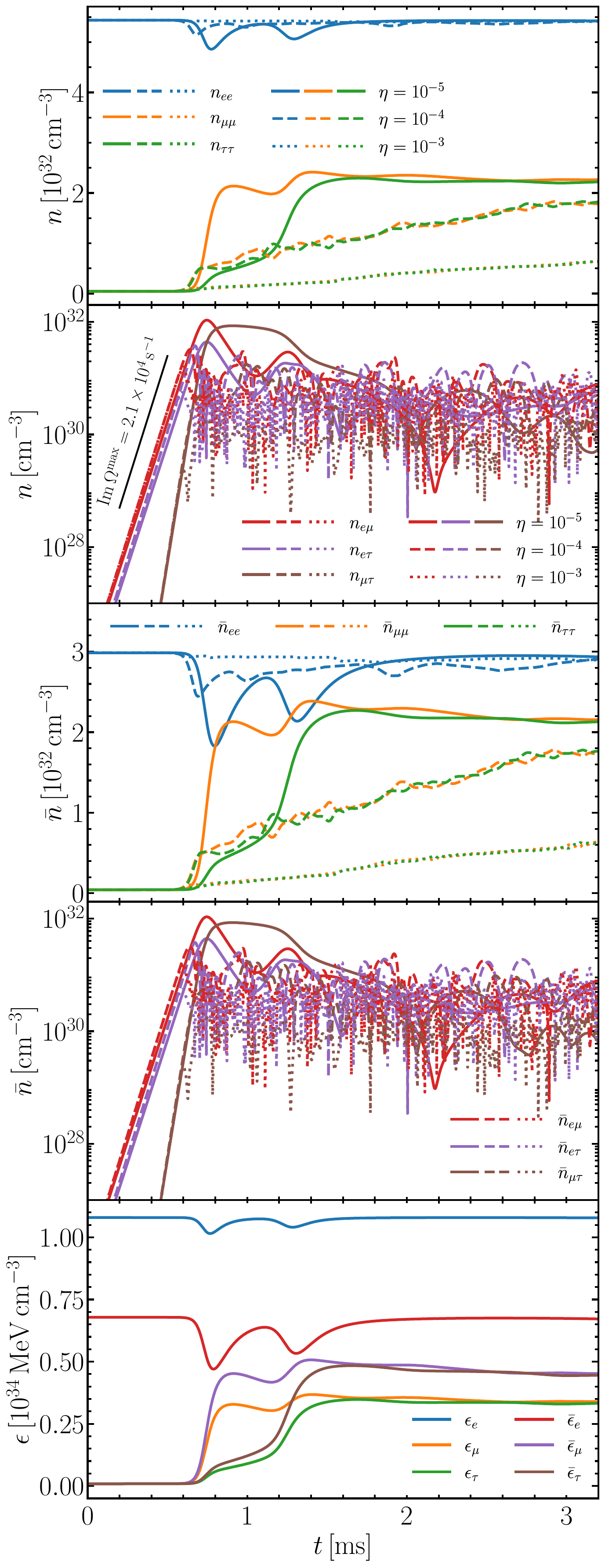}
                    \caption{
                        \label{fig:n_local_cfi}
                        Number and energy density evolution under the CFI for the point marked with a green upsidedown triangle in the third top panel of Fig.~\ref{fig:instabilities}. Initial conditions were extracted from the {\tt HAR-class} simulation, with parameters listed in Table~\ref{tab:resolution}. Collisional flavor conversion redefines the equilibrium number densities, increasing the population of heavy flavors. The growth rate of the largest isotropy preserving unstable mode is similar for all $\eta$. However, small attenuation factors make flavor conversion faster in the nonlinear regime. The heavy-antineutrino energy density surpasses that of neutrinos, even though their number densities are identical.
                    }
                \end{figure}
                
                \begin{figure}[!htbp]
                    \centering
                    \includegraphics[width=0.428\textwidth]{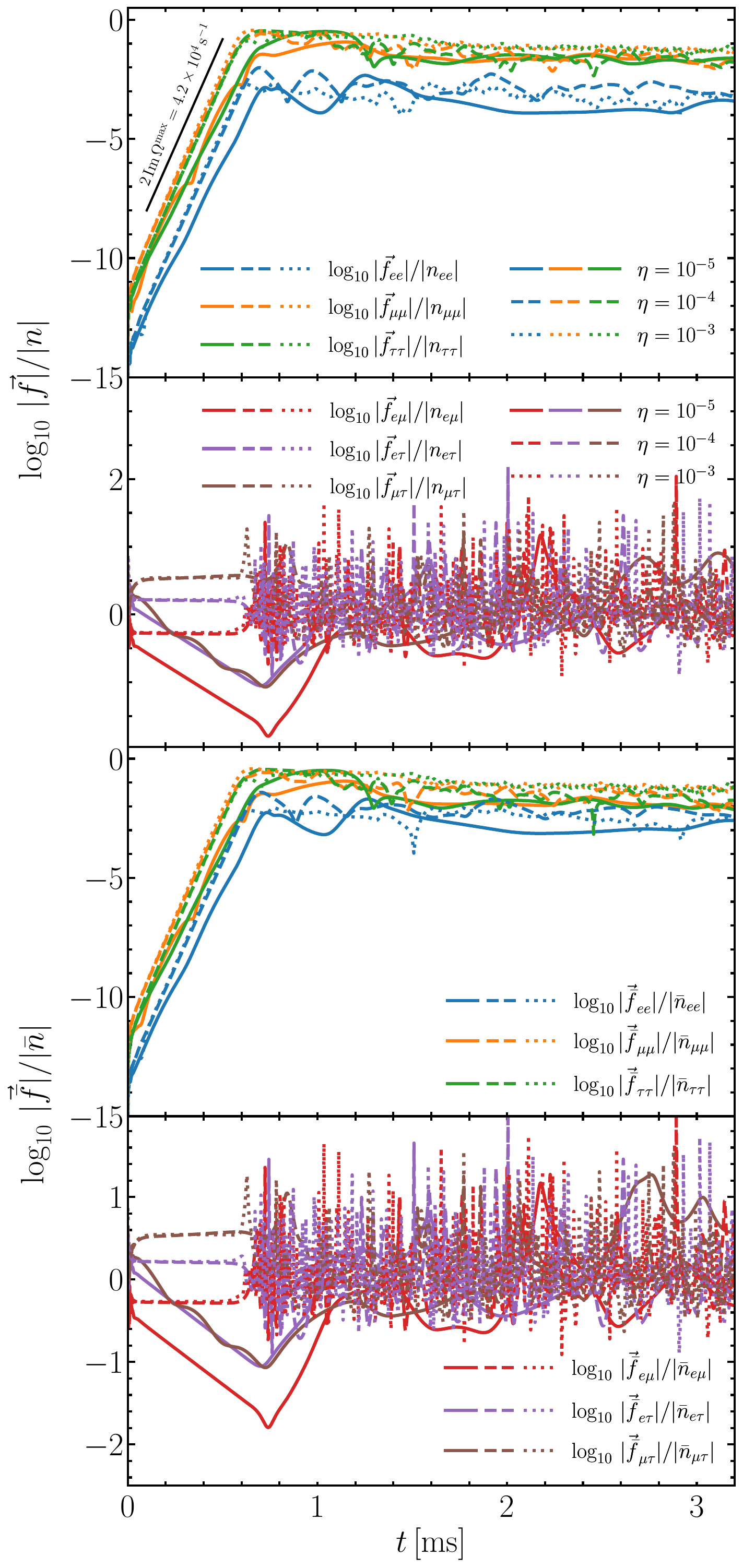}
                    \caption{
                        \label{fig:f_local_cfi}
                        Flux factors evolution under the CFI for the point marked with a green upsidedown triangle in the third top panel of Fig.~\ref{fig:instabilities}. Initial conditions were extracted from the {\tt HAR-class} simulation, with parameters listed in Table~\ref{tab:resolution}. For the most attenuated case ($\eta=10^{-5}$), the isotropy-breaking modes that drive the flux evolution grow more slowly than the isotropy-preserving modes that drive the number-density evolution. This yields to exponentially decreasing off-diagonal flux factors and an evolution close to the isotropic limit (solid red, purple, and brown curves). For weaker attenuation ($\eta=10^{-4}$ and $10^{-3}$), the isotropy-breaking and isotropy-preserving modes grow at comparable rates, so the off-diagonal flux factors remain large and approximately constant during the linear regime (dotted and dashed red, purple, and brown curves). 
                    }
                \end{figure}

            \paragraph{Evolution and attenuation factor}
                
                Fig.~\ref{fig:n_local_cfi} shows the time evolution of the neutrino and antineutrino number densities for each flavor. Collisional flavor conversion modifies the asymptotic number densities of each flavor, leading to higher amounts of heavy flavors. For $\eta=10^{-5}$, the heavy flavor number densities grow by two orders of magnitudes. The most prominent CFI modes in our simulation ($\eta = 10^{-5}$) have growth rates of $\mathrm{Im}\,\omega_{e\mu} = 2.1 \times 10^{4}\,\mathrm{s}^{-1}$. This shows excellent agreement with a multi-energy CFI analysis (see details in~\cite{froustey25cfi}), which predicts a gapless mode with growth rate $\mathrm{Im}\,\omega_{ex} = 2.1 \times 10^{4}\,\mathrm{s}^{-1}$. The monochromatic CFI growth rate at this point (see second column of Fig.~\ref{fig:instabilities}) is $\mathrm{Im}\,\omega_{ex} = 4.2 \times 10^{4}\,\mathrm{s}^{-1}$. This matches the expectation from Ref.~\cite{froustey25cfi} that the monoenergetic approximation overpredicts the growth rates of gapless modes by approximately a factor of two in postmerger environments.
                
                The exponential growth of the number density off-diagonal components in Fig.~\ref{fig:n_local_cfi} shows that varying the attenuation factor $\eta$ has little impact on the unstable-mode growth rates. This can be understood simply in the monochromatic case. Different attenuation factors modify the dispersion relation~\eqref{mode_iso_pres} via $G \to \eta G$ and $A \to \eta A$. As long as $|\eta G \alpha| \ll (\eta A)^2$ (which is true for $\eta = 1$ and remains valid except if $\eta$ is too small), one can expand the square root in equation \eqref{mode_iso_pres}. The resulting growth rate is found to be independent of $\eta$, consistent with the numerical results (see also Eq.~(A7) in Ref.~\cite{froustey25cfi} or Eqs.~(4) and (5) in Ref.~\cite{nagakura2025neutrino}). In other words, with the range of attenuation factors used, we do not artificially create a regime of resonant CFI.
                
                However, in the nonlinear regime, the attenuation factor $\eta$ becomes important: larger $\eta$ (i.e., closer to $1$) slows flavor conversion, whereas smaller $\eta$ accelerates it. This differs from the results of Ref.~\cite{Froustey:2025nbi}, which showed that the evolution is insensitive to the Hamiltonian attenuation factor in a fully isotropic setup. Therefore, the dependence on the attenuation factor must enter primarily through the development of anisotropies in the radiation field. Since the initial distributions in this calculation are isotropic, the anisotropy arises from the unstable isotropy-breaking modes. We note that in the actual disk, nonzero fluxes are traditionally generated by global properties (geometry, gradients, advection...), while in our calculation this is a purely local effect due to CFI modes, studied in our limited effective homogeneous setup. Figure~\ref{fig:f_local_cfi} shows the time evolution of the magnitudes of the neutrino and antineutrino flux factors for each flavor. For the most attenuated case, $\eta=10^{-5}$, the isotropy-breaking growth rate that drives the $\vec{f}$ evolution is smaller than the isotropy-preserving growth rate that drives the $n$ evolution. Accordingly, the slope of the off-diagonal flux factors, $\omega^{\mathrm{break}}-\omega^{\mathrm{pres}}$, is negative (see the solid red, purple, and brown curves). For the less attenuated cases, $\eta=10^{-4}$ and $10^{-3}$, the isotropy-breaking and isotropy-preserving modes grow at comparable rates. This yields nearly constant off-diagonal flux factors (see the dashed and dotted red, purple, and brown curves). Although $n_{aa}(t)-n_{aa}(0)$ grows exponentially, driven by the isotropy-preserving mode, the diagonal number densities do not change appreciably because the heavy-flavor neutrino populations are already nonzero and the exponential contribution remains negligible at early times. By contrast, the diagonal fluxes, which begin at zero, grow twice as fast as the isotropy-breaking modes. We can understand this behavior in the polarization-vector picture, as we did for the local study of the FFI. The instability develops in the off-diagonal sector through the exponential growth of the transverse component of the polarization vector, that is, of $\sin\theta \sim \theta$, where $\theta$ is the angle with respect to the vertical direction. If the length of the polarization vector is approximately conserved in the linear regime, then the flavor-diagonal densities evolve as $1-\cos\theta \propto \theta^2$ \cite{Particle-in-cell}. As a consequence, the exponential contributions to the diagonal fluxes and densities, which correspond to the vertical projections of the polarization vector, grow twice as fast as the transverse projection. The exponential growth of the diagonal flux factors in the linear regime is set by the flux evolution, and the growth of the diagonal flux factors is twice that of the isotropy-breaking mode. This behavior is reflected by the slope $2\,\mathrm{Im}\,\Omega^{\max}=4.2\times 10^{4}\,\mathrm{s}^{-1}$. Overall, the amplitudes of the diagonal flux factors remain small for strong attenuation, for example $\eta=10^{-5}$, and the evolution closely resembles the isotropic limit. This behavior appears clearly in Fig.~\ref{fig:f_local_cfi}, where the solid blue, orange, and green curves for $\eta=10^{-5}$ remain smaller than the dashed and dotted curves for $\eta=10^{-4}$ and $10^{-3}$, respectively. The remaining difference between the number-density and flux-density evolution in the $\eta=10^{-3}$ and $\eta=10^{-4}$ simulations during the nonlinear regime is consistent with an attenuation-dependent stability boundary. A multiangle LSA performed using the instantaneous moments $(n,\vec{f})$ shows that the $\eta=10^{-3}$ configuration becomes less unstable near saturation than the $\eta=10^{-4}$ configuration. Consequently, the $\eta=10^{-4}$ case continues to grow and undergoes stronger flavor conversion, whereas the $\eta=10^{-3}$ case quenches earlier.  This trend opens the possibility that, in the limit $\eta\to 1$, flavor conversion becomes negligible, which in turn implies that CFI may have little to no effect under realistic conditions. Given the computational cost of a CFI calculation with $\eta=1$, we do not attempt to verify this trend here and leave this question for future work.
                
\begin{figure*}[!htbp]
                    \centering
                    \includegraphics[width=0.8\textwidth]{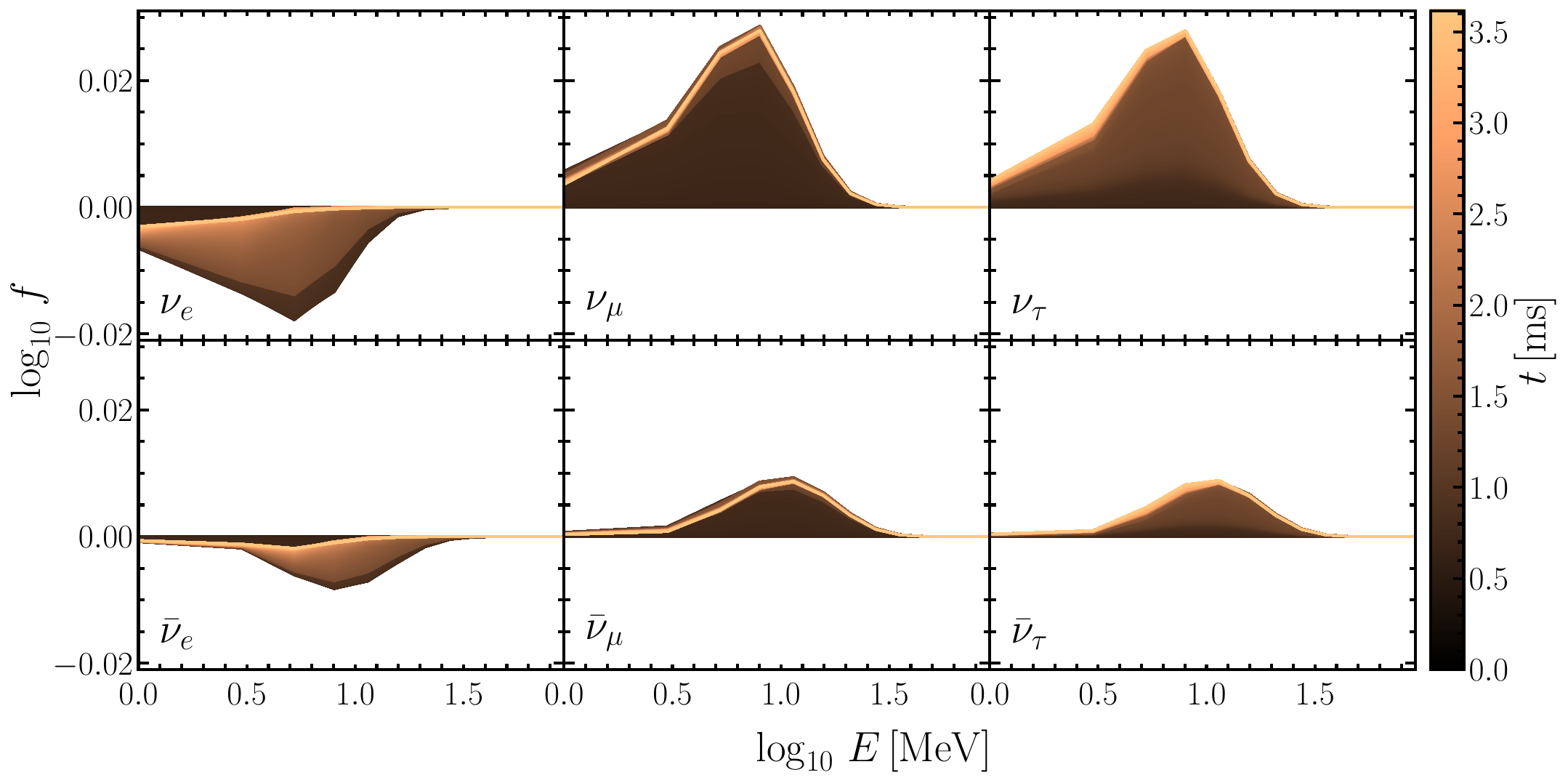}
                    \caption{
                        \label{fig:f_diff_spectrum}
                        Evolution of the difference between the initial and final distribution functions driven by the CFI in the point marked with a green upsidedown triangle in the third top panel of Fig.~\ref{fig:instabilities} for an attenuation factor of $\eta=10^{-5}$ (same as the solid lines in Fig.~\ref{fig:n_local_cfi}). Initial conditions were extracted from the {\tt HAR-class} simulation, with parameters listed in Table~\ref{tab:resolution}. By $3.2$ ms, heavy neutrinos gain more occupied states than heavy antineutrinos, but the heavy-antineutrino distribution peaks at higher energies. This gives heavy antineutrinos a larger total energy density.
                    }
                \end{figure*}

                Even though the asymptotic number densities of heavy lepton neutrinos and antineutrinos produced by the CFI are nearly identical, their energetics differ. The lower panel of Fig.~\ref{fig:n_local_cfi} shows the time evolution of the total energy density for each flavor, where the heavy-lepton antineutrino energy density surpasses that of heavy-lepton neutrinos. This is further clarified in the time evolution of the energy spectra shown in Fig.~\ref{fig:f_spectrum} and Fig.~\ref{fig:f_diff_spectrum}. By $3.2$ ms, the occupation numbers of electron neutrinos and antineutrinos at low energies have decreased, while those at higher energies remain unchanged. The change in occupied states is larger for heavy-lepton neutrinos than for heavy-lepton antineutrinos. However, the peak of the heavy-lepton antineutrinos distribution shifts to higher energies compared to heavy-lepton neutrinos. As a result, heavy-lepton antineutrinos reach a higher total energy density than heavy-lepton neutrinos.
                
                We find that a small attenuation parameter ($\eta = 10^{-5}$) makes the flavor evolution sensitive to the initial random perturbations in the off-diagonal quantum-coherence terms. These perturbations are rapidly amplified, leading to an early separation between the $\nu_\mu$ ($\bar{\nu}_\mu$) and $\nu_\tau$ ($\bar{\nu}_\tau$) evolutions because they initially amplify the muon neutrino and antineutrino components more strongly than the tau neutrino and antineutrino components. As a result, muon neutrinos and antineutrinos reach their asymptotic number densities and energy spectra sooner than the corresponding tau species. We corroborate that, in the opposite case, when the random perturbations initially amplify the tau neutrino and antineutrino components more strongly than the muon neutrino and antineutrino components, the tau families reach their asymptotic number densities and energy spectra sooner than the corresponding muon families, but the asymptotic state is independent of the initial perturbations. Increasing the attenuation to $\eta = 10^{-4}$ and $10^{-3}$ reduces this heavy-flavor splitting, suggesting that for $\eta = 1$ the $\nu_\mu$ ($\bar{\nu}_\mu$) and $\nu_\tau$ ($\bar{\nu}_\tau$) evolutions become similar.
                            
                \paragraph{Impact on observables}
                
                \begin{figure}[!htbp]
                    \centering
                    \includegraphics[width=0.45\textwidth]{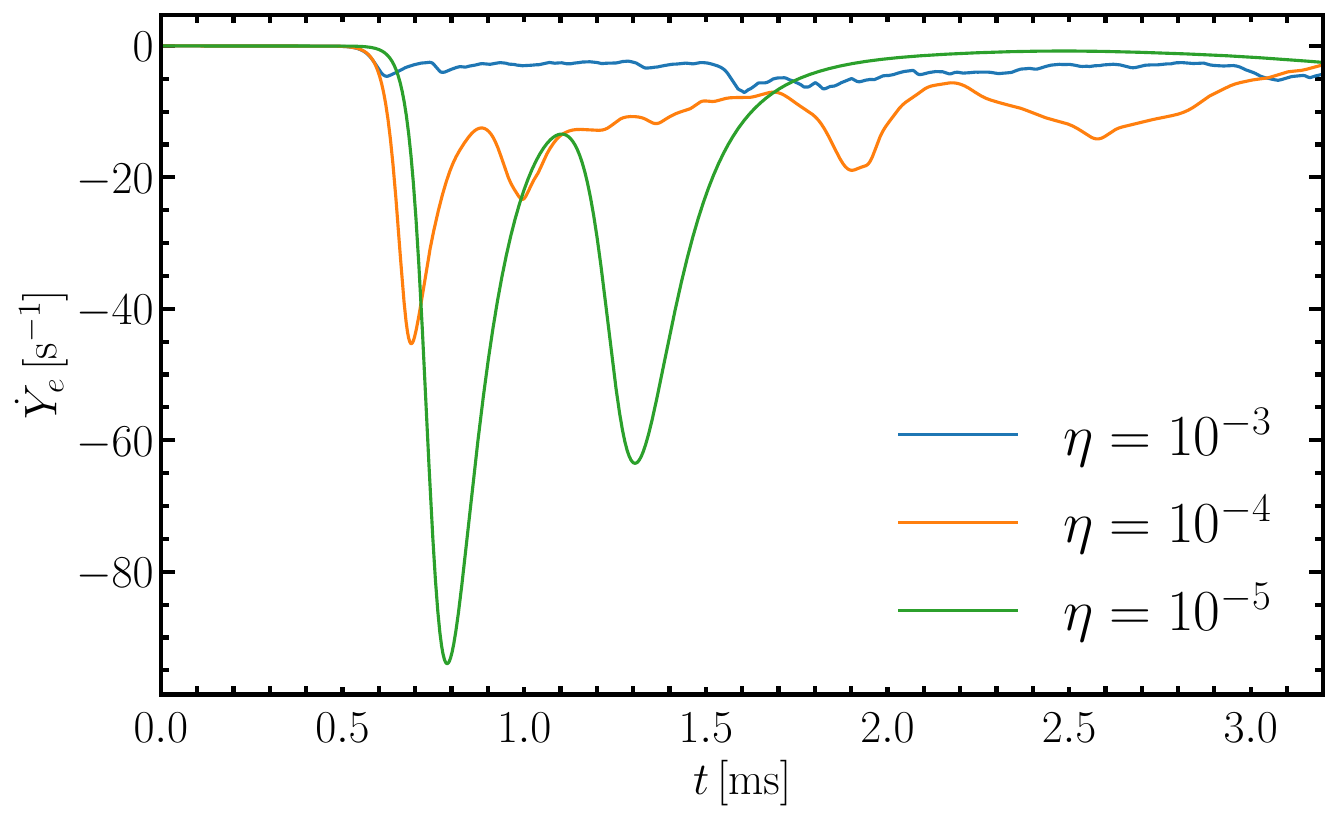}
                    \caption{
                        \label{fig:yedot}
                        $\dot{Y}_e$ evolution at the point marked with a green upside-down triangle in the third panel of the top row of Fig.~\ref{fig:instabilities} which correspond to the {\tt HAR-class} simulation, with parameters listed in Table~\ref{tab:resolution}, when CFI is isolated and relaxes under different attenuation factors $\eta$ in Eq.~\eqref{eq:QKE}. The changes in $Y_e$ follow from $\dot{Y}_e n_b = \dot{n}_{\bar{\nu}_e}^{\mathrm{coll}} - \dot{n}_{\nu_e}^{\mathrm{coll}}$, that is, from the asymmetry between the changes in the electron-neutrino and electron-antineutrino number densities due to absorption and emission.
                        We do not include feedback on the fluid, since our fluid profile remains static, even though $Y_e$ should evolve. The rate of change of the electron fraction decreases as $\eta$ approaches 1 (i.e., no attenuation).
                    }
                \end{figure}

                Our local, multi-energy quantum-kinetic simulations of a small region taken from a GW170817-like post-merger disk show that unstable modes saturate on a timescale of $\sim 10^{-3}\,\mathrm{s}$, corresponding to neutrino path lengths of $\sim 10^{2}\,\mathrm{km}$. Since trapped neutrinos do not participate in the instability (as suggested by the energy spectra in Figs.~\ref{fig:f_spectrum} and \ref{fig:f_diff_spectrum}), CFI is most likely to saturate only for neutrinos that traverse the dense region of the disk. We analyzed the rate of change of the electron fraction, $\dot{Y}_e$, for the CFI-only evolution and found that, in the $\eta=10^{-5}$ simulation, $\dot{Y}_e$ is negative throughout. This implies an overall neutronization of the disk fluid. But for attenuation factors closer to unity ($\eta = 10^{-4}$ and $10^{-3}$), the system undergoes less flavor conversion (see Fig.~\ref{fig:n_local_cfi}) and $\dot{Y}_e$ remains small, as shown in Fig.~\ref{fig:yedot}, suggesting a limited impact on the fluid conditions that determine subsequent nucleosynthesis. Our analysis would thus suggest that it is unlikely that CFI alone affects the dynamics or nucleosynthesis in post-merger disks that promptly form black holes. There are nevertheless some significant caveats in our analysis. Most importantly, our effective homogeneous setup \emph{mimics} the global effects of advection by introducing an artificial ``reservoir'' driving the system to the classical steady state. It is thus possible that some of the effects we observe could be artifacts of this setup.
                In our local simulations, the existence of a CFI and its unstable homogeneous modes are independent of the approximation of global advection by an effective equilibrium. The equilibrium emissivity term does not control the CFI unstable modes since it does not contribute to the off-diagonal emission and absorption terms and therefore does not appear in Eqs.~\eqref{mode_iso_pres} and \eqref{mode_iso_break}. This does not necessarily mean the unstable modes will persist inside the accretion disk, since the CFI modes might change if we relax the assumption of homogeneity and impose boundary conditions other than periodic. The nonlinear saturation, sustained flavor conversion, and resulting changes in neutrino fluxes depend on how the reservoir replenishes the local neutrino distributions. As mentioned previously, CFI saturates at a path length of $\sim 10^{2}\,\mathrm{km}$, much longer than the homogeneity scale of our local simulation $\sim 1\,\mathrm{km}$. At these scales, matter inhomogeneity and global advection are important, but we neglect them in our local simulation. We therefore cannot establish which of these nonlinear outcomes would persist in the disk, or use them to place a firm bound on the CFI's global impact. Resolving this question requires a global calculation that maintains the advective steady state while allowing the instabilities to grow and saturate with unattenuated flavor dynamics. Such a first-principles quantum kinetic calculation is not yet computationally feasible.
                On top of the CFI, neutrinos are also affected by FFI on shorter timescales. Ejecta in the disk plane are therefore likely to carry imprints of both CFI and more importantly FFI. The asymptotic state produced by their combined nonlinear evolution remains an open question, as does the impact of the CFI when the fluid $Y_e$ and opacities evolve dynamically with the neutrino radiation field, with both components feeding back on each other. Ref.~\cite{takahashi2026dynamic} analyzed the interplay between FFI and CFI in an inhomogeneous, anisotropic, single-energy set-up, relaxing the assumption of steady-state advection equilibrium imposed in our simulation and driving the system toward neutrino-fluid thermal equilibrium. They found diverse evolutionary flavor pathways between the initial conditions and the expected asymptotic state in thermal equilibrium.                
                More generally, providing definitive answers to these questions, especially regarding the interplay of actual global advection, FFI and CFI, will require global QKE simulations, which face significant computational challenges. This is the topic of the next section.
                
\SwitchableClearpage
\section{Global Quantum Kinetics\label{sec:global_qke_simulations}}

    \begin{figure*}[!htbp] 
        \centering
        \includegraphics[width=0.6\textwidth]{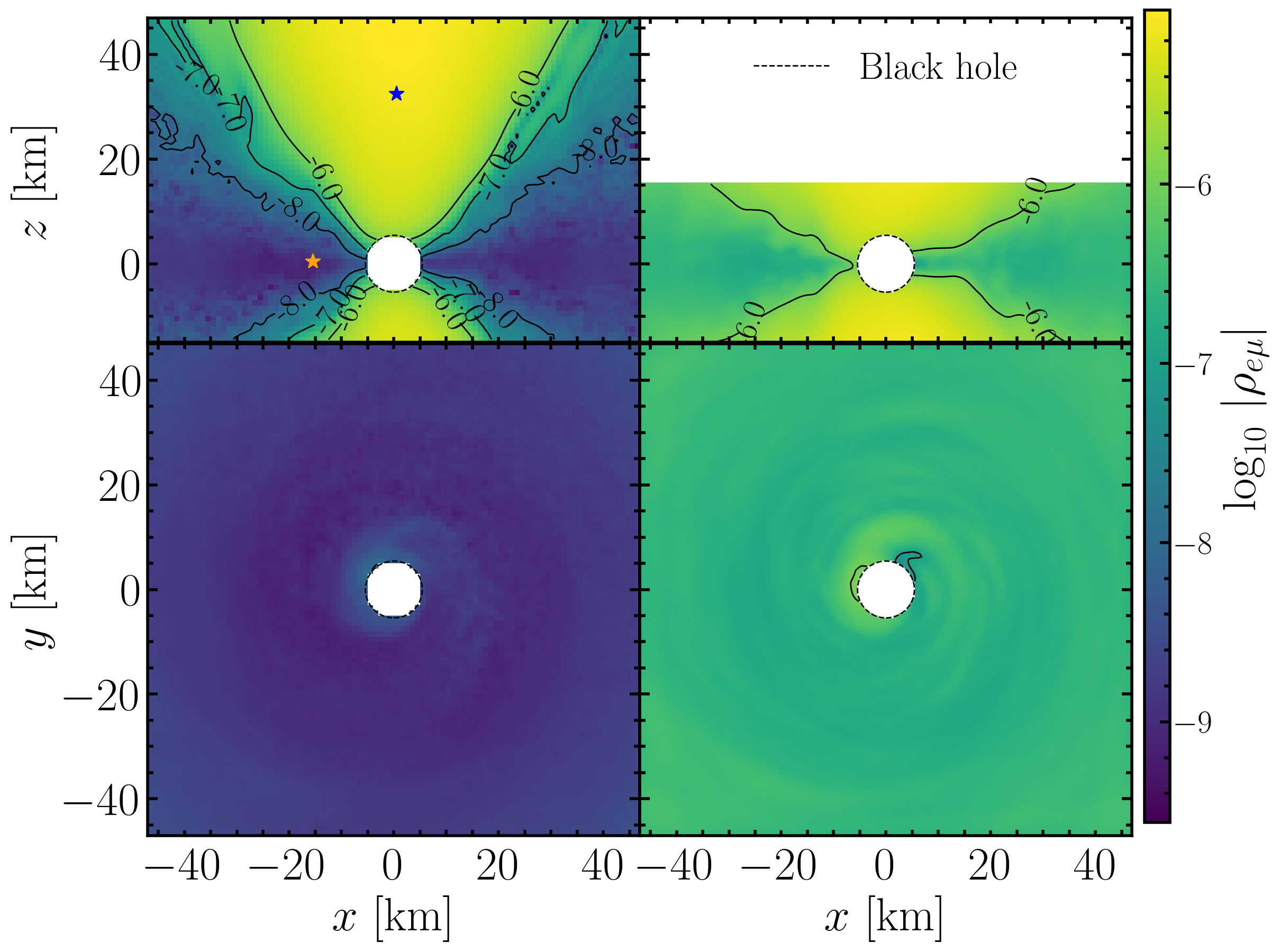}
        \caption{
            \label{fig:global_qke_rho_eu} 
            Asymptotic electron–muon neutrino quantum coherence $\rho_{e\mu} = n_{e\mu} / \mathrm{Tr}(n)$ in the \texttt{HAR-QKE} (left) and \texttt{HSR-QKE-reducedmatter} (right) simulations (see the simulations parameters in Table~\ref{tab:resolution}). The upper (lower) panels show polar (equatorial) slices. 
        }
    \end{figure*}

    \begin{figure}[!htbp]
        \centering
        \includegraphics[width=0.49\textwidth]{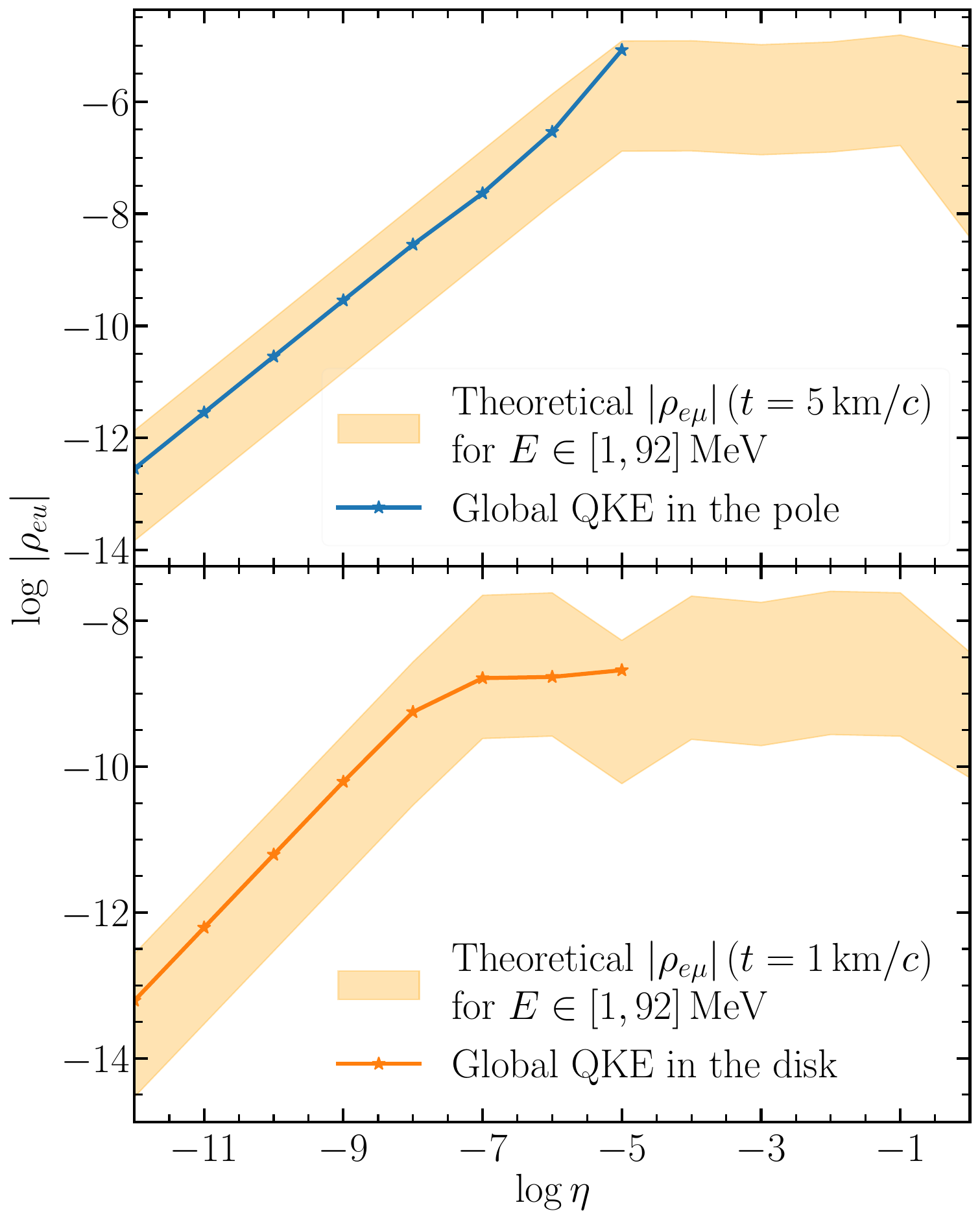}
        \caption{
            \label{fig:eta_vs_rho_eu_pole_and_disk}
            Comparison of the energy-integrated quantum coherence, $\rho_{e\mu}$, from the global QKE simulations (\texttt{LAR-LSR-QKE-*} for $\eta$ between $10^{-12}$ and $10^{-6}$, and \texttt{HAR-QKE}, which has higher angular resolution, for $\eta=10^{-5}$; see the simulation parameters in Table~\ref{tab:resolution}) with the theoretical prediction (beige bands), as a function of the attenuation factor, $\eta$. The upper and lower panels show the coherence in the pole and the disk, evaluated at the locations of the blue and orange stars in Fig.~\ref{fig:global_qke_rho_eu}, respectively. The theoretical bands assume evolution under a constant in-medium effective mixing angles and mass splittings [Eq.~\eqref{eq:eff_rho_eu_t}] over propagation distances of $5~\mathrm{km}$ in the pole and $1~\mathrm{km}$ in the disk.
        }
    \end{figure}
    
    We previously introduced an attenuation factor $\eta$ in the QKE [Eq.~\eqref{eq:QKE}] for our local CFI calculations (Sec.~\ref{sec:global_cfi}) to reduce the discrepancy between the timescales of neutrino interactions and flavor transformations, thereby making the problem numerically tractable. This is a common strategy in global astrophysical neutrino QKE simulations (see, e.g.,~\cite{xiong2024robust, nagakura2023global}). Physically, $\eta$ rescales the microscopic flavor-conversion length and timescales to larger effective values, so that flavor evolution occurs over macroscopic distances and times that can be probed concurrently with the bulk fluid evolution. For example, in our disk, choosing $\eta \sim 10^{-2}$ extends the characteristic fast-flavor-conversion path length from $10\,\mathrm{m}$ to $1\,\mathrm{km}$, but a consistent treatment of flavor conversion at this value of $\eta$ still requires resolving spatial flavor waves on subgrid scales in order to reach consistent asymptotic states. Three-dimensional neutrino quantum-kinetic simulations of post-merger accretion disks at this resolution, or at higher resolution corresponding to $\eta$ closer to unity, that resolve and saturate FFI and CFI, have not yet been performed, and we do not attempt them in this work.

    Instead, we run two simulations with an attenuation factor of $\eta = 10^{-5}$, one with higher spatial resolution (\texttt{HSR-QKE-reducedmatter}) and one with higher angular resolution (\texttt{HAR-QKE}) together with six low-angular-resolution-low-spatial-resolution-QKE simulations (\texttt{LAR-LSR-QKE-*}) with attenuation factors spanning from $10^{-12}$ to $10^{-6}$. Table~\ref{tab:resolution} summarizes the simulation setups. Unlike the local QKE simulations, in which we were able to observe the growth and saturation of FFI and CFI, in our attenuated global QKE simulations no substantial flavor conversion occurs because neutrinos advect out of the disk before any instability saturates. Nevertheless, these simulations serve to visualize global features of the in-medium neutrino quantum coherence produced by vacuum, matter, and neutrino-neutrino potentials in a collisional, energy-dependent, inhomogeneous, and anisotropic matter snapshot, even if qualitatively different from those that would arise from $\eta=1$.
    
    The \texttt{HAR-QKE} simulation is equivalent to the spatial resolution of the dotted local FFI simulation in Fig.~\ref{fig:ffi_number_densities} of $1~\mathrm{cm}$, but for $\eta = 10^{-5}$ we enlarge the domain to effective resolution of $1~\mathrm{km}$. The \texttt{HSR-QKE-reducedmatter} simulation is equivalent to the dot-dashed resolution case of $0.5~\mathrm{cm}$, but with an attenuation factor of $10^{-5}$ the effective simulation resolution increases to $0.5~\mathrm{km}$. This approximately ensures that fast flavor waves are resolved, since the asymptotic states of the dotted and dot-dashed simulations in Fig.~\ref{fig:ffi_number_densities} converge approximately toward the higher-resolution solid simulation. Even though no substantial flavor conversion occurs because neutrinos advect out of the disk before any instability saturates, these simulations serve to visualize the global features of neutrino quantum coherence in a collisional, energy-dependent, inhomogeneous, and anisotropic matter snapshot.
   
    In the {\tt HSR-QKE-reducedmatter} simulation the matter Hamiltonian is attenuated by a factor of $\eta_{\mathrm{matter}} = 10^{-2}$ in addition to the global attenuation factor $\eta = 10^{-5}$. We attenuate the matter Hamiltonian to ensure that FFI-unstable modes are resolved in the global calculation. After finding little FFI development in the \texttt{HAR-QKE} simulation, we attenuate the matter Hamiltonian to make it comparable to the self-interaction Hamiltonian and ensure that we resolve any fast mode in the \texttt{HSR-QKE-reducedmatter} simulation.
    
    Figure~\ref{fig:global_qke_rho_eu} shows the electron-muon component of the density matrix for the \texttt{HAR-QKE} and \texttt{HSR-QKE-reducedmatter} simulations. Within the disk, the in-medium effective mixing angle, set by the combination of the vacuum, matter and neutrino-neutrino potentials, varies with energy. In the \texttt{HAR-QKE} simulation, $\sin 2\theta_{\mathrm{eff}}\sim 10^{-8}$ for the lowest energy bin at $1$ MeV and decreases with increasing energy down to $10^{-10}$ for the highest energy bin at $92$ MeV. In the \texttt{HSR-QKE-reducedmatter} simulation, $\sin 2\theta_{\mathrm{eff} }$ varies approximately from $10^{-6}$ to $10^{-8}$. This agrees with the approximate amplitude of the electron-muon flavor coherence within the disk in the respective global QKE simulations shown in Fig.~\ref{fig:global_qke_rho_eu} (see the $10^{-8}$ and $10^{-6}$ contours). In the polar regions, both simulations have an effective mixing angle of $\sin 2\theta_{\mathrm{eff} } \sim 10^{-6}$ for the lowest energy bin, which decreases with increasing energy down to $\sim 10^{-8}$ due to the self-interaction and vacuum potentials, while the matter potential remains subdominant. Thus, the effective mixing angle is suppressed in the disk, whereas in the polar regions, where the matter potential is subdominant, it is larger and the coherence is correspondingly stronger. This approximately matches the order of magnitude of the electron-muon coherence in the polar region in Fig.~\ref{fig:global_qke_rho_eu}, although it falls somewhat short. The in-medium effective mixing angles and mass splittings, which combine the effects of the vacuum, matter, and neutrino--neutrino potentials, govern the flavor evolution in our attenuated QKE simulations.

    Figure~\ref{fig:eta_vs_rho_eu_pole_and_disk} compares the energy-integrated coherence $\rho_{e\mu}$  in the disk and at the poles (corresponding to the orange and blue stars locations in Fig.~\ref{fig:global_qke_rho_eu}, respectively) for attenuation factors ranging from $10^{-5}$ to $10^{-12}$ (labeled as \texttt{LAR-LSR-QKE-*} in Table~\ref{tab:resolution}) with the theoretical evolution of a neutrino initially produced in the electron-flavor state and propagating under constant effective mixing angles and mass splittings in a two-flavor system,
    \begin{equation}
    \begin{aligned}
    \rho_{e\mu}(t)
    =
    \frac{1}{2}
    \Biggl(
    &-\sin 4\theta_{\mathrm{eff}}
    \sin^2\frac{\omega_{\mathrm{eff}}t}{2}
    \\
    \qquad
    &+i\,\sin 2\theta_{\mathrm{eff}}
    \sin\omega_{\mathrm{eff}}t
    \Biggr),
    \end{aligned}
    \label{eq:eff_rho_eu_t}
    \end{equation}
    where $\theta_{\mathrm{eff}}$ and $\omega_{\mathrm{eff}}$ satisfy
    %\begin{eqnarray}
    %    H^{\mathrm{eff}} = \frac{\omega_{\mathrm{eff}}}{2}
    %    \begin{pmatrix}
    %    -\cos\,2\theta_{\mathrm{eff}} &
    %    \sin\,2\theta_{\mathrm{eff}} \\
    %    \sin\,2\theta_{\mathrm{eff}} &
    %    \cos\,2\theta_{\mathrm{eff}}
    %    \end{pmatrix}
    %\end{eqnarray}
    %\begin{equation}
    %\begin{aligned}
    %H^{\mathrm{eff}}(\mathbf{x},\mathbf{p})
    %= \eta \Big[
    %    &H^{\mathrm{vacuum}}(\mathbf{p})
    %    + H^{\mathrm{matter}}(\mathbf{x},\mathbf{p})
    %    \\
    %    &+ H^{\nu-\nu}(\mathbf{x},\mathbf{p})
    %\Big].
    %\end{aligned}
    %\end{equation}
    %
    \begin{align}
        H^{\mathrm{eff}}(\mathbf{x},\mathbf{p}) &= \eta \Big[
        H^{\mathrm{vacuum}}(\mathbf{p})
        + H^{\mathrm{matter}}(\mathbf{x},\mathbf{p})
        \\
        &\qquad \quad + \, H^{\nu-\nu}(\mathbf{x},\mathbf{p})
    \Big] \nonumber \\
        &\equiv \frac{\omega_{\mathrm{eff}}}{2}
        \begin{pmatrix}
        -\cos\,2\theta_{\mathrm{eff}} &
        \sin\,2\theta_{\mathrm{eff}} \\
        \sin\,2\theta_{\mathrm{eff}} &
        \cos\,2\theta_{\mathrm{eff}}
        \end{pmatrix} \, . 
    \end{align}
    The analytical solution for $\theta_{\mathrm{eff}}$ (effective in-medium mixing angle) and $\omega_{\mathrm{eff}}$ (difference of Hamiltonian eigenvalues) is
    \begin{align}
    \tan\!\left(2\theta_{\mathrm{eff}}\right)
    &=
    \frac{2\left|H^{\mathrm{eff}}_{e\mu}\right|}
    {H^{\mathrm{eff}}_{\mu\mu}-H^{\mathrm{eff}}_{ee}},
    \\[4pt]
    \omega_{\mathrm{eff}}
    %&=
    %\lambda_{+}-\lambda_{-}
    %\nonumber\\
    &=
    \sqrt{
    \left(H^{\mathrm{eff}}_{\mu\mu}
          -H^{\mathrm{eff}}_{ee}\right)^2
    +4\left|H^{\mathrm{eff}}_{e\mu}\right|^2
    }.
    \end{align}
    %where $\lambda_+$ and $\lambda_-$ are the eigenvalues of $H^{\mathrm{eff}}$.
    
    For the vacuum Hamiltonian, we consider the two-flavor case with mixing angle $\theta_{12}=33.82^\circ$ and the normal-hierarchy mass-squared difference $\Delta m^2_{21}=7.39\times10^{-5}\,\mathrm{eV}^2$. Only the electron and muon components are retained in the matter and neutrino self-interaction Hamiltonians, resulting in two-by-two matrices. For the points in the disk (lower panel) and at the pole (upper panel), $\mathbf{p}$ is chosen along $-\hat{x}$ and $\hat{z}$, respectively. These directions align approximately with the local total neutrino flux (see Fig.~\ref{fig:fhat_diagonal_densities}). At each selected spatial location $\mathbf{x}$, propagation direction $\hat{\mathbf{p}}$, and energy-bin center $E_i$, we construct and diagonalize the local two-flavor effective Hamiltonian. Both $\omega_{\mathrm{eff}}$ and $\theta_{\mathrm{eff}}$ therefore include the vacuum, matter, and neutrino self-interaction contributions and are evaluated separately for each energy bin between $1$ and $92~\mathrm{MeV}$. The resulting local values of $\omega_{\mathrm{eff}}$ and $\theta_{\mathrm{eff}}$ are held constant when evaluating the analytical solution in Eq.~(22). The theoretical prediction for the time evolution of the off-diagonal density-matrix component $\rho_{e\mu}$ is shown as a beige band because it spans the predictions obtained across the full neutrino energy range.

    The effective mixing angle is independent of the attenuation factor, whereas
    the effective oscillation frequency scales linearly with it. Consequently,
    for sufficiently small attenuation factors,
    $\sin(\omega_{\mathrm{eff}}t/2)\approx\omega_{\mathrm{eff}}t/2$,
    producing the linear trend observed in
    Fig.~\ref{fig:eta_vs_rho_eu_pole_and_disk}
    for $\eta \in [10^{-12},\,10^{-8}]$ in the lower panel and
    $\eta \in [10^{-12},\,10^{-5}]$ in the upper panel.
    For the theoretical curves, we use propagation distances of
    $5~\mathrm{km}$ at the poles and $1~\mathrm{km}$ in the disk, representative
    of regions over which the fluid profile is approximately homogeneous.
    These distances set the oscillation phase and therefore affect the value
    of the predicted coherence at a given $\eta$, as well as the value of
    $\eta$ at which the linear approximation ceases to apply.
    We also evaluated Eq.~\eqref{eq:eff_rho_eu_t} along paths ranging
    from $1$ to $25~\mathrm{km}$ and checked its agreement with the
    corresponding global QKE results. The qualitative behavior described
    below persists over these path lengths, although the location of the
    turnover and the coherence at individual values of $\eta$ depend on
    the chosen path. Nevertheless, the theoretical band of $ \rho_ {e\mu} $ across energy robustly remains between the QKE global simulation data.
    As the attenuation factor increases,
    the linear approximation breaks down. The sine factor can then
    reach values of order unity, so the upper limit of the coherence is
    of order $\sin2\theta_{\mathrm{eff}}$; its value at a fixed propagation
    length can still vary with the oscillation phase.
    In the disk, the plotted coherence reaches values of this order
    at $\eta=10^{-7}$, $10^{-6}$, and $10^{-5}$; at the poles, it begins
    to do so at $\eta=10^{-5}$. Extrapolating the beige theoretical band to
    $\eta>10^{-5}$ illustrates the approximately $\eta$-independent limit. The flattening in Fig.~\ref{fig:eta_vs_rho_eu_pole_and_disk} thus reflects the bounded,
    small-amplitude in-medium coherence described by
    Eq.~\eqref{eq:eff_rho_eu_t}. It does not establish convergence of the
    global QKE results with respect to $\eta$ or indicate nonlinear
    saturation of a FFI or CFI.
    
    This behavior explains the results of the simulations shown in Fig.~\ref{fig:global_qke_rho_eu} for $\eta=10^{-5}$. Most neutrinos and antineutrinos are produced in the dense, hot regions of the accretion disk, where they rapidly reach the regime in which $\sin(\omega_{\mathrm{eff}}t/2)$ is close to unity. As a result, the coherence reaches values of order $\sin 2\theta_{\mathrm{eff}}$. Within the disk, the high electron-neutrino density suppresses the effective mixing angle. However, as neutrinos advect out of the disk and into the polar region, they enter an environment where $\sin 2\theta_{\mathrm{eff}}$ is approximately two orders of magnitude larger than in the disk. Consequently, the coherence is further enhanced in the polar region, producing a larger quantum coherence within the polar cones, as shown in Fig.~\ref{fig:global_qke_rho_eu}. Correcting the interchanged assignments of $\theta_{13}$ and $\theta_{23}$ (see Sec.~\ref{sec:quantum_kinetics}) would, in principle, change the effective mixing angles and, consequently, the coherence amplitudes $\sin(\omega_{\mathrm{eff}}t/2)$ and the effective frequencies $\omega_{\mathrm{eff}}$. 
    Nevertheless, in our accretion disk, we are in the limit of high matter and neutrino fields, and the effective mixing angles and frequencies will be driven to limiting values, making the original decision about the vacuum angles not really matter \cite{BLENNOW2005330, PhysRevD.80.053002, XING2019114803}. The system then behaves approximately as a two-flavor system \cite{BLENNOW2005330}, in agreement with Fig.~\ref{fig:eta_vs_rho_eu_pole_and_disk}. The total quantum coherence in Figs.~\ref{fig:global_qke_rho_eu} and \ref{fig:eta_vs_rho_eu_pole_and_disk} would therefore remain unchanged. The trend of enhanced coherence in the polar region due to strong electron densities, along with the theoretical understanding of this coherence as the in-medium effective mixing upper-bound coherent amplitude, would remain.

    Even though our global simulations capture the linear in-medium quantum coherence, they fail to grow and saturate flavor instabilities. Quantum correlations in the \texttt{HAR-QKE} and \texttt{HSR-QKE-reducedmatter} the QKE simulations are smaller than the diagonal number densities, so no significant flavor conversion occurs. The local FFI simulation in Fig.~\ref{fig:ffi_number_densities} shows that a representative point inside our accretion disk would saturate the FFI on timescales of $\sim 10^{-7}\,\mathrm{s}$, corresponding to path lengths of $\sim 10\,\mathrm{m}$. In contrast, applying the attenuation factor $\eta = 10^{-5}$ increases the flavor-conversion path to $\sim 10^{6}\,\mathrm{m}$, while our accretion disk is only $\sim 10^{5}\,\mathrm{m}$ thick. Neutrinos are therefore advected out of the disk before substantial quantum coherence due to FFI can develop. As presented in Fig.~\ref{fig:n_local_cfi}, the flavor-conversion paths associated with CFI are even longer than for FFI, suggesting that their impact on our disk is also negligible.    
    
\SwitchableClearpage
\section{Summary and conclusions\label{sec:sumcon}}

    We upgrade incoherent neutrino collisions (emission, absorption, and pair annihilation) in \texttt{Emu} and perform non-general relativistic $6{+}1$-dimensional classical neutrino simulations on a three-dimensional snapshot of a post-merger accretion disk designed to reproduce observables from GW170817. Using the steady state of the neutrino radiation field from the classical global simulation, we address the emergence of ELN angular crossings and show how they arises in this environment. We carry out a linear stability analysis to identify regions where CFIs can occur and to estimate their growth rates. To shed light on the post-saturation asymptotic state of FFIs and CFIs and their possible implications in the post-merger accretion disk, we perform local simulations at selected points within the accretion disk. Finally, we perform quantum kinetic neutrino simulations on the same post-merger accretion-disk snapshot with attenuated Hamiltonians.

    Our calculations show that, at this cooling stage, the disk contains mildly degenerate electrons and is deleptonizing. Its emission is dominated by electron neutrinos, followed by electron antineutrinos and heavy-lepton neutrinos (see Fig.~\ref{fig:n_diagonal_densities}). A large fraction of electron neutrinos couple to the disk fluid and fill the disk, leading to low flux factors (i.e., less forward-peaked angular distributions) for this neutrino flavor. In contrast, electron antineutrinos are more weakly coupled to the disk fluid and propagate through it more easily, yielding higher flux factors (i.e., more forward-peaked angular distributions), see Fig.~\ref{fig:fhat_diagonal_densities}. Electron neutrinos decouple from the disk only at low energies, while higher-energy $\nu_e$ remain efficiently coupled to the fluid. Electron antineutrinos are free streaming to substantially higher energies (see Fig.~\ref{fig:ffi_point_energies}). This hierarchy in decoupling is conducive to ELN angular crossings and favors the development of FFI.
   
    We map the regions that exhibit a FFI and estimate growth rates from the steady-state angular distributions of each neutrino flavor (see left panels of Fig.~\ref{fig:instabilities}). FFIs are strongest inside the disk and weaken toward the poles, consistent with previous relativistic transport studies~\cite{mukhopadhyay2024time}. In the disk, the $\bar{\nu}_e$ distribution from localized emission hotspots remains strongly forward-peaked because it undergoes relatively few interactions, while the $\nu_e$ distribution is more isotropic due to stronger coupling to the bulk fluid. As a result, sight lines toward $\bar{\nu}_e$ hotspots can become antineutrino dominated, producing ELN crossings and triggering FFIs (see top panels of Fig.~\ref{fig:ffi_angular_distributions}). In the polar funnel, both $\nu_e$ and $\bar{\nu}_e$ become increasingly forward-peaked, and there is an intermediate region where their angular peaks nearly coincide, so the ELN nearly cancels and FFI is suppressed. Farther above the disk plane, directional competition between polar $\bar{\nu}_e$ emission and the broader disk-emitted $\nu_e$ field yields a more complex, multi-crossing ELN structure. Overall, spatially inhomogeneous emission and energy-dependent coupling to the fluid naturally generate ELN crossings and favorable conditions for FFI.

    To deepen our understanding of fast flavor conversion in the interior of the accretion disk, we evolve the neutrino flavor field at a representative location in a periodic box until the instability saturates and the number densities of the flavors approach their asymptotic values. The conversion follows the expected asymptotic behavior: angular regions on the shallow side of the ELN--XLN crossing undergo strong conversion toward flavor equipartition, while the remaining directions adjust to satisfy conservation laws (see the lower panels of Fig.~\ref{fig:ffi_angular_distributions}). This conversion is strongest in outward-going beams with negative ELN--XLN, where it transfers $\nu_e$ and $\bar{\nu}_e$ into heavy flavors and increases the heavy-flavor flux emerging from the disk, thereby accelerating disk cooling \cite{qiu2025impact, qiu2025neutrino, lund2025angle}. Moreover, fast flavor conversion occurs over propagation distances that are far smaller than the hydrodynamic scales resolved in current merger simulations (see Fig.~\ref{fig:ffi_number_densities}).
    
    We use the steady-state neutrino radiation field from our global transport simulation to predict where the disk is unstable to CFI using both monochromatic and multi-energy LSA (see the middle columns of Fig.~\ref{fig:instabilities}). The CFI-unstable region spans most of the accretion disk and weakens toward the polar funnel. We confirm that monochromatic estimates reproduce the overall morphology of the unstable region but systematically overpredict the growth rates relative to the multi-energy calculation (see Fig.~\ref{fig:instabilities}). While CFI is generally subdominant to FFI, we identify localized regions in the innermost disk where CFI occurs without FFI, suggesting that collisions can set the initial inter-flavor coherence in specific parts of the flow (see right panels of Fig.~\ref{fig:instabilities}).

    Similarly to the FFI case, we perform a local multi-energy QKE evolution initialized from a collisional flavor-unstable point in the disk. The CFI can substantially reshape the asymptotic flavor content, transferring $\nu_e$ and $\bar{\nu}_e$ into heavy flavors (see Fig.~\ref{fig:n_local_cfi}), while leaving the effectively trapped parts of the spectrum largely unchanged (see Fig.~\ref{fig:f_spectrum} and Fig.~\ref{fig:f_diff_spectrum}). The CFI also restructures the heavy-flavor spectra such that heavy-lepton antineutrinos carry more energy than heavy-lepton neutrinos despite nearly equal number densities, breaking the heavy-flavor neutrino--antineutrino symmetry that is often imposed in global radiation-hydrodynamic simulations. We highlight that the CFI is most relevant for neutrinos that traverse dense regions of the disk, which are also susceptible to the FFI, but the combined nonlinear asymptotic state remains an open question.
            
    Finally, we use the strategy of adding an attenuation factor to the QKE Hamiltonian to be able to resolve flavor-coherent waves in the disk. We perform two global QKE simulations with different resolutions to probe how and where quantum coherence develops in a three-dimensional inhomogeneous, anisotropic, collisional disk snapshot. We find that, in both the accretion disk and the polar regions, the electron-muon coherence corresponds to the in-medium effective mixing angle. The local matter and neutrino densities set this angle through the combined vacuum, matter, and self-interaction potentials. In the disk, the effective mixing angle is strongly suppressed, so the coherence remains weak. In the polar regions, where the matter potential is subdominant, the effective mixing angle is larger and the coherence is correspondingly stronger. This suggests that the coherence observed in our attenuated simulations is primarily controlled by the local effective mixing angle. Because of the attenuation factor $\eta$, neutrinos are advected out before instabilities can build appreciable coherence and saturate to produce significant flavor conversion. Future work should adopt an attenuation factor large enough to shift FFI and CFI to macroscopic scales, while still resolving the resulting spatial flavor waves at subgrid hydrodynamic resolution. This is necessary to allow the neutrino field to saturate before advection becomes dominant. Such simulations would enable a more direct study of FFI and CFI in the disk and clarify how their nonlinear interplay affects the neutrino field.

    As mentioned previously, our analysis focuses on a single snapshot of a post-neutron-star-merger accretion disk following the rapid collapse of the compact-object remnant into a black hole. Rather than studying the time evolution of the instabilities, we focus on a snapshot in which both FFI and CFI are present and aim to characterize the angular distributions robustly to assess the presence or absence of instabilities. Our system contains a massive accretion disk with high neutrino opacities and mildly degenerate electrons, in which differences between neutrino and antineutrino transport generate angular structure in the ELN distribution and promote the formation of ELN crossings, FFI, and CFI. This stage of a neutron-star merger is not generic, and our conclusions must therefore be extrapolated to other systems with caution. The qualitative mechanism by which neutrino opacities and differences between electron-neutrino and electron-antineutrino transport generate angular structure and ELN crossings may be broadly applicable. However, the presence, location, and strength of the crossings, as well as the associated instability conditions, depend on the disk mass, evolutionary time, remnant type, and transport treatment. In the system considered here, the high disk mass enhances neutrino opacities and transport asymmetries, favoring both FFI and CFI. Less massive disks may not have sufficient densities and neutrino emissivities to generate comparably strong asymmetries between electron neutrinos and electron antineutrinos. Long-lived hypermassive-neutron-star–accretion-disk systems also possess characteristics that differ substantially from those studied in our simulations. Our results therefore cannot be directly extrapolated to such systems, because electron antineutrinos emitted during the cooling of the hypermassive neutron star can play a major role in generating flavor instabilities. We refer the reader to Ref.~\cite{nagakura2025neutrino} for an extended analysis of FFI and CFI in these systems.

\SwitchableClearpage
\begin{acknowledgments}

    E.U. is supported by the Dr. Elizabeth M. Bains and Dr. James A. Bains Graduate Research Fellowship. J.F. acknowledges support from the Severo Ochoa Excellence Grant CEX2023-001292-S funded by MICIU/AEI/10.13039/501100011033. This research used resources of the National Energy Research Scientific Computing Center (NERSC), a Department of Energy User Facility using NERSC award NP-ERCAP0033473. The U.S.\ DOE partially supported this work through contract numbers DE-FG0202ER41216 (G.C.M., J.P. K.), DE-SC0024388 (ENAF - G.C.M., J.P.K.), and DE-SC0020435 (F.F). F.F. and G.C.M. acknowledge support from the Network for Neutrinos, Nuclear Astrophysics and Symmetries (N3AS), through the National Science Foundation Physics Frontier Center award No.\ PHY-2020275.  In addition, this work was supported by the Office of Defense Nuclear Nonproliferation Research and Development (DNN R\&D), National Nuclear Security Administration, U.S.\ DOE (G.C.M) under contract number LA22-ML-DE-FOA-2440 and by Lawrence Livermore National Laboratory under Contract DE-AC52-107NA27344, with support from LDRD project 24-ERD-023 (G. C. M.). S.S. acknowledges the support by the European Research Council (ERC) Advanced Grant INSPIRATION under the European Union’s Horizon 2020 research and innovation programme (Grant agreement No. 101053985) and by Deutsche Forschungsgemeinschaft (DFG, German Research Foundation) under Germany's Excellence Strategy – EXC~2121 ``Quantum Universe'' – 390833306. We thank the anonymous referee for their valuable comments and suggestions, which have significantly improved the quality and clarity of this manuscript.

\end{acknowledgments}

\appendix
\SwitchableClearpage

\begin{figure*}[!htbp]
        \centering
        \includegraphics[width=0.7\linewidth]{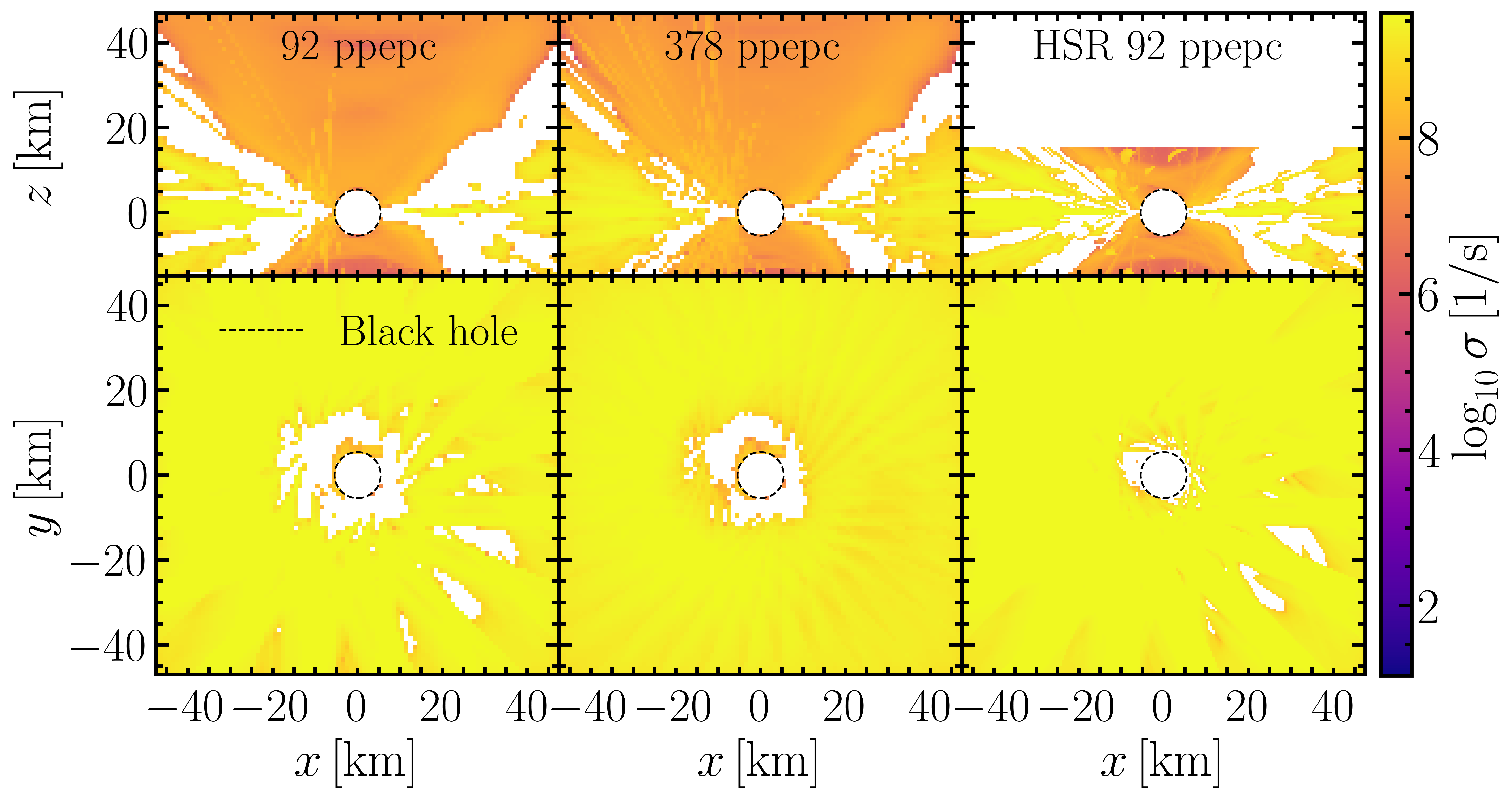}
        \caption{
            \label{fig:instabilities_growth_rate_apendix} 
            Fast flavor instability growth rates in the GW170817 post-merger disk simulations with angular resolutions of $92$ (left), $378$ (center). All other parameters match those of the {\tt HAR-class} simulation (Table~\ref{tab:resolution}). The right panels show the FFI growth rate for the high spatial resolution {\tt HSR-class} simulation (Table~\ref{tab:resolution}). The upper (lower) panels show polar (equatorial) slices.
        }
\end{figure*}

\section{Impact of Angular and Spatial Resolution on ELN Distributions and FFI Growth Rates\label{con_num_par_ffi}}

    Since an accurate description of each neutrino flavor angular distribution is crucial for assessing the emergence of ELN angular crossings, we ran simulations with angular resolutions of 92, 378, and 1506 particles per energy bin per cell (ppepc), with isotropically distributed momentum directions. Other parameters are set as in the {\tt HAR-class} simulation (see Table~\ref{tab:resolution}). These runs allow us to test the numerical robustness of the angular distributions that determine the emergence of FFI against the angular resolution used. Fig.~\ref{fig:instabilities_growth_rate_apendix} shows the FFI growth rates for angular resolutions of 92 (left) and 378 (center) ppepc. The 1506 ppepc case is shown in the right panels of Fig.~\ref{fig:instabilities} as the {\tt HAR-class}. Overall, the order of magnitude of the FFI growth rate is robust across all simulations. However, the lower angular resolution runs exhibit ray-like numerical artifacts, most prominently in the accretion-disk regions. These patterns likely arise from the limited angular resolution, combined with the non-fixed domain subdivision used in the angular integrals of the ELN distributions when computing the FFI growth rate [see Eq.~\eqref{eq:sigma}], because the positive and negative ELN--XLN angular domains differ from one location to another. We found that angular-integrated quantities such as neutrino number densities and fluxes converge reliably across resolutions. 
    
\begin{figure*}[!htbp]
        \centering
        \includegraphics[width=0.8\textwidth]{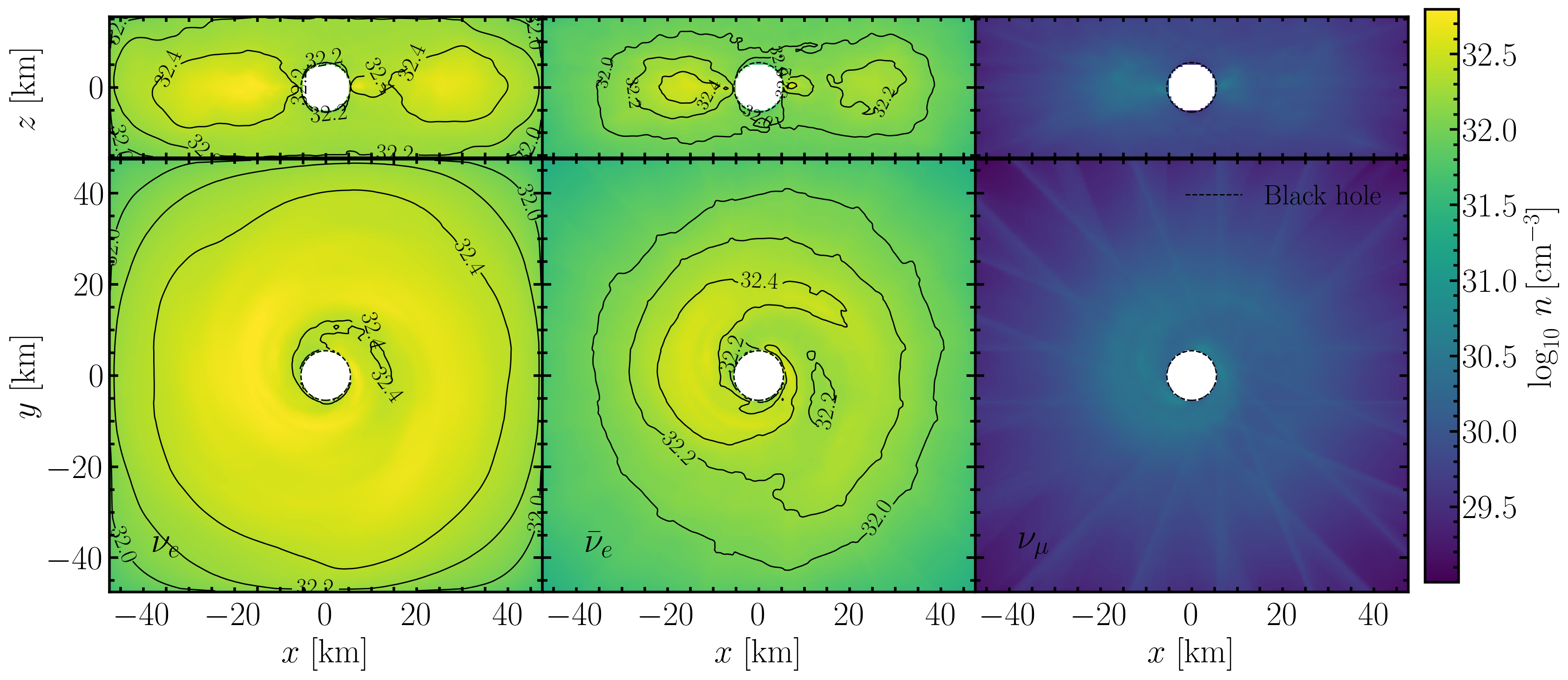}
        \caption{\label{fig:hsr_n_diagonal_densities} 
            Number densities of electron neutrinos (left), electron antineutrinos (center), and muon neutrinos (right) for the {\tt HSR-class} simulation (see parameters in Table~\ref{tab:resolution}). Other heavy-flavor neutrinos and antineutrinos follow the same trend as the muon-neutrino panel.
            The upper (lower) panels show polar (equatorial) slices.
        }
    \end{figure*}

\begin{figure*}[!htbp]
        \centering
        \includegraphics[width=0.8\textwidth]{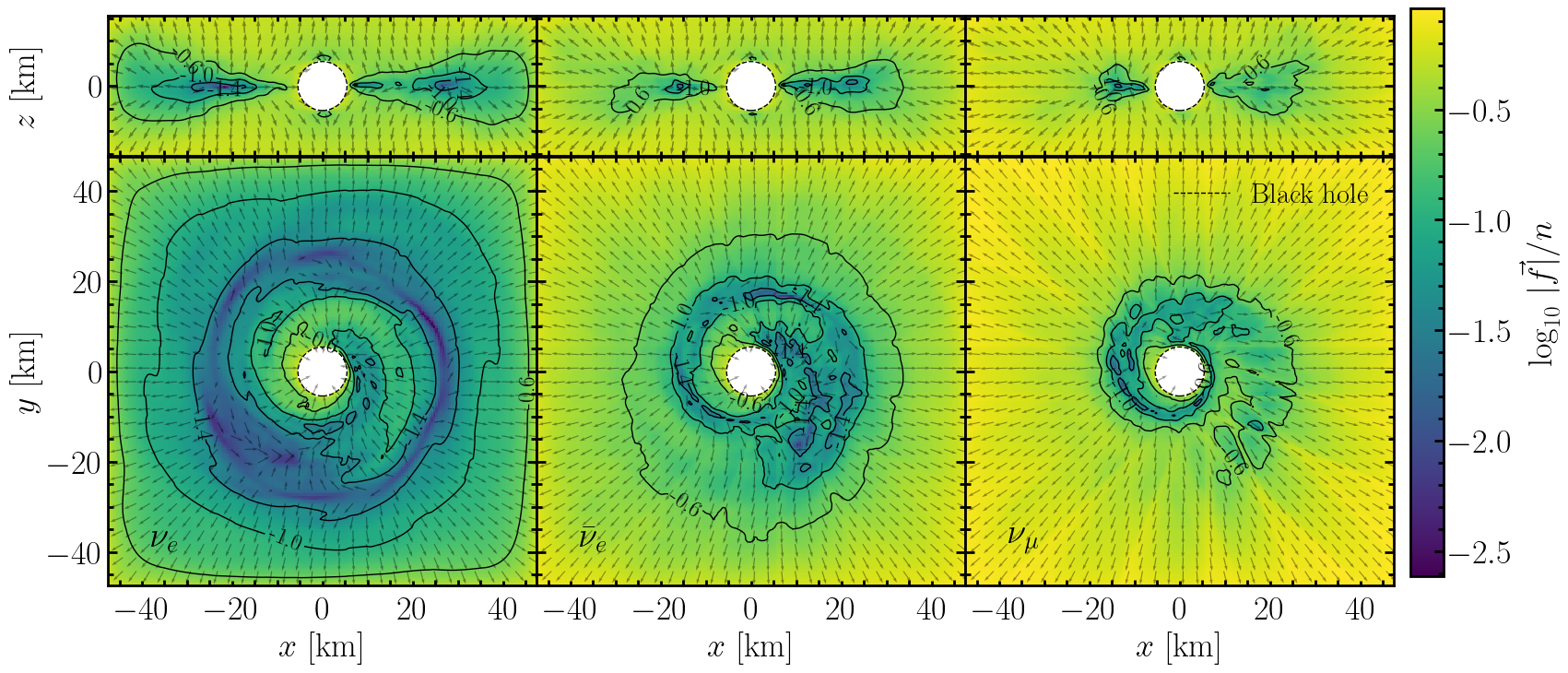}
        \caption{\label{fig:hsr_fhat_diagonal_densities} 
            Flux factors of electron (left), electron antineutrino (center), and muon neutrinos (right) for the {\tt HSR-class} simulation (see simulation parameters in Table~\ref{tab:resolution}). Other heavy-flavor neutrinos and antineutrinos follow the same trend as the muon-neutrino panel.
            The upper (lower) panels show polar (equatorial) slices.
        }
    \end{figure*}

    To test spatial-resolution convergence, we also ran a simulation with twice the spatial resolution of {\tt HAR-class}. This simulation, denoted {\tt HSR-class}, uses 92~ppepc. The order of magnitude of the FFI growth rate (see right panels of Fig.~\ref{fig:instabilities_growth_rate_apendix}) remains robust across both spatial resolutions. Fig.~\ref{fig:hsr_n_diagonal_densities} and \ref{fig:hsr_fhat_diagonal_densities} show the number densities and flux factors for the {\tt HSR-class} simulation. The description given in Sec.~\ref{sec:global_description_class} for {\tt HAR-class} still applies to {\tt HSR-class}; the main difference is that the {\tt HSR-class} simulation reveals sharper features in the contour lines as a result of the lower angular resolution. In addition, the heavy-lepton neutrino number densities exhibit more pronounced ray-like beam artifacts originating from the emission hot spots, which are not visible in {\tt HAR-class} (see Fig.~\ref{fig:n_diagonal_densities}).
    
\SwitchableClearpage
\bibliographystyle{apsrev4-2}
\bibliography{refs}

@article{fiorillo2024fast,
    author = "Fiorillo, Damiano F. G. and Raffelt, Georg G.",
    title = "{Fast Flavor Conversions at the Edge of Instability in a Two-Beam Model}",
    eprint = "2403.12189",
    archivePrefix = "arXiv",
    primaryClass = "hep-ph",
    doi = "10.1103/PhysRevLett.133.221004",
    journal = "Phys. Rev. Lett.",
    volume = "133",
    number = "22",
    pages = "221004",
    year = "2024"
}

@article{Fiorillo:2024uki,
    author = "Fiorillo, Damiano F. G. and Raffelt, Georg G.",
    title = "{Theory of neutrino fast flavor evolution. Part II. Solutions at the edge of instability}",
    eprint = "2409.17232",
    archivePrefix = "arXiv",
    primaryClass = "hep-ph",
    doi = "10.1007/JHEP12(2024)205",
    journal = "JHEP",
    volume = "12",
    pages = "205",
    year = "2024"
}

@article{Fiorillo:2025zio,
    author = "Fiorillo, Damiano F. G. and Raffelt, Georg G.",
    title = "{Dispersion relation of the neutrino plasma: unifying fast, slow, and collisional instabilities}",
    eprint = "2505.20389",
    archivePrefix = "arXiv",
    primaryClass = "hep-ph",
    doi = "10.1007/JHEP01(2026)147",
    journal = "JHEP",
    volume = "01",
    pages = "147",
    year = "2026"
}

@article{nagakura2025neutrino,
    author = "Nagakura, Hiroki and Sumiyoshi, Kohsuke and Fujibayashi, Sho and Sekiguchi, Yuichiro and Shibata, Masaru",
    title = "{Neutrino flavor instabilities in a binary neutron star merger remnant: Roles of a long-lived hypermassive neutron star}",
    eprint = "2504.20143",
    archivePrefix = "arXiv",
    primaryClass = "astro-ph.HE",
    doi = "10.1103/kf8q-d67t",
    journal = "Phys. Rev. D",
    volume = "112",
    number = "4",
    pages = "043029",
    year = "2025"
}

@article{mukhopadhyay2024time,
doi = {10.3847/1538-4357/ad6c42},
url = {https://doi.org/10.3847/1538-4357/ad6c42},
year = {2024},
month = {oct},
publisher = {The American Astronomical Society},
volume = {974},
number = {1},
pages = {110},
author = {Mukhopadhyay, Payel and Miller, Jonah and McLaughlin, Gail C.},
title = {The Time Evolution of Fast Flavor Crossings in Postmerger Disks around a Black Hole Remnant},
journal = {The Astrophysical Journal}
}

@article{Miller:2019gig,
    author = "Miller, Jonah M. and Ryan, Ben R. and Dolence, Joshua C.",
    title = "{$\nu\texttt{bhlight}$: Radiation GRMHD for Neutrino-Driven Accretion Flows}",
    eprint = "1903.09273",
    archivePrefix = "arXiv",
    primaryClass = "astro-ph.IM",
    reportNumber = "LA-UR-19-20336",
    doi = "10.3847/1538-4365/ab09fc",
    journal = "Astrophys. J. Suppl.",
    volume = "241",
    number = "2",
    pages = "30",
    year = "2019"
}

@article{Particle-in-cell,
    author = "Richers, Sherwood and Willcox, Don E. and Ford, Nicole M. and Myers, Andrew",
    title = "{Particle-in-cell Simulation of the Neutrino Fast Flavor Instability}",
    eprint = "2101.02745",
    archivePrefix = "arXiv",
    primaryClass = "astro-ph.HE",
    reportNumber = "N3AS-21-001",
    doi = "10.1103/PhysRevD.103.083013",
    journal = "Phys. Rev. D",
    volume = "103",
    number = "8",
    pages = "083013",
    year = "2021"
}

@article{wang2025effectcfi,
    author = "Wang, Tianshu and Nagakura, Hiroki and Johns, Lucas and Burrows, Adam",
    title = "{Effect of the collisional flavor instability on core-collapse supernova models}",
    eprint = "2507.01100",
    archivePrefix = "arXiv",
    primaryClass = "astro-ph.HE",
    doi = "10.1103/z3qh-nj18",
    journal = "Phys. Rev. D",
    volume = "112",
    number = "6",
    pages = "063039",
    year = "2025"
}

@article{PhysRevD.107.103022,
    author = "Zaizen, Masamichi and Nagakura, Hiroki",
    title = "{Simple method for determining asymptotic states of fast neutrino-flavor conversion}",
    eprint = "2211.09343",
    archivePrefix = "arXiv",
    primaryClass = "astro-ph.HE",
    doi = "10.1103/PhysRevD.107.103022",
    journal = "Phys. Rev. D",
    volume = "107",
    number = "10",
    pages = "103022",
    year = "2023"
}

@article{PhysRevD.110.123018,
  title = {Evolution and the quasistationary state of collective fast neutrino flavor conversion in three dimensions without axisymmetry},
  author = {George, Manu and Xiong, Zewei and Wu, Meng-Ru and Lin, Chun-Yu},
  journal = {Phys. Rev. D},
  volume = {110},
  issue = {12},
  pages = {123018},
  numpages = {12},
  year = {2024},
  month = {Dec},
  publisher = {American Physical Society},
  doi = {10.1103/PhysRevD.110.123018},
  url = {https://link.aps.org/doi/10.1103/PhysRevD.110.123018}
}

@article{PhysRevD.108.063003,
  title = {Evaluating approximate asymptotic distributions for fast neutrino flavor conversions in a periodic 1D box},
  author = {Xiong, Zewei and Wu, Meng-Ru and Abbar, Sajad and Bhattacharyya, Soumya and George, Manu and Lin, Chun-Yu},
  journal = {Phys. Rev. D},
  volume = {108},
  issue = {6},
  pages = {063003},
  numpages = {15},
  year = {2023},
  month = {Sep},
  publisher = {American Physical Society},
  doi = {10.1103/PhysRevD.108.063003},
  url = {https://link.aps.org/doi/10.1103/PhysRevD.108.063003}
}

@article{PhysRevD.107.123021,
  title = {Characterizing quasisteady states of fast neutrino-flavor conversion by stability and conservation laws},
  author = {Zaizen, Masamichi and Nagakura, Hiroki},
  journal = {Phys. Rev. D},
  volume = {107},
  issue = {12},
  pages = {123021},
  numpages = {10},
  year = {2023},
  month = {Jun},
  publisher = {American Physical Society},
  doi = {10.1103/PhysRevD.107.123021},
  url = {https://link.aps.org/doi/10.1103/PhysRevD.107.123021}
}

@article{Zaizen:2023ihz,
    author = "Zaizen, Masamichi and Nagakura, Hiroki",
    title = "{Characterizing quasisteady states of fast neutrino-flavor conversion by stability and conservation laws}",
    eprint = "2304.05044",
    archivePrefix = "arXiv",
    primaryClass = "astro-ph.HE",
    doi = "10.1103/PhysRevD.107.123021",
    journal = "Phys. Rev. D",
    volume = "107",
    number = "12",
    pages = "123021",
    year = "2023"
}

@article{Richers:2024zit,
    author = "Richers, Sherwood and Froustey, Julien and Ghosh, Somdutta and Foucart, Francois and Gomez, Javier",
    title = "{Asymptotic-state prediction for fast flavor transformation in neutron star mergers}",
    eprint = "2409.04405",
    archivePrefix = "arXiv",
    primaryClass = "astro-ph.HE",
    reportNumber = "N3AS-24-030",
    doi = "10.1103/PhysRevD.110.103019",
    journal = "Phys. Rev. D",
    volume = "110",
    number = "10",
    pages = "103019",
    year = "2024"
}

@article{PhysRevResearch.2.012046,
  title = {Fast neutrino-flavor conversion in the preshock region of core-collapse supernovae},
  author = {Morinaga, Taiki and Nagakura, Hiroki and Kato, Chinami and Yamada, Shoichi},
  journal = {Phys. Rev. Res.},
  volume = {2},
  issue = {1},
  pages = {012046},
  numpages = {6},
  year = {2020},
  month = {Feb},
  publisher = {American Physical Society},
  doi = {10.1103/PhysRevResearch.2.012046},
  url = {https://link.aps.org/doi/10.1103/PhysRevResearch.2.012046}
}

@article{urquilla2025testingcommonapproximationsneutrino,
    author = "Urquilla, Erick and Johns, Lucas",
    title = "{Testing common approximations of neutrino fast flavor conversion}",
    eprint = "2510.23917",
    archivePrefix = "arXiv",
    primaryClass = "astro-ph.HE",
    doi = "10.1103/xmxv-9sb2",
    journal = "Phys. Rev. D",
    volume = "113",
    number = "10",
    pages = "103013",
    year = "2026"
}

@article{Kajita_1999,
	title = {Atmospheric neutrino results from Super-Kamiokande and Kamiokande - Evidence for $\nu_{\mu}$ oscillations},
    doi = {10.1016/s0920-5632(99)00407-7},
	url = {https://doi.org/10.1016/s0920-5632(99)00407-7},
	year = {1999},
	month = {may},
	publisher = {Elsevier {BV}},
	volume = {77},
	number = {1-3},
	pages = {123-132},
	author = {Takaaki Kajita},
	journal = {Nuclear Physics B - Proceedings Supplements}
}

@article{SNO2001,
  title = "{Measurement of the Rate of ${\ensuremath{\nu}}_{e}+\mathit{d}\ensuremath{\rightarrow}\mathit{p}+\mathit{p}+{\mathit{e}}^{\ensuremath{-}}$ Interactions Produced by $^{8}B$ Solar Neutrinos at the Sudbury Neutrino Observatory}",
  author = {Ahmad, Q. R. and Allen, R. C. and Andersen, T. C. and others},
  collaboration = {SNO Collaboration},
  journal = {Phys. Rev. Lett.},
  volume = {87},
  issue = {7},
  pages = {071301},
  numpages = {6},
  year = {2001},
  month = {Jul},
  publisher = {American Physical Society},
  doi = {10.1103/PhysRevLett.87.071301},
  url = {https://link.aps.org/doi/10.1103/PhysRevLett.87.071301}
}

@article{Mikheyev_and_Smirnov,
title = {Resonant amplification of $\nu$ oscillations in matter and solar-neutrino spectroscopy},
author = {Mikheyev, S. P. and Smirnov, A. Yu.},
journal = {Il Nuovo Cimento C},
volume = {9},
issue = {1},
year = {1986},
moth = {Jan},
doi = {10.1007/BF02508049},
url = {https://doi.org/10.1007/BF02508049},
issn = {0390-5551}
}

@article{Wolfenstein1978,
  title = {Neutrino oscillations in matter},
  author = {Wolfenstein, L.},
  journal = {Phys. Rev. D},
  volume = {17},
  issue = {9},
  pages = {2369--2374},
  numpages = {0},
  year = {1978},
  month = {May},
  publisher = {American Physical Society},
  doi = {10.1103/PhysRevD.17.2369},
  url = {https://link.aps.org/doi/10.1103/PhysRevD.17.2369}
}

@incollection{MikheevSmirnov_1985,
  title={Resonance enhancement of oscillations in matter and solar neutrino spectroscopy},
  author={Mikheev, SP and Smirnov, A Yu},
  booktitle={Solar Neutrinos},
  pages={305--309},
  year={2018},
  publisher={CRC Press}
}

@article{Pantaleone:1992eq,
    author = "Pantaleone, James T.",
    title = "{Neutrino oscillations at high densities}",
    reportNumber = "DOE-ER-40561-056, INT-92-07-01",
    doi = "10.1016/0370-2693(92)91887-F",
    journal = "Phys. Lett. B",
    volume = "287",
    pages = "128--132",
    year = "1992"
}

@article{sigl1993general,
    author = "Sigl, G. and Raffelt, G.",
    title = "{General kinetic description of relativistic mixed neutrinos}",
    reportNumber = "MPI-PH-92-112",
    doi = "10.1016/0550-3213(93)90175-O",
    journal = "Nucl. Phys. B",
    volume = "406",
    pages = "423--451",
    year = "1993",
    number ="1-2",
    publisher= "Elsevier"
}

@article{Vlasenko:2013fja,
    author = "Vlasenko, Alexey and Fuller, George M. and Cirigliano, Vincenzo",
    title = "{Neutrino Quantum Kinetics}",
    eprint = "1309.2628",
    archivePrefix = "arXiv",
    primaryClass = "hep-ph",
    reportNumber = "LA-UR-13-27035",
    doi = "10.1103/PhysRevD.89.105004",
    journal = "Phys. Rev. D",
    volume = "89",
    number = "10",
    pages = "105004",
    year = "2014"
}

@article{Richers:2019grc,
    author = "Richers, Sherwood A. and McLaughlin, Gail C. and Kneller, James P. and Vlasenko, Alexey",
    title = "{Neutrino Quantum Kinetics in Compact Objects}",
    eprint = "1903.00022",
    archivePrefix = "arXiv",
    primaryClass = "astro-ph.HE",
    doi = "10.1103/PhysRevD.99.123014",
    journal = "Phys. Rev. D",
    volume = "99",
    number = "12",
    pages = "123014",
    year = "2019",
    note = "[Erratum: Phys.Rev.D 109, 129902 (2024)]"
}

@article{xiong2024robust,
    author = "Xiong, Zewei and Wu, Meng-Ru and George, Manu and Lin, Chun-Yu",
    title = "{Robust Integration of Fast Flavor Conversions in Classical Neutrino Transport}",
    eprint = "2403.17269",
    archivePrefix = "arXiv",
    primaryClass = "astro-ph.HE",
    doi = "10.1103/PhysRevLett.134.051003",
    journal = "Phys. Rev. Lett.",
    volume = "134",
    number = "5",
    pages = "051003",
    year = "2025"
}

@article{akaho2026bifurcatedimpactneutrinofast,
    author = "Akaho, Ryuichiro and Nagakura, Hiroki and Iwakami, Wakana and Furusawa, Shun and Harada, Akira and Okawa, Hirotada and Matsufuru, Hideo and Sumiyoshi, Kohsuke and Yamada, Shoichi",
    title = "{Bifurcated Impact of Neutrino Fast Flavor Conversion on Core-Collapse Supernovae Informed by Multiangle Neutrino Radiation Hydrodynamics}",
    eprint = "2601.08269",
    archivePrefix = "arXiv",
    primaryClass = "astro-ph.HE",
    doi = "10.1103/fksy-1jtw",
    journal = "Phys. Rev. Lett.",
    volume = "136",
    number = "19",
    pages = "191002",
    year = "2026"
}

@article{just2022,
    author = "Just, Oliver and Abbar, Sajad and Wu, Meng-Ru and Tamborra, Irene and Janka, Hans-Thomas and Capozzi, Francesco",
    title = "{Fast neutrino conversion in hydrodynamic simulations of neutrino-cooled accretion disks}",
    eprint = "2203.16559",
    archivePrefix = "arXiv",
    primaryClass = "astro-ph.HE",
    doi = "10.1103/PhysRevD.105.083024",
    journal = "Phys. Rev. D",
    volume = "105",
    number = "8",
    pages = "083024",
    year = "2022"
}

@article{lund2025angle,
    author = "Lund, Kelsey A. and Mukhopadhyay, Payel and Miller, Jonah M. and McLaughlin, G. C.",
    title = "{Angle-dependent in Situ Fast Flavor Transformations in Post-neutron-star-merger Disks}",
    eprint = "2503.23727",
    archivePrefix = "arXiv",
    primaryClass = "astro-ph.HE",
    reportNumber = "LA-UR-25-23002, INT-PUB-25-005, N3AS-25-005",
    doi = "10.3847/2041-8213/add0a7",
    journal = "Astrophys. J. Lett.",
    volume = "985",
    number = "1",
    pages = "L9",
    year = "2025"
}

@article{mori2025,
	author = {Mori, Kanji and Takiwaki, Tomoya and Kotake, Kei and Horiuchi, Shunsaku},
	doi = {10.1093/pasj/psaf007},
	issn = {2053-051X},
	journal = {Publ. Astron. Soc. Jpn.},
	month = {03},
	number = {2},
	pages = {L9-L15},
	title = {Three-dimensional core-collapse supernova models with phenomenological treatment of neutrino flavor conversions},
	url = {https://doi.org/10.1093/pasj/psaf007},
	volume = {77},
	year = {2025}
}

@article{nagakura2024bhatnagar,
  title = {Bhatnagar-Gross-Krook subgrid model for neutrino quantum kinetics},
  author = {Nagakura, Hiroki and Johns, Lucas and Zaizen, Masamichi},
  journal = {Phys. Rev. D},
  volume = {109},
  issue = {8},
  pages = {083013},
  numpages = {18},
  year = {2024},
  month = {Apr},
  publisher = {American Physical Society},
  doi = {10.1103/PhysRevD.109.083013},
  url = {https://link.aps.org/doi/10.1103/PhysRevD.109.083013}
}

@article{qiu2025neutrino,
    author = "Qiu, Yi and Radice, David and Richers, Sherwood and Bhattacharyya, Maitraya",
    title = "{Neutrino Flavor Transformation in Neutron Star Mergers}",
    eprint = "2503.11758",
    archivePrefix = "arXiv",
    primaryClass = "astro-ph.HE",
    reportNumber = "INT-PUB-25-004",
    doi = "10.1103/h2q7-kn3v",
    journal = "Phys. Rev. Lett.",
    volume = "135",
    number = "9",
    pages = "091401",
    year = "2025"
}

@article{qiu2025impact,
    author = "Qiu, Yi and Radice, David and Richers, Sherwood and Guercilena, Federico Maria and Perego, Albino and Bhattacharyya, Maitraya",
    title = "{Impact of neutrino flavor conversions on neutron star merger dynamics, ejecta, nucleosynthesis, and multimessenger signals}",
    eprint = "2510.15028",
    archivePrefix = "arXiv",
    primaryClass = "astro-ph.HE",
    doi = "10.1103/qckq-78gt",
    journal = "Phys. Rev. D",
    volume = "112",
    number = "12",
    pages = "123039",
    year = "2025"
}

@article{Morinaga:2021vmc,
    author = "Morinaga, Taiki",
    title = "{Fast neutrino flavor instability and neutrino flavor lepton number crossings}",
    eprint = "2103.15267",
    archivePrefix = "arXiv",
    primaryClass = "hep-ph",
    doi = "10.1103/PhysRevD.105.L101301",
    journal = "Phys. Rev. D",
    volume = "105",
    number = "10",
    pages = "L101301",
    year = "2022"
}

@article{Dasgupta:2021gfs,
    author = "Dasgupta, Basudeb",
    title = "{Collective Neutrino Flavor Instability Requires a Crossing}",
    eprint = "2110.00192",
    archivePrefix = "arXiv",
    primaryClass = "hep-ph",
    reportNumber = "TIFR/TH/21-15",
    doi = "10.1103/PhysRevLett.128.081102",
    journal = "Phys. Rev. Lett.",
    volume = "128",
    number = "8",
    pages = "081102",
    year = "2022"
}

@article{Fiorillo:2024bzm,
    author = "Fiorillo, Damiano F. G. and Raffelt, Georg G.",
    title = "{Theory of neutrino fast flavor evolution. Part I. Linear response theory and stability conditions.}",
    eprint = "2406.06708",
    archivePrefix = "arXiv",
    primaryClass = "hep-ph",
    doi = "10.1007/JHEP08(2024)225",
    journal = "JHEP",
    volume = "08",
    pages = "225",
    year = "2024"
}

@article{Nagakura:2025brr,
    author = "Nagakura, Hiroki and Zaizen, Masamichi and Liu, Jiabao and Johns, Lucas",
    title = "{Resolution requirements for numerical modeling of neutrino quantum kinetics}",
    eprint = "2501.14145",
    archivePrefix = "arXiv",
    primaryClass = "astro-ph.HE",
    reportNumber = "LA-UR-24-33051",
    doi = "10.1103/PhysRevD.111.043028",
    journal = "Phys. Rev. D",
    volume = "111",
    number = "4",
    pages = "043028",
    year = "2025"
}

@article{Liu:2023pjw,
    author = "Liu, Jiabao and Zaizen, Masamichi and Yamada, Shoichi",
    title = "{Systematic study of the resonancelike structure in the collisional flavor instability of neutrinos}",
    eprint = "2302.06263",
    archivePrefix = "arXiv",
    primaryClass = "hep-ph",
    doi = "10.1103/PhysRevD.107.123011",
    journal = "Phys. Rev. D",
    volume = "107",
    number = "12",
    pages = "123011",
    year = "2023"
}

@article{Froustey:2025nbi,
    author = "Froustey, Julien",
    title = "{Predicting the outcome of collisional neutrino flavor conversion}",
    eprint = "2505.16961",
    archivePrefix = "arXiv",
    primaryClass = "hep-ph",
    reportNumber = "N3AS-25-008",
    doi = "10.1103/syxg-sqfn",
    journal = "Phys. Rev. D",
    volume = "112",
    number = "2",
    pages = "023029",
    year = "2025"
}

@article{froustey25cfi,
    author = "Froustey, Julien and Foucart, Francois and Hall, Christian and Kneller, James P. and Kundu, Debraj and Lin, Zidu and McLaughlin, Gail C. and Richers, Sherwood",
    title = "{Neutrino flavor instabilities in neutron star mergers with moment transport: Slow, fast, and collisional modes}",
    eprint = "2601.02461",
    archivePrefix = "arXiv",
    primaryClass = "astro-ph.HE",
    reportNumber = "N3AS-25-020",
    doi = "10.1103/htq2-d1t7",
    journal = "Phys. Rev. D",
    volume = "113",
    number = "6",
    pages = "063050",
    year = "2026"
}

@article{burns2020neutron,
    author = "Burns, Eric",
    title = "{Neutron Star Mergers and How to Study Them}",
    eprint = "1909.06085",
    archivePrefix = "arXiv",
    primaryClass = "astro-ph.HE",
    doi = "10.1007/s41114-020-00028-7",
    journal = "Living Rev. Rel.",
    volume = "23",
    number = "1",
    pages = "4",
    year = "2020"
}

@article{fernandez2016electro,
    author = "Fern{\'a}ndez, Rodrigo and Metzger, Brian D.",
    title = "{Electromagnetic Signatures of Neutron Star Mergers in the Advanced LIGO Era}",
    eprint = "1512.05435",
    archivePrefix = "arXiv",
    primaryClass = "astro-ph.HE",
    doi = "10.1146/annurev-nucl-102115-044819",
    journal = "Ann. Rev. Nucl. Part. Sci.",
    volume = "66",
    pages = "23--45",
    year = "2016"
}

@article{Margutii2021firstmul,
    author = "Margutti, Raffaella and Chornock, Ryan",
    title = "{First Multimessenger Observations of a Neutron Star Merger}",
    eprint = "2012.04810",
    archivePrefix = "arXiv",
    primaryClass = "astro-ph.HE",
    doi = "10.1146/annurev-astro-112420-030742",
    journal = "Ann. Rev. Astron. Astrophys.",
    volume = "59",
    pages = "155--202",
    year = "2021"
}

@article{radice_dynamics_2020,
    author = "Radice, David and Bernuzzi, Sebastiano and Perego, Albino",
    title = "{The Dynamics of Binary Neutron Star Mergers and GW170817}",
    eprint = "2002.03863",
    archivePrefix = "arXiv",
    primaryClass = "astro-ph.HE",
    doi = "10.1146/annurev-nucl-013120-114541",
    journal = "Ann. Rev. Nucl. Part. Sci.",
    volume = "70",
    pages = "95--119",
    year = "2020"
}

@article{Radice_2018,
    author = "Radice, David and Perego, Albino and Zappa, Francesco and Bernuzzi, Sebastiano",
    title = "{GW170817: Joint Constraint on the Neutron Star Equation of State from Multimessenger Observations}",
    eprint = "1711.03647",
    archivePrefix = "arXiv",
    primaryClass = "astro-ph.HE",
    reportNumber = "LIGO-P1700421, VIR-0894A-17",
    doi = "10.3847/2041-8213/aaa402",
    journal = "Astrophys. J. Lett.",
    volume = "852",
    number = "2",
    pages = "L29",
    year = "2018"
}

@article{Margalit_2019,
    author = "Margalit, Ben and Metzger, Brian D.",
    title = "{The Multi-Messenger Matrix: the Future of Neutron Star Merger Constraints on the Nuclear Equation of State}",
    eprint = "1904.11995",
    archivePrefix = "arXiv",
    primaryClass = "astro-ph.HE",
    doi = "10.3847/2041-8213/ab2ae2",
    journal = "Astrophys. J. Lett.",
    volume = "880",
    number = "1",
    pages = "L15",
    year = "2019"
}

@article{Foucart2021estimating,
    author = "Foucart, Francois and Moesta, Philipp and Ramirez, Teresita and Wright, Alex James and Darbha, Siva and Kasen, Daniel",
    title = "{Estimating outflow masses and velocities in merger simulations: Impact of r-process heating and neutrino cooling}",
    eprint = "2109.00565",
    archivePrefix = "arXiv",
    primaryClass = "astro-ph.HE",
    doi = "10.1103/PhysRevD.104.123010",
    journal = "Phys. Rev. D",
    volume = "104",
    number = "12",
    pages = "123010",
    year = "2021"
}

@article{Mosta_2020_amagnetar,
    author = {M{\"o}sta, Philipp and Radice, David and Haas, Roland and Schnetter, Erik and Bernuzzi, Sebastiano},
    title = "{A magnetar engine for short GRBs and kilonovae}",
    eprint = "2003.06043",
    archivePrefix = "arXiv",
    primaryClass = "astro-ph.HE",
    doi = "10.3847/2041-8213/abb6ef",
    journal = "Astrophys. J. Lett.",
    volume = "901",
    pages = "L37",
    year = "2020"
}

@article{Miller2019,
    author = "Miller, Jonah M. and Ryan, Benjamin R. and Dolence, Joshua C. and Burrows, Adam and Fontes, Christopher J. and Fryer, Christopher L. and Korobkin, Oleg and Lippuner, Jonas and Mumpower, Matthew R. and Wollaeger, Ryan T.",
    title = "{Full Transport Model of GW170817-Like Disk Produces a Blue Kilonova}",
    eprint = "1905.07477",
    archivePrefix = "arXiv",
    primaryClass = "astro-ph.HE",
    reportNumber = "LA-UR-19-22623",
    doi = "10.1103/PhysRevD.100.023008",
    journal = "Phys. Rev. D",
    volume = "100",
    number = "2",
    pages = "023008",
    year = "2019"
}

@article{Nedora_2022,
    author = "Nedora, Vsevolod and Schianchi, Federico and Bernuzzi, Sebastiano and Radice, David and Daszuta, Boris and Endrizzi, Andrea and Perego, Albino and Prakash, Aviral and Zappa, Francesco",
    title = "{Mapping dynamical ejecta and disk masses from numerical relativity simulations of neutron star mergers}",
    eprint = "2011.11110",
    archivePrefix = "arXiv",
    primaryClass = "astro-ph.HE",
    doi = "10.1088/1361-6382/ac35a8",
    journal = "Class. Quant. Grav.",
    volume = "39",
    number = "1",
    pages = "015008",
    year = "2022"
}

@article{Zhu2021fullygeneral,
    author = "Zhu, Zhenyu and Rezzolla, Luciano",
    title = "{Fully general-relativistic simulations of isolated and binary strange quark stars}",
    eprint = "2102.07721",
    archivePrefix = "arXiv",
    primaryClass = "astro-ph.HE",
    doi = "10.1103/PhysRevD.104.083004",
    journal = "Phys. Rev. D",
    volume = "104",
    number = "8",
    pages = "083004",
    year = "2021"
}

@article{Metzger_2021,
    author = "Metzger, Brian D. and Fernandez, Rodrigo",
    title = "{From Neutrino- to Photon-cooled in Three Years: Can Fallback Accretion Explain the X-Ray Excess in GW170817?}",
    eprint = "2106.02052",
    archivePrefix = "arXiv",
    primaryClass = "astro-ph.HE",
    doi = "10.3847/2041-8213/ac1169",
    journal = "Astrophys. J. Lett.",
    volume = "916",
    number = "1",
    pages = "L3",
    year = "2021"
}

@article{Li2021neutrinofast,
    author = "Li, Xinyu and Siegel, Daniel M.",
    title = "{Neutrino Fast Flavor Conversions in Neutron-Star Postmerger Accretion Disks}",
    eprint = "2103.02616",
    archivePrefix = "arXiv",
    primaryClass = "astro-ph.HE",
    doi = "10.1103/PhysRevLett.126.251101",
    journal = "Phys. Rev. Lett.",
    volume = "126",
    number = "25",
    pages = "251101",
    year = "2021"
}

@article{just2022neutrino,
    author = "Just, Oliver and Goriely, Stephane and Janka, Hans-Thomas and Nagataki, Shigehiro and Bauswein, Andreas",
    title = "{Neutrino absorption and other physics dependencies in neutrino-cooled black hole accretion discs}",
    eprint = "2102.08387",
    archivePrefix = "arXiv",
    primaryClass = "astro-ph.HE",
    doi = "10.1093/mnras/stab2861",
    journal = "Mon. Not. Roy. Astron. Soc.",
    volume = "509",
    number = "1",
    pages = "1377--1412",
    year = "2021"
}

@article{Abbott2017gw170817,
    author = "Abbott, B. P. and others",
    collaboration = "LIGO Scientific, Virgo",
    title = "{GW170817: Observation of Gravitational Waves from a Binary Neutron Star Inspiral}",
    eprint = "1710.05832",
    archivePrefix = "arXiv",
    primaryClass = "gr-qc",
    reportNumber = "LIGO-P170817",
    doi = "10.1103/PhysRevLett.119.161101",
    journal = "Phys. Rev. Lett.",
    volume = "119",
    number = "16",
    pages = "161101",
    year = "2017"
}

@article{Abbott2017multimessengerobs,
    author = "Abbott, B. P. and others",
    title = "{Multi-messenger Observations of a Binary Neutron Star Merger}",
    eprint = "1710.05833",
    archivePrefix = "arXiv",
    primaryClass = "astro-ph.HE",
    reportNumber = "LIGO-P1700294, VIR-0802A-17, FERMILAB-PUB-17-478-A-AE-CD",
    doi = "10.3847/2041-8213/aa91c9",
    journal = "Astrophys. J. Lett.",
    volume = "848",
    number = "2",
    pages = "L12",
    year = "2017"
}

@article{Cowperthwaite_2017,
    author = "Cowperthwaite, P. S. and others",
    title = "{The Electromagnetic Counterpart of the Binary Neutron Star Merger LIGO/Virgo GW170817. II. UV, Optical, and Near-infrared Light Curves and Comparison to Kilonova Models}",
    eprint = "1710.05840",
    archivePrefix = "arXiv",
    primaryClass = "astro-ph.HE",
    reportNumber = "FERMILAB-PUB-17-470-A-AE-CD-PPD",
    doi = "10.3847/2041-8213/aa8fc7",
    journal = "Astrophys. J. Lett.",
    volume = "848",
    number = "2",
    pages = "L17",
    year = "2017"
}

@article{metzger2017kilonovae,
    author = "Metzger, Brian D.",
    title = "{Kilonovae}",
    eprint = "1910.01617",
    archivePrefix = "arXiv",
    primaryClass = "astro-ph.HE",
    doi = "10.1007/s41114-019-0024-0",
    journal = "Living Rev. Rel.",
    volume = "23",
    number = "1",
    pages = "1",
    year = "2020"
}

@article{barnes2020physics,
    author = "Barnes, Jennifer",
    title = "{The Physics of Kilonovae}",
    doi = "10.3389/fphy.2020.00355",
    journal = "Front. in Phys.",
    volume = "8",
    pages = "355",
    year = "2020"
}

@article{Radice2020thedynamics,
    author = "Radice, David and Bernuzzi, Sebastiano and Perego, Albino",
    title = "{The Dynamics of Binary Neutron Star Mergers and GW170817}",
    eprint = "2002.03863",
    archivePrefix = "arXiv",
    primaryClass = "astro-ph.HE",
    doi = "10.1146/annurev-nucl-013120-114541",
    journal = "Ann. Rev. Nucl. Part. Sci.",
    volume = "70",
    pages = "95--119",
    year = "2020"
}

@article{curtis2023nucleosynthesis,
    author = "Curtis, Sanjana and Miller, Jonah M. and Frohlich, Carla and Sprouse, Trevor and Lloyd-Ronning, Nicole and Mumpower, Matthew",
    title = "{Nucleosynthesis in Outflows from Black Hole{\textendash}Neutron Star Merger Disks with Full GR({\ensuremath{\nu}})RMHD}",
    eprint = "2212.10691",
    archivePrefix = "arXiv",
    primaryClass = "astro-ph.HE",
    reportNumber = "LA-UR-22-32395",
    doi = "10.3847/2041-8213/acba16",
    journal = "Astrophys. J. Lett.",
    volume = "945",
    number = "1",
    pages = "L13",
    year = "2023"
}

@article{wu2015effects,
    author = "Wu, Meng-Ru and Qian, Yong-Zhong and Martinez-Pinedo, Gabriel and Fischer, Tobias and Huther, Lutz",
    title = "{Effects of neutrino oscillations on nucleosynthesis and neutrino signals for an 18 M supernova model}",
    eprint = "1412.8587",
    archivePrefix = "arXiv",
    primaryClass = "astro-ph.HE",
    doi = "10.1103/PhysRevD.91.065016",
    journal = "Phys. Rev. D",
    volume = "91",
    number = "6",
    pages = "065016",
    year = "2015"
}

@article{Lippuner_2017,
    author = "Lippuner, Jonas and Fern{\'a}ndez, Rodrigo and Roberts, Luke F. and Foucart, Francois and Kasen, Daniel and Metzger, Brian D. and Ott, Christian D.",
    title = "{Signatures of hypermassive neutron star lifetimes on r-process nucleosynthesis in the disc ejecta from neutron star mergers}",
    eprint = "1703.06216",
    archivePrefix = "arXiv",
    primaryClass = "astro-ph.HE",
    reportNumber = "YITP-17-26",
    doi = "10.1093/mnras/stx1987",
    journal = "Mon. Not. Roy. Astron. Soc.",
    volume = "472",
    number = "1",
    pages = "904--918",
    year = "2017"
}

@incollection{Foucart:2024cjr,
    title = {Colliding neutron stars and black holes},
    booktitle = {Reference Module in Materials Science and Materials Engineering},
    publisher = {Elsevier},
    year = {2025},
    isbn = {978-0-12-803581-8},
    doi = {https://doi.org/10.1016/B978-0-443-26598-3.00004-3},
    url = {https://www.sciencedirect.com/science/article/pii/B9780443265983000043},
    author = {Francois Foucart},
    eprint = "2410.03646",
    archivePrefix = "arXiv",
    primaryClass = "astro-ph.HE",
}

@article{Nagakura:2022kic,
    author = "Nagakura, Hiroki and Zaizen, Masamichi",
    title = "{Time-Dependent and Quasisteady Features of Fast Neutrino-Flavor Conversion}",
    eprint = "2206.04097",
    archivePrefix = "arXiv",
    primaryClass = "astro-ph.HE",
    doi = "10.1103/PhysRevLett.129.261101",
    journal = "Phys. Rev. Lett.",
    volume = "129",
    number = "26",
    pages = "261101",
    year = "2022"
}

@article{nagakura2023global,
    author = "Nagakura, Hiroki",
    title = "{Global features of fast neutrino-flavor conversion in binary neutron star mergers}",
    eprint = "2306.10108",
    archivePrefix = "arXiv",
    primaryClass = "astro-ph.HE",
    doi = "10.1103/PhysRevD.108.103014",
    journal = "Phys. Rev. D",
    volume = "108",
    number = "10",
    pages = "103014",
    year = "2023"
}

@article{sawyer2016neutrino,
    author = "Sawyer, R. F.",
    title = "{Neutrino cloud instabilities just above the neutrino sphere of a supernova}",
    eprint = "1509.03323",
    archivePrefix = "arXiv",
    primaryClass = "astro-ph.HE",
    doi = "10.1103/PhysRevLett.116.081101",
    journal = "Phys. Rev. Lett.",
    volume = "116",
    number = "8",
    pages = "081101",
    year = "2016"
}

@article{wu2017fast,
    author = "Wu, Meng-Ru and Tamborra, Irene",
    title = "{Fast neutrino conversions: Ubiquitous in compact binary merger remnants}",
    eprint = "1701.06580",
    archivePrefix = "arXiv",
    primaryClass = "astro-ph.HE",
    doi = "10.1103/PhysRevD.95.103007",
    journal = "Phys. Rev. D",
    volume = "95",
    number = "10",
    pages = "103007",
    year = "2017"
}

@article{morinaga2020fast,
    author = "Morinaga, Taiki and Nagakura, Hiroki and Kato, Chinami and Yamada, Shoichi",
    title = "{Fast neutrino-flavor conversion in the preshock region of core-collapse supernovae}",
    eprint = "1909.13131",
    archivePrefix = "arXiv",
    primaryClass = "astro-ph.HE",
    doi = "10.1103/PhysRevResearch.2.012046",
    journal = "Phys. Rev. Res.",
    volume = "2",
    number = "1",
    pages = "012046",
    year = "2020"
}

@article{george2020fast,
    author = "George, Manu and Wu, Meng-Ru and Tamborra, Irene and Ardevol-Pulpillo, Ricard and Janka, Hans-Thomas",
    title = "{Fast neutrino flavor conversion, ejecta properties, and nucleosynthesis in newly-formed hypermassive remnants of neutron-star mergers}",
    eprint = "2009.04046",
    archivePrefix = "arXiv",
    primaryClass = "astro-ph.HE",
    doi = "10.1103/PhysRevD.102.103015",
    journal = "Phys. Rev. D",
    volume = "102",
    number = "10",
    pages = "103015",
    year = "2020"
}

@article{nagakura2021where,
    author = "Nagakura, Hiroki and Johns, Lucas and Burrows, Adam and Fuller, George M.",
    title = "{Where, when, and why: Occurrence of fast-pairwise collective neutrino oscillation in three-dimensional core-collapse supernova models}",
    eprint = "2108.07281",
    archivePrefix = "arXiv",
    primaryClass = "astro-ph.HE",
    doi = "10.1103/PhysRevD.104.083025",
    journal = "Phys. Rev. D",
    volume = "104",
    number = "8",
    pages = "083025",
    year = "2021"
}

@article{tamborra2021new,
    author = "Tamborra, Irene and Shalgar, Shashank",
    title = "{New Developments in Flavor Evolution of a Dense Neutrino Gas}",
    eprint = "2011.01948",
    archivePrefix = "arXiv",
    primaryClass = "astro-ph.HE",
    doi = "10.1146/annurev-nucl-102920-050505",
    journal = "Ann. Rev. Nucl. Part. Sci.",
    volume = "71",
    pages = "165--188",
    year = "2021"
}

@article{fernandez2022,
    author = "Fern{\'a}ndez, Rodrigo and Richers, Sherwood and Mulyk, Nicole and Fahlman, Steven",
    title = "{Fast flavor instability in hypermassive neutron star disk outflows}",
    eprint = "2207.10680",
    archivePrefix = "arXiv",
    primaryClass = "astro-ph.HE",
    doi = "10.1103/PhysRevD.106.103003",
    journal = "Phys. Rev. D",
    volume = "106",
    number = "10",
    pages = "103003",
    year = "2022"
}

@article{ehring2023fast,
    author = "Ehring, Jakob and Abbar, Sajad and Janka, Hans-Thomas and Raffelt, Georg and Tamborra, Irene",
    title = "{Fast neutrino flavor conversion in core-collapse supernovae: A parametric study in 1D models}",
    eprint = "2301.11938",
    archivePrefix = "arXiv",
    primaryClass = "astro-ph.HE",
    doi = "10.1103/PhysRevD.107.103034",
    journal = "Phys. Rev. D",
    volume = "107",
    number = "10",
    pages = "103034",
    year = "2023"
}

@article{akaho2024collisional,
    author = "Akaho, Ryuichiro and Liu, Jiabao and Nagakura, Hiroki and Zaizen, Masamichi and Yamada, Shoichi",
    title = "{Collisional and fast neutrino flavor instabilities in two-dimensional core-collapse supernova simulation with Boltzmann neutrino transport}",
    eprint = "2311.11272",
    archivePrefix = "arXiv",
    primaryClass = "astro-ph.HE",
    doi = "10.1103/PhysRevD.109.023012",
    journal = "Phys. Rev. D",
    volume = "109",
    number = "2",
    pages = "023012",
    year = "2024"
}

@article{Wang_2026,
    author = "Wang, Tianshu and Burrows, Adam",
    title = "{The Effect of the Fast-flavor Instability on Core-collapse Supernova Models. II. Quasi-equipartition and the Impact of Various Angular Reconstruction Methods}",
    eprint = "2511.20767",
    archivePrefix = "arXiv",
    primaryClass = "astro-ph.HE",
    doi = "10.3847/1538-4357/ae3244",
    journal = "Astrophys. J.",
    volume = "997",
    number = "2",
    pages = "325",
    year = "2026"
}

@article{wang2025,
    author = "Wang, Tianshu and Burrows, Adam",
    title = "{The Effect of the Fast-Flavor Instability on Core-Collapse Supernova Models}",
    journal = "ApJ",
    doi = "10.3847/1538-4357/add889",
    eprint = "2503.04896",
    archivePrefix = "arXiv",
    primaryClass = "astro-ph.HE",
    year = 2025,
    month = jun,
    volume = {986},
    number = {2},
    pages = {153}
}

@article{duan2006collective,
    author = "Duan, Huaiyu and Fuller, George M. and Qian, Yong-Zhong",
    title = "{Collective neutrino flavor transformation in supernovae}",
    eprint = "astro-ph/0511275",
    archivePrefix = "arXiv",
    doi = "10.1103/PhysRevD.74.123004",
    journal = "Phys. Rev. D",
    volume = "74",
    pages = "123004",
    year = "2006"
}

@article{duan2010collective,
    author = "Duan, Huaiyu and Fuller, George M. and Qian, Yong-Zhong",
    title = "{Collective Neutrino Oscillations}",
    eprint = "1001.2799",
    archivePrefix = "arXiv",
    primaryClass = "hep-ph",
    reportNumber = "LA-UR-09-08309, INT-PUB-10-001",
    doi = "10.1146/annurev.nucl.012809.104524",
    journal = "Ann. Rev. Nucl. Part. Sci.",
    volume = "60",
    pages = "569--594",
    year = "2010"
}

@article{dasgupta2015temporal,
    author = "Dasgupta, Basudeb and Mirizzi, Alessandro",
    title = "{Temporal Instability Enables Neutrino Flavor Conversions Deep Inside Supernovae}",
    eprint = "1509.03171",
    archivePrefix = "arXiv",
    primaryClass = "hep-ph",
    doi = "10.1103/PhysRevD.92.125030",
    journal = "Phys. Rev. D",
    volume = "92",
    number = "12",
    pages = "125030",
    year = "2015"
}

@article{chakraborty2016collective,
    author = "Chakraborty, Sovan and Hansen, Rasmus and Izaguirre, Ignacio and Raffelt, Georg",
    title = "{Collective neutrino flavor conversion: Recent developments}",
    eprint = "1602.02766",
    archivePrefix = "arXiv",
    primaryClass = "hep-ph",
    doi = "10.1016/j.nuclphysb.2016.02.012",
    journal = "Nucl. Phys. B",
    volume = "908",
    pages = "366--381",
    year = "2016"
}

@article{shalgar2024neutrino,
    author = "Shalgar, Shashank and Tamborra, Irene",
    title = "{Neutrino quantum kinetics in a core-collapse supernova}",
    eprint = "2406.09504",
    archivePrefix = "arXiv",
    primaryClass = "astro-ph.HE",
    doi = "10.1088/1475-7516/2024/09/021",
    journal = "J. Cosmol. Astropart. Phys.",
    volume = "09",
    pages = "021",
    year = "2024"
}

@article{fiorillo2025theoryslow,
    author = "Fiorillo, Damiano F. G. and Raffelt, Georg G.",
    title = "{Theory of neutrino slow flavor evolution. Part I. Homogeneous medium}",
    eprint = "2412.02747",
    archivePrefix = "arXiv",
    primaryClass = "hep-ph",
    doi = "10.1007/JHEP04(2025)146",
    journal = "JHEP",
    volume = "04",
    pages = "146",
    year = "2025"
}

@article{fiorillo2025theoryslow2,
    author = "Fiorillo, Damiano F. G. and Raffelt, Georg G.",
    title = "{Theory of neutrino slow flavor evolution. Part II. Space-time evolution of linear instabilities}",
    eprint = "2501.16423",
    archivePrefix = "arXiv",
    primaryClass = "hep-ph",
    doi = "10.1007/JHEP06(2025)146",
    journal = "JHEP",
    volume = "06",
    pages = "146",
    year = "2025"
}

@article{fiorillo2025first,
    author = "Fiorillo, Damiano F. G. and Janka, Hans-Thomas and Raffelt, Georg G.",
    title = "{Neutrino-Mass-Driven Instabilities as the Earliest Flavor Conversion in Supernovae}",
    eprint = "2507.22985",
    archivePrefix = "arXiv",
    primaryClass = "hep-ph",
    doi = "10.1103/jbmx-rbzt",
    journal = "Phys. Rev. Lett.",
    volume = "135",
    number = "23",
    pages = "231003",
    year = "2025"
}

@article{johns2021collisional,
    author = "Johns, Lucas",
    title = "{Collisional Flavor Instabilities of Supernova Neutrinos}",
    eprint = "2104.11369",
    archivePrefix = "arXiv",
    primaryClass = "hep-ph",
    doi = "10.1103/PhysRevLett.130.191001",
    journal = "Phys. Rev. Lett.",
    volume = "130",
    number = "19",
    pages = "191001",
    year = "2023"
}

@article{johns2022collisional,
    author = "Johns, Lucas and Xiong, Zewei",
    title = "{Collisional instabilities of neutrinos and their interplay with fast flavor conversion in compact objects}",
    eprint = "2208.11059",
    archivePrefix = "arXiv",
    primaryClass = "hep-ph",
    doi = "10.1103/PhysRevD.106.103029",
    journal = "Phys. Rev. D",
    volume = "106",
    number = "10",
    pages = "103029",
    year = "2022"
}

@article{xiong2022evolution,
    author = "Xiong, Zewei and Wu, Meng-Ru and Mart\'\i{}nez-Pinedo, Gabriel and Fischer, Tobias and George, Manu and Lin, Chun-Yu and Johns, Lucas",
    title = "{Evolution of collisional neutrino flavor instabilities in spherically symmetric supernova models}",
    eprint = "2210.08254",
    archivePrefix = "arXiv",
    primaryClass = "astro-ph.HE",
    doi = "10.1103/PhysRevD.107.083016",
    journal = "Phys. Rev. D",
    volume = "107",
    number = "8",
    pages = "083016",
    year = "2023"
}

@article{xiong2022collisional,
    author = "Xiong, Zewei and Johns, Lucas and Wu, Meng-Ru and Duan, Huaiyu",
    title = "{Collisional flavor instability in dense neutrino gases}",
    eprint = "2212.03750",
    archivePrefix = "arXiv",
    primaryClass = "hep-ph",
    doi = "10.1103/PhysRevD.108.083002",
    journal = "Phys. Rev. D",
    volume = "108",
    number = "8",
    pages = "083002",
    year = "2023"
}

@article{liu2023systematic,
    author = "Liu, Jiabao and Zaizen, Masamichi and Yamada, Shoichi",
    title = "{Systematic study of the resonancelike structure in the collisional flavor instability of neutrinos}",
    eprint = "2302.06263",
    archivePrefix = "arXiv",
    primaryClass = "hep-ph",
    doi = "10.1103/PhysRevD.107.123011",
    journal = "Phys. Rev. D",
    volume = "107",
    number = "12",
    pages = "123011",
    year = "2023"
}

@article{liu2023universality,
    author = "Liu, Jiabao and Nagakura, Hiroki and Akaho, Ryuichiro and Ito, Akira and Zaizen, Masamichi and Yamada, Shoichi",
    title = "{Universality of the neutrino collisional flavor instability in core-collapse supernovae}",
    eprint = "2310.05050",
    archivePrefix = "arXiv",
    primaryClass = "astro-ph.HE",
    doi = "10.1103/PhysRevD.108.123024",
    journal = "Phys. Rev. D",
    volume = "108",
    number = "12",
    pages = "123024",
    year = "2023"
}

@article{akaho2023collisional,
    author = "Akaho, Ryuichiro and Liu, Jiabao and Nagakura, Hiroki and Zaizen, Masamichi and Yamada, Shoichi",
    title = "{Collisional and fast neutrino flavor instabilities in two-dimensional core-collapse supernova simulation with Boltzmann neutrino transport}",
    eprint = "2311.11272",
    archivePrefix = "arXiv",
    primaryClass = "astro-ph.HE",
    doi = "10.1103/PhysRevD.109.023012",
    journal = "Phys. Rev. D",
    volume = "109",
    number = "2",
    pages = "023012",
    year = "2024"
}

@article{shalgar2023neutrinos,
    author = "Shalgar, Shashank and Tamborra, Irene",
    title = "{Do neutrinos become flavor unstable due to collisions with matter in the supernova decoupling region?}",
    eprint = "2307.10366",
    archivePrefix = "arXiv",
    primaryClass = "astro-ph.HE",
    doi = "10.1103/PhysRevD.109.103011",
    journal = "Phys. Rev. D",
    volume = "109",
    number = "10",
    pages = "103011",
    year = "2024"
}

@article{kato2023collisional,
    author = "Kato, Chinami and Nagakura, Hiroki and Johns, Lucas",
    title = "{Collisional flavor swap with neutrino self-interactions}",
    eprint = "2309.02619",
    archivePrefix = "arXiv",
    primaryClass = "astro-ph.HE",
    doi = "10.1103/PhysRevD.109.103009",
    journal = "Phys. Rev. D",
    volume = "109",
    number = "10",
    pages = "103009",
    year = "2024"
}

@article{froustey2024neutrino,
    author = "Froustey, Julien and Richers, Sherwood and Grohs, Evan and Flynn, Samuel D. and Foucart, Francois and Kneller, James P. and McLaughlin, Gail C.",
    title = "{Neutrino fast flavor oscillations with moments: Linear stability analysis and application to neutron star mergers}",
    eprint = "2311.11968",
    archivePrefix = "arXiv",
    primaryClass = "astro-ph.HE",
    reportNumber = "N3AS-23-032",
    doi = "10.1103/PhysRevD.109.043046",
    journal = "Phys. Rev. D",
    volume = "109",
    number = "4",
    pages = "043046",
    year = "2024"
}

@article{zaizen2025spectral,
    author = "Zaizen, Masamichi",
    title = "{Spectral diversity in collisional neutrino-flavor conversion: Flavor equipartition or swap}",
    eprint = "2502.09260",
    archivePrefix = "arXiv",
    primaryClass = "hep-ph",
    doi = "10.1103/PhysRevD.111.103029",
    journal = "Phys. Rev. D",
    volume = "111",
    number = "10",
    pages = "103029",
    year = "2025"
}

@article{vaananen2016uncovering,
    author = "Vaananen, D. and McLaughlin, G. C.",
    title = "{Uncovering the Matter-Neutrino Resonance}",
    eprint = "1510.00751",
    archivePrefix = "arXiv",
    primaryClass = "hep-ph",
    doi = "10.1103/PhysRevD.93.105044",
    journal = "Phys. Rev. D",
    volume = "93",
    number = "10",
    pages = "105044",
    year = "2016"
}

@article{zhu2016matter,
    author = "Zhu, Yong-Lin and Perego, Albino and McLaughlin, Gail C.",
    title = "{Matter Neutrino Resonance Transitions above a Neutron Star Merger Remnant}",
    eprint = "1607.04671",
    archivePrefix = "arXiv",
    primaryClass = "hep-ph",
    doi = "10.1103/PhysRevD.94.105006",
    journal = "Phys. Rev. D",
    volume = "94",
    number = "10",
    pages = "105006",
    year = "2016"
}

@article{vlasenko2018matter,
    author = "Vlasenko, A. and McLaughlin, G. C.",
    title = "{Matter-neutrino resonance in a multiangle neutrino bulb model}",
    eprint = "1801.07813",
    archivePrefix = "arXiv",
    primaryClass = "astro-ph.HE",
    doi = "10.1103/PhysRevD.97.083011",
    journal = "Phys. Rev. D",
    volume = "97",
    number = "8",
    pages = "083011",
    year = "2018"
}

@article{malkus2016symmetric,
    author = "Malkus, A. and McLaughlin, G. C. and Surman, R.",
    title = "{Symmetric and Standard Matter-Neutrino Resonances Above Merging Compact Objects}",
    eprint = "1507.00946",
    archivePrefix = "arXiv",
    primaryClass = "hep-ph",
    doi = "10.1103/PhysRevD.93.045021",
    journal = "Phys. Rev. D",
    volume = "93",
    number = "4",
    pages = "045021",
    year = "2016"
}

@ARTICLE{2012PhRvD..86h5015M,
       author = {{Malkus}, A. and {Kneller}, J.~P. and {McLaughlin}, G.~C. and {Surman}, R.},
        title = "{Neutrino oscillations above black hole accretion disks: Disks with electron-flavor emission}",
      journal = {\prd},
         year = 2012,
        month = oct,
       volume = {86},
       number = {8},
          eid = {085015},
        pages = {085015},
          doi = {10.1103/PhysRevD.86.085015},
archivePrefix = {arXiv},
       eprint = {1207.6648},
 primaryClass = {hep-ph},
       adsurl = {https://ui.adsabs.harvard.edu/abs/2012PhRvD..86h5015M}
}

@article{wu2016physics,
  title={Physics of neutrino flavor transformation through matter--neutrino resonances},
  author={Wu, Meng-Ru and Duan, Huaiyu and Qian, Yong-Zhong},
  journal={Physics Letters B},
  volume={752},
  pages={89--94},
  year={2016},
  publisher={Elsevier}
}

@article{padilla2024symmetry,
  title={Symmetry breaking due to multi-angle matter-neutrino resonance in neutron star merger remnants},
  author={Padilla-Gay, Ian and Shalgar, Shashank and Tamborra, Irene},
  journal={JCAP},
  volume={2024},
  number={05},
  pages={037},
  year={2024},
  publisher={IOP Publishing},
    eprint = "2403.15532",
    archivePrefix = "arXiv",
    primaryClass = "astro-ph.HE",
    reportNumber = "SLAC-PUB-17761",
    doi = "10.1088/1475-7516/2024/05/037",
}

@article{faiz2025can,
    author = "Faiz, Owais Ullah and Hussain, Mushahid and Shalgar, Shashank",
    title = "{Can we ignore the time dependence in matter-neutrino resonance?}",
    eprint = "2510.19485",
    archivePrefix = "arXiv",
    primaryClass = "astro-ph.HE",
    doi = "10.1142/s0218271826500331",
    journal = "Int. J. Mod. Phys. D",
    volume = "35",
    number = "10",
    pages = "2650033",
    year = "2026"
}

@article{richers2021neutrino,
    author = "Richers, Sherwood and Willcox, Donald and Ford, Nicole",
    title = "{Neutrino fast flavor instability in three dimensions}",
    eprint = "2109.08631",
    archivePrefix = "arXiv",
    primaryClass = "astro-ph.HE",
    reportNumber = "N3AS-21-013",
    doi = "10.1103/PhysRevD.104.103023",
    journal = "Phys. Rev. D",
    volume = "104",
    number = "10",
    pages = "103023",
    year = "2021"
}

@article{richers2022code,
    author = "Richers, Sherwood and Duan, Huaiyu and Wu, Meng-Ru and Bhattacharyya, Soumya and Zaizen, Masamichi and George, Manu and Lin, Chun-Yu and Xiong, Zewei",
    title = "{Code comparison for fast flavor instability simulations}",
    eprint = "2205.06282",
    archivePrefix = "arXiv",
    primaryClass = "astro-ph.HE",
    reportNumber = "N3AS-22-009",
    doi = "10.1103/PhysRevD.106.043011",
    journal = "Phys. Rev. D",
    volume = "106",
    number = "4",
    pages = "043011",
    year = "2022"
}

@article{Grohs:2023pgq,
    author = "Grohs, Evan and Richers, Sherwood and Couch, Sean M. and Foucart, Francois and Froustey, Julien and Kneller, James P. and McLaughlin, Gail C.",
    title = "{Two-moment Neutrino Flavor Transformation with Applications to the Fast Flavor Instability in Neutron Star Mergers}",
    eprint = "2309.00972",
    archivePrefix = "arXiv",
    primaryClass = "astro-ph.HE",
    reportNumber = "N3AS-23-029",
    doi = "10.3847/1538-4357/ad13f2",
    journal = "Astrophys. J.",
    volume = "963",
    number = "1",
    pages = "11",
    year = "2024"
}

@article{Padilla-Gay:2025tko,
    author = "Padilla-Gay, Ian and Chen, Heng-Hao and Abbar, Sajad and Wu, Meng-Ru and Xiong, Zewei",
    title = "{Flavor equilibration of supernova neutrinos: Exploring the dynamics of slow modes}",
    eprint = "2505.11588",
    archivePrefix = "arXiv",
    primaryClass = "astro-ph.HE",
    reportNumber = "SLAC-PUB-250421",
    doi = "10.1103/jg14-8p4l",
    journal = "Phys. Rev. D",
    volume = "112",
    number = "4",
    pages = "043039",
    year = "2025"
}

@article{OConnor:2014sgn,
    author = "O'Connor, Evan",
    title = "{An Open-Source Neutrino Radiation Hydrodynamics Code for Core-Collapse Supernovae}",
    eprint = "1411.7058",
    archivePrefix = "arXiv",
    primaryClass = "astro-ph.HE",
    doi = "10.1088/0067-0049/219/2/24",
    journal = "Astrophys. J. Suppl.",
    volume = "219",
    number = "2",
    pages = "24",
    year = "2015"
}

@article{Steiner:2012rk,
    author = "Steiner, Andrew W. and Hempel, Matthias and Fischer, Tobias",
    title = "{Core-collapse supernova equations of state based on neutron star observations}",
    eprint = "1207.2184",
    archivePrefix = "arXiv",
    primaryClass = "astro-ph.SR",
    reportNumber = "INT-PUB-12-033",
    doi = "10.1088/0004-637X/774/1/17",
    journal = "Astrophys. J.",
    volume = "774",
    pages = "17",
    year = "2013"
}

@article{particle2020review,
    author = "Zyla, P. A. and others",
    collaboration = "Particle Data Group",
    title = "{Review of Particle Physics}",
    doi = "10.1093/ptep/ptaa104",
    journal = "PTEP",
    volume = "2020",
    number = "8",
    pages = "083C01",
    year = "2020"
}

@article{de2021igniting,
    author = "De, Soumi and Siegel, Daniel M.",
    title = "{Igniting Weak Interactions in Neutron Star Postmerger Accretion Disks}",
    eprint = "2011.07176",
    archivePrefix = "arXiv",
    primaryClass = "astro-ph.HE",
    reportNumber = "LA-UR-20-28607",
    doi = "10.3847/1538-4357/ac110b",
    journal = "Astrophys. J.",
    volume = "921",
    number = "1",
    pages = "94",
    year = "2021"
}

@article{ng2025accurate,
    author = "Ng, Harry Ho-Yin and Musolino, Carlo and Tootle, Samuel D. and Rezzolla, Luciano",
    title = "{Accurate Muonic Interactions in Neutron Star Mergers and Impact on Heavy-element Nucleosynthesis}",
    eprint = "2411.19178",
    archivePrefix = "arXiv",
    primaryClass = "astro-ph.HE",
    doi = "10.3847/2041-8213/add324",
    journal = "Astrophys. J. Lett.",
    volume = "985",
    number = "2",
    pages = "L36",
    year = "2025"
}

@article{gieg2025role,
    author = "Gieg, Henrique and Schianchi, Federico and Ujevic, Maximiliano and Dietrich, Tim",
    title = "{Role of muons in binary neutron star mergers: First simulations}",
    eprint = "2409.04420",
    archivePrefix = "arXiv",
    primaryClass = "gr-qc",
    doi = "10.1103/52ph-53yw",
    journal = "Phys. Rev. D",
    volume = "112",
    number = "2",
    pages = "023036",
    year = "2025"
}

@article{volpe2015neutrino,
    author = "Volpe, Cristina",
    title = "{Neutrino Quantum Kinetic Equations}",
    eprint = "1506.06222",
    archivePrefix = "arXiv",
    primaryClass = "astro-ph.SR",
    doi = "10.1142/S0218301315410098",
    journal = "Int. J. Mod. Phys. E",
    volume = "24",
    number = "09",
    pages = "1541009",
    year = "2015"
}

@article{blaschke2016neutrino,
    author = "Blaschke, Daniel N. and Cirigliano, Vincenzo",
    title = "{Neutrino Quantum Kinetic Equations: The Collision Term}",
    eprint = "1605.09383",
    archivePrefix = "arXiv",
    primaryClass = "hep-ph",
    reportNumber = "LA-UR-15-25029",
    doi = "10.1103/PhysRevD.94.033009",
    journal = "Phys. Rev. D",
    volume = "94",
    number = "3",
    pages = "033009",
    year = "2016"
}

@article{qianfuller,
    author = "Qian, Yong Zhong and Fuller, George M.",
    title = "{Neutrino-neutrino scattering and matter enhanced neutrino flavor transformation in Supernovae}",
    eprint = "astro-ph/9406073",
    archivePrefix = "arXiv",
    reportNumber = "DOE-ER-40561-150, INT-94-00-63",
    doi = "10.1103/PhysRevD.51.1479",
    journal = "Phys. Rev. D",
    volume = "51",
    pages = "1479--1494",
    year = "1995"
}

@article{PhysRevD.110.083028,
    author = "Foucart, Francois and Cheong, Patrick Chi-Kit and Duez, Matthew D. and Kidder, Lawrence E. and Pfeiffer, Harald P. and Scheel, Mark A.",
    title = "{Robustness of neutron star merger simulations to changes in neutrino transport and neutrino-matter interactions}",
    eprint = "2407.15989",
    archivePrefix = "arXiv",
    primaryClass = "astro-ph.HE",
    doi = "10.1103/PhysRevD.110.083028",
    journal = "Phys. Rev. D",
    volume = "110",
    number = "8",
    pages = "083028",
    year = "2024"
}

@article{PhysRevD.94.123016,
    author = "Foucart, Francois and O'Connor, Evan and Roberts, Luke and Kidder, Lawrence E. and Pfeiffer, Harald P. and Scheel, Mark A.",
    title = "{Impact of an improved neutrino energy estimate on outflows in neutron star merger simulations}",
    eprint = "1607.07450",
    archivePrefix = "arXiv",
    primaryClass = "astro-ph.HE",
    doi = "10.1103/PhysRevD.94.123016",
    journal = "Phys. Rev. D",
    volume = "94",
    number = "12",
    pages = "123016",
    year = "2016"
}

@article{Banerjee:2011fj,
    author = "Banerjee, Arka and Dighe, Amol and Raffelt, Georg",
    title = "{Linearized flavor-stability analysis of dense neutrino streams}",
    eprint = "1107.2308",
    archivePrefix = "arXiv",
    primaryClass = "hep-ph",
    reportNumber = "MPP-2011-81, TIFR-TH-11-30",
    doi = "10.1103/PhysRevD.84.053013",
    journal = "Phys. Rev. D",
    volume = "84",
    pages = "053013",
    year = "2011"
}

@article{Izaguirre:2016gsx,
    author = "Izaguirre, Ignacio and Raffelt, Georg and Tamborra, Irene",
    title = "{Fast Pairwise Conversion of Supernova Neutrinos: A Dispersion-Relation Approach}",
    eprint = "1610.01612",
    archivePrefix = "arXiv",
    primaryClass = "hep-ph",
    reportNumber = "INT-PUB-16-023, MPP-2016-266",
    doi = "10.1103/PhysRevLett.118.021101",
    journal = "Phys. Rev. Lett.",
    volume = "118",
    number = "2",
    pages = "021101",
    year = "2017"
}

@article{Bhattacharyya:2021klg,
    author = "Bhattacharyya, Soumya and Dasgupta, Basudeb",
    title = "{Fast flavor oscillations of astrophysical neutrinos with 1, 2, {\textellipsis}, {\ensuremath{\infty}} crossings}",
    eprint = "2101.01226",
    archivePrefix = "arXiv",
    primaryClass = "hep-ph",
    reportNumber = "TIFR/TH/20-47",
    doi = "10.1088/1475-7516/2021/07/023",
    journal = "JCAP",
    volume = "07",
    pages = "023",
    year = "2021"
}

@article{takahashi2026dynamic,
    author = "Takahashi, Shota and Nagakura, Hiroki and Zaizen, Masamichi and Kato, Chinami and Liu, Jiabao",
    title = "{Dynamic competition of fast and collisional neutrino flavor instabilities with collisional damping in spatially inhomogeneous systems}",
    eprint = "2605.10435",
    archivePrefix = "arXiv",
    primaryClass = "astro-ph.HE",
    doi = "10.1103/7c32-rfjb",
    journal = "Phys. Rev. D",
    volume = "114",
    number = "4",
    pages = "043059",
    year = "2026"
}

@article{PhysRevD.80.053002,
  title = {Three flavor neutrino oscillations in matter: Flavor diagonal potentials, the adiabatic basis, and the $CP$ phase},
  author = {Kneller, James P. and McLaughlin, Gail C.},
  journal = {Phys. Rev. D},
  volume = {80},
  issue = {5},
  pages = {053002},
  numpages = {18},
  year = {2009},
  month = {Sep},
  publisher = {American Physical Society},
  doi = {10.1103/PhysRevD.80.053002},
  url = {https://link.aps.org/doi/10.1103/PhysRevD.80.053002}
}

@article{XING2019114803,
title = {Sum rules and asymptotic behaviors of neutrino mixing in dense matter},
journal = {Nuclear Physics B},
volume = {949},
pages = {114803},
year = {2019},
issn = {0550-3213},
doi = {https://doi.org/10.1016/j.nuclphysb.2019.114803},
url = {https://www.sciencedirect.com/science/article/pii/S0550321319302895},
author = {Zhi-zhong Xing and Jing-yu Zhu}
}

@article{BLENNOW2005330,
title = {Effective neutrino mixing and oscillations in dense matter},
journal = {Physics Letters B},
volume = {609},
number = {3},
pages = {330-338},
year = {2005},
issn = {0370-2693},
doi = {https://doi.org/10.1016/j.physletb.2005.01.063},
url = {https://www.sciencedirect.com/science/article/pii/S0370269305001292},
author = {Mattias Blennow and Tommy Ohlsson}
}

\end{document}